\documentclass[universe,article,accept,pdftex,moreauthors]{Definitions/mdpi} 
\firstpage{1} 
\pubvolume{11}
\issuenum{10}
\articlenumber{325}
\pubyear{2025}
\copyrightyear{2025}
\externaleditor{João Luís de Figueiredo Rosa and Lorenzo Iorio} 
\datereceived{11 July 2025} 
\daterevised{10 September 2025} 
\dateaccepted{22 September 2025} 
\datepublished{24 September 2025 } 
\hreflink{https://doi.org/ \linebreak 10.3390/universe11100325} 

\newcommand{\nn}{\nonumber}
\usepackage[normalem]{ulem}

\def\highlighting{} 

\Title{Quadrupole Perturbations of Slowly Spinning \linebreak  Ellis--Bronnikov 
	Wormholes}

\TitleCitation{Quadrupole Perturbations of Slowly Spinning Ellis--Bronnikov Wormholes}

\Author{{Bahareh Azad} 
 $^{1}$\orcidA{}, Jose Luis Bl\'azquez-Salcedo $^{2}$\orcidB{}, Fech Scen Khoo $^{2,}$*\orcidC{}, Jutta Kunz $^{1}$\orcidD{} \linebreak  and Francisco Navarro-L\'erida $^{2}$\orcidE{}}

\AuthorNames{Bahareh Azad, Jose Luis Bl\'azquez-Salcedo, Fech Scen Khoo, Jutta Kunz and Francisco Navarro-L\'erida}

         {%
        {
        \AuthorCitation{Azad, B.; Bl\'azquez-\linebreak  Salcedo, J.L.; Khoo, F.S.; Kunz, J.; Navarro-L\'erida, F.}
        }
}

\address{%
$^{1}$ \quad Institut f\"ur  Physik, Universit\"at Oldenburg, Postfach 2503,
D-26111 Oldenburg, Germany; {bahareh.azad@uni-oldenburg.de (B.A.); jutta.kunz@uni-oldenburg.de (J.K.)}
 \\
$^{2}$ \quad Departamento de F\'isica Te\'orica and IPARCOS, Facultad de Ciencias F\'isicas, Universidad Complutense de Madrid, 28040 Madrid, Spain; {jlblaz01@ucm.es (J.L.B.-S.); fnavarro@ucm.es (F.N.-L.)}}

\corres{Correspondence: fkhoo@ucm.es 
}

\abstract{We study the axial and polar perturbations of slowly rotating Ellis--Bronnikov wormholes in General Relativity, applying a perturbative double expansion.
In particular, we derive the equations for $l=2$, $M_z=2$ perturbations of these objects,
which are parametrized by an asymmetry parameter. The equations constitute an astrophysically
interesting sector of the perturbations that contribute dominantly to the gravitational wave
radiation.
Moreover, calculation of these modes may exhibit potential instabilities in the quadrupole sector.}


\begin{document}


\section{Introduction}

The detection of gravitational waves from the inspiral, merger, and ringdown of compact objects provides an excellent tool to learn about both gravitation in the strong gravity regime and the compact objects themselves
\cite{LIGOScientific:2016aoc,LIGOScientific:2017vwq,KAGRA:2013rdx,Cahillane:2022pqm,Punturo:2010zz,Dwyer:2015,Colpi:2024xhw}.
Besides black holes and neutron stars, a variety of further compact objects are being discussed as hypothetical astrophysical objects, whose signatures might be observable.
Many of these objects feature as {{black hole mimickers}
}, making their study an interesting endeavor \cite{Bambi:2025wjx}.

One type of {black hole mimicker} is represented by wormholes \cite{Damour:2007ap}.
In General Relativity (GR), wormholes need exotic matter 
\cite{Morris:1988cz,Visser:1995cc,Lobo:2017,Ellis:1973yv,Bronnikov:1973fh,Ellis:1979bh}
or quantum matter \cite{Blazquez-Salcedo:2020czn,Konoplya:2021hsm,Blazquez-Salcedo:2021udn} for their support.
In contrast, in alternative theories of gravity, the energy conditions may also be violated by the effective stress--energy tensor arising from the modified gravitational interaction~\cite{Lobo:2017,Kanti:2011jz,Kanti:2011yv,Antoniou:2019awm,Bakopoulos:2021liw,Naseer:2025flw}.

Numerous potentially observable features of wormholes have already been discussed (see e.g.,~gravitational lensing by wormholes 
\cite{Cramer:1994qj,Safonova:2001vz,Perlick:2003vg,Nandi:2006ds,Abe:2010ap,Toki:2011zu,Nakajima:2012pu,Tsukamoto:2012xs,Kuhfittig:2013hva,Bambi:2013nla,Takahashi:2013jqa,Tsukamoto:2016zdu},
shadows of wormholes \cite{Bambi:2013nla,Nedkova:2013msa,Ohgami:2015nra,Shaikh:2018kfv,Gyulchev:2018fmd,Bouhmadi-Lopez:2021zwt,Guerrero:2022qkh,Huang:2023yqd},
or accretion disks surrounding wormholes
\cite{Harko:2008vy,Harko:2009xf,Bambi:2013jda,Zhou:2016koy,Lamy:2018zvj,Deligianni:2021ecz,Deligianni:2021hwt,Abdulkhamidov:2024lvp}).
Concerning the ringdown of wormholes, so far, mainly quasinormal modes have been studied for static wormholes  \cite{Konoplya:2005et,Kim:2008zzj,Konoplya:2010kv,Konoplya:2016hmd,Volkel:2018hwb,Aneesh:2018hlp,Konoplya:2018ala,Blazquez-Salcedo:2018ipc,Konoplya:2019hml,Churilova:2019qph,Jusufi:2020mmy,Bronnikov:2021liv,Gonzalez:2022ote,Azad:2022qqn,Maji:2025qpp},
and only recently has the exploration of quasinormal modes of rotating wormholes begun~\cite{Khoo:2024yeh}.
Besides those quasinormal modes, echoes of Kerr-like wormholes have also been addressed~\cite{Bueno:2017hyj}.

Here we focus on the ringdown of slowly rotating wormholes, choosing Ellis--Bronnikov wormholes \cite{Ellis:1973yv,Bronnikov:1973fh} as the background solutions.
Slowly rotating Ellis--Bronnikov wormholes have been constructed perturbatively up to second order in rotation in closed form {\highlighting{\cite{Kashargin:2007mm,Kashargin:2008pk,Azad:2023iju,Azad:2024axu}.} 
} 
For rapid rotation, no closed-form solutions have been obtained so far~\cite{Volkov:2021blw}, while, numerically, some sets of solutions are available \cite{Kleihaus:2014dla,Chew:2016epf}.
The analysis of the quasinormal modes in the slow rotation case usually proves rather valuable, since on the one hand, important features of the presence of rotation arise already, while on the other hand, the modes provide a crucial limiting test for the rapidly rotating case (see e.g.,~\cite{Blazquez-Salcedo:2024oek,Khoo:2025qjc}).

In this paper, we derive the complete set of quadrupole perturbation equations for the quasinormal modes of second-order Ellis--Bronnikov background solutions. 
In \mbox{{Section} 
 \ref{theory-bckg}}, we present the theoretical settings together with the background solutions. 
Section \ref{sec3} provides the general ansatz for the perturbations and discusses the derivation in the set of perturbation equations.
The derivation makes large use of previous work for quasinormal modes of slowly rotating black holes \cite{Blazquez-Salcedo:2022eik}, which showed that the exact quasinormal mode spectrum was approximated with rather good precision at least up to 50\% of the extremal angular momentum.
We conclude in Section \ref{sec4} and provide some of the lengthy equations in the {Appendix} 
 \ref{appA}.

\section{Theoretical Settings}
\label{theory-bckg}

Ellis--Bronnikov wormholes are based on the action 
\begin{eqnarray}
			S[g,\Phi] &=& \frac{1}{16 \pi G}\int d^4x \sqrt{-g} 
		\Big[\mathrm{R} + 2 \partial_\mu \Phi \, \partial^\mu \Phi 
		 \Big] \, ,
   \label{eq:ellis} 
\end{eqnarray}
where a phantom scalar field $\Phi$ is coupled minimally to GR.
Thus, $\mathrm{R}$ is the curvature scalar, $G$ is Newton's constant, and the kinetic term of the scalar field has the sign reversed as compared to an ordinary scalar field.
Variation of the action leads to the field equations of the theory
\begin{eqnarray}
    \mathrm{R}_{\mu\nu} = -2 \partial_{\mu}\Phi\partial_{\nu}\Phi \, , \quad 
    \nabla_{\mu} \nabla^{\mu}\Phi = 0 \, .
    \label{eqs}
\end{eqnarray}

The resulting Ellis--Bronnikov wormhole solutions are well-known static spherically symmetric solutions that can be expressed in the form
\begin{eqnarray}
   ds^2 &=& g_{\mu\nu}
   dx^{\mu} dx^{\nu} = 
   -e^f dt^2+ 
   {e^{-f}}\Big[dr^2+\left(r^2+\mathit{r_0}^2\right)\left(d\theta^2+\sin^2\theta d\varphi^2\right)\Big], \\
   \Phi(r) &=& \frac{Q_0 f(r)}{C} \,
   \label{Phi_static}
\end{eqnarray}
with
  \begin{equation} 
  f=\frac{C}{\mathit{r_0}}\Big[\tan^{-1}\left(\frac{r}{\mathit{r_0}}\right)-\frac{\pi}{2}\Big] \,.
  \end{equation}
{This}
 solution contains several parameters: $C$ is related to the symmetry of the wormholes with respect to reflections $r \to - r$.
Since only $C=0$ leads to a symmetric wormhole, $C$ is referred to as the asymmetry parameter.
The mass $M_0$ at (radial) plus infinity of the static wormholes is also given in terms of the asymmetry parameter $C$, $M_0 = C/2$; the symmetric wormholes thus have vanishing mass, $M_0=0$.
The corresponding scalar charge $Q_0$ of the static wormholes is given by $Q_{0} = \sqrt{C^2/4 + \mathit{r_0}^2}$.
Here, $\mathit{r_0}$ is a free parameter that determines the size of the throat of symmetric wormholes.
Note that for $C=0$, the scalar charge is simply given by this free parameter.
The circumferential radius ${\cal R}$ is obtained from $g_{\phi\phi}$,
\begin{equation}
    {\cal R}^2 = e^{-f} (r^2 + \mathit{r_0}^2) \, .
    \label{circ}
\end{equation}
Clearly, for symmetric wormholes, it reaches its minimum for $r=0$.
However, for asymmetric wormholes, the minimum, and thus the throat, is located at $r=C/2$.
The area $A$ of the throat is therefore given by 
\begin{equation}
  \left.  A= 4 \pi {\cal R}^2 \right|_{r=C/2} \, .
\end{equation}
Note that this expression corrects the expression for the area given in \cite{Azad:2023iju}.

\section{Perturbative Background Solutions}
\label{sec3}

We now briefly recall the background solutions which are given up to a second order in rotation by \cite{Hartle:1967he,Hartle:1968si,Azad:2023iju,Azad:2024axu}
\begin{eqnarray}
ds^2 &=& -e^{f}\left[1+\epsilon_r^2 \,2\left(h_0(r)+h_2(r)P_2(\theta)\right)\right]dt^2 
+ e^{-f}\left[1+\epsilon_r^2 \, 2\left(b_0(r)+b_2(r)P_2(\theta)\right)\right]dr^2 \nonumber \\ 
&+& 
{e^{-f}} R^2
\left[1+\epsilon_r^2 \, 2\left(k_0(r)+k_2(r)P_2(\theta)\right)\right]
\times 
\left[
d\theta^2+\sin^2{(\theta)}\left[d\varphi-\epsilon_r \, w(r)dt\right]^2 
\right]
\ ,
\label{metric_1}
\end{eqnarray}
where 
$R^2=(r^2+\mathit{r_0}^2)$, 
and  $P_2(\theta) = \left(3\cos^2{(\theta)}-1\right)/2$ denotes the Legendre polynomial.
With the parameter $\epsilon_r \ll 1 $,
we keep track of the order of the slow rotation perturbation contributions.
Up to 
second order, we need to introduce seven radial perturbation functions,
$ h_0, h_2, b_0, b_2, k_0, k_2, w$, that can, in general, be reduced to six by choosing a gauge $k_0(r)=0$ and redefining $k_2=h_2-\nu_2$.
At the same time, the phantom field is given by 
\begin{eqnarray}
\Phi = \phi(r) + \epsilon_r^2 \left(\phi_{20}(r)+\phi_{22}(r)P_2(\theta)\right) \ ,
\label{scalar_1}
\end{eqnarray}
for the second order in rotation, with the two perturbation functions $\phi_{20}, \phi_{22}$. 
$\phi(r)$ is the static background scalar given by Equation (\ref{Phi_static}).
Thus, together there are eight unknown functions,
$w, h_0, h_2, b_0, b_2, \nu_2, \phi_{20}, \phi_{22}$,
that have to be determined by solving the field equations to obtain the desired background metric.

\begin{figure}[H]
	
\includegraphics[width=0.3\textwidth,angle=-90]{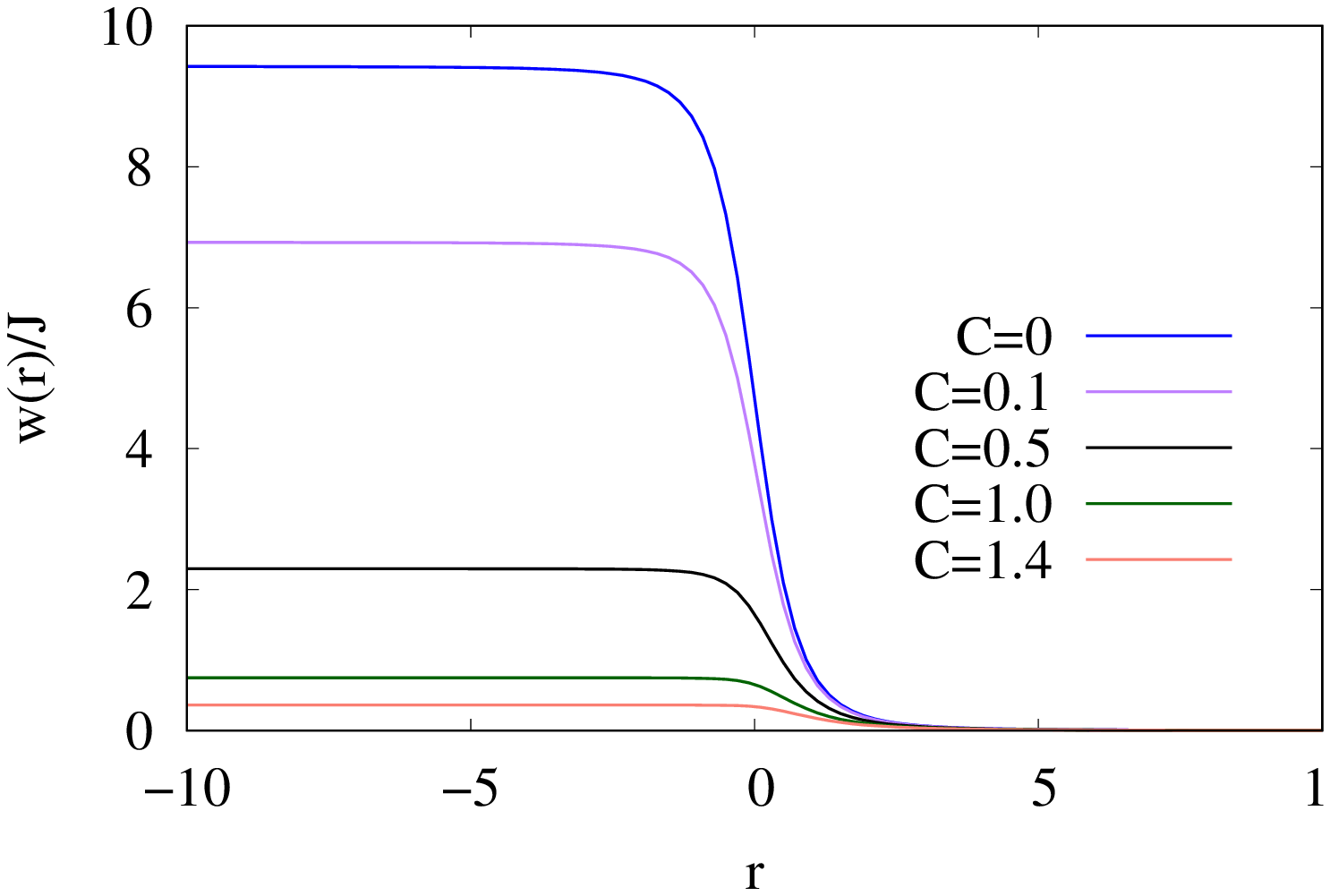}
\includegraphics[width=0.3\textwidth,angle=-90]{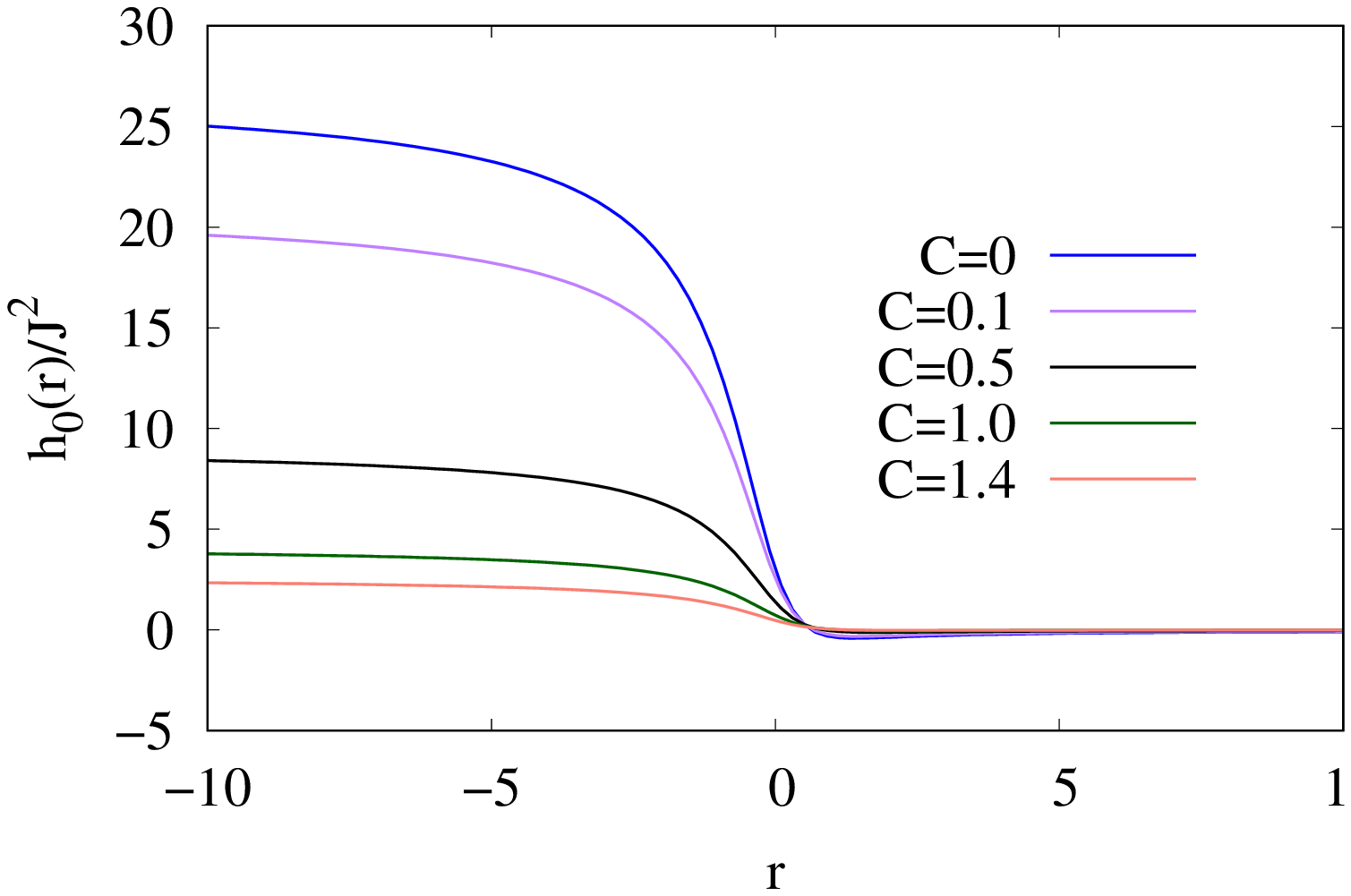}\\
\includegraphics[width=0.3\textwidth,angle=-90]{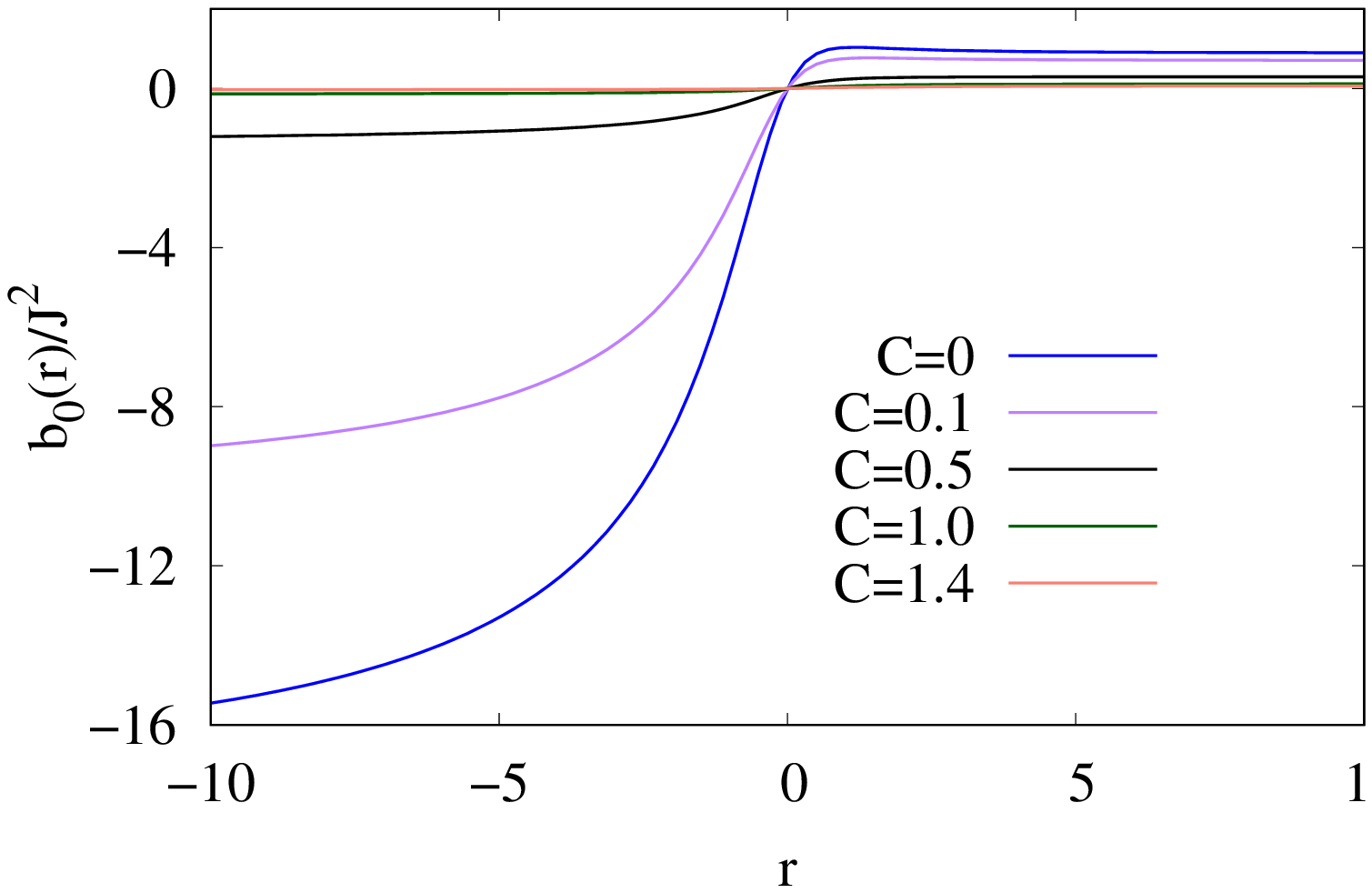}
\includegraphics[width=0.3\textwidth,angle=-90]{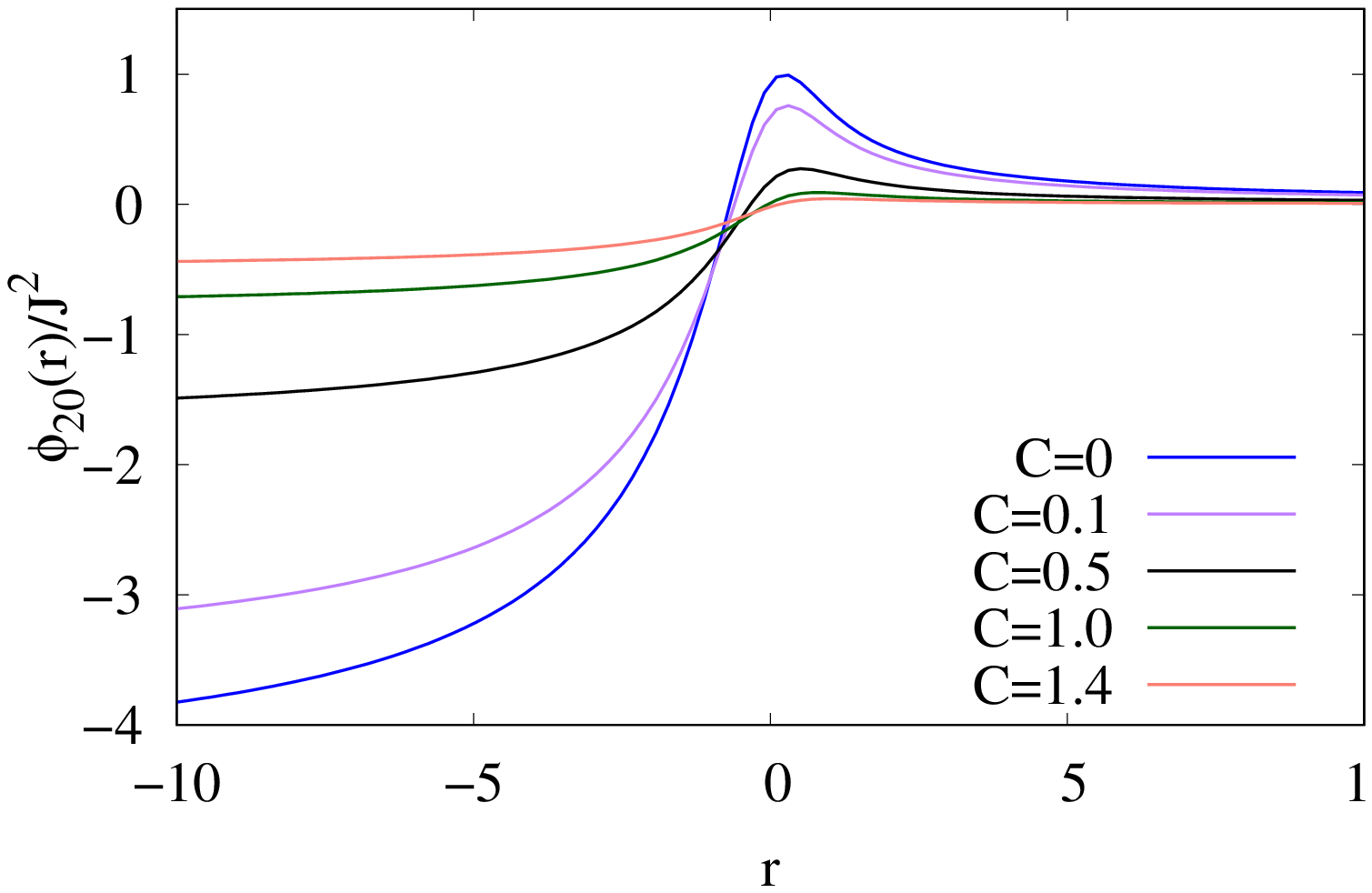}\\
\includegraphics[width=0.3\textwidth,angle=-90]{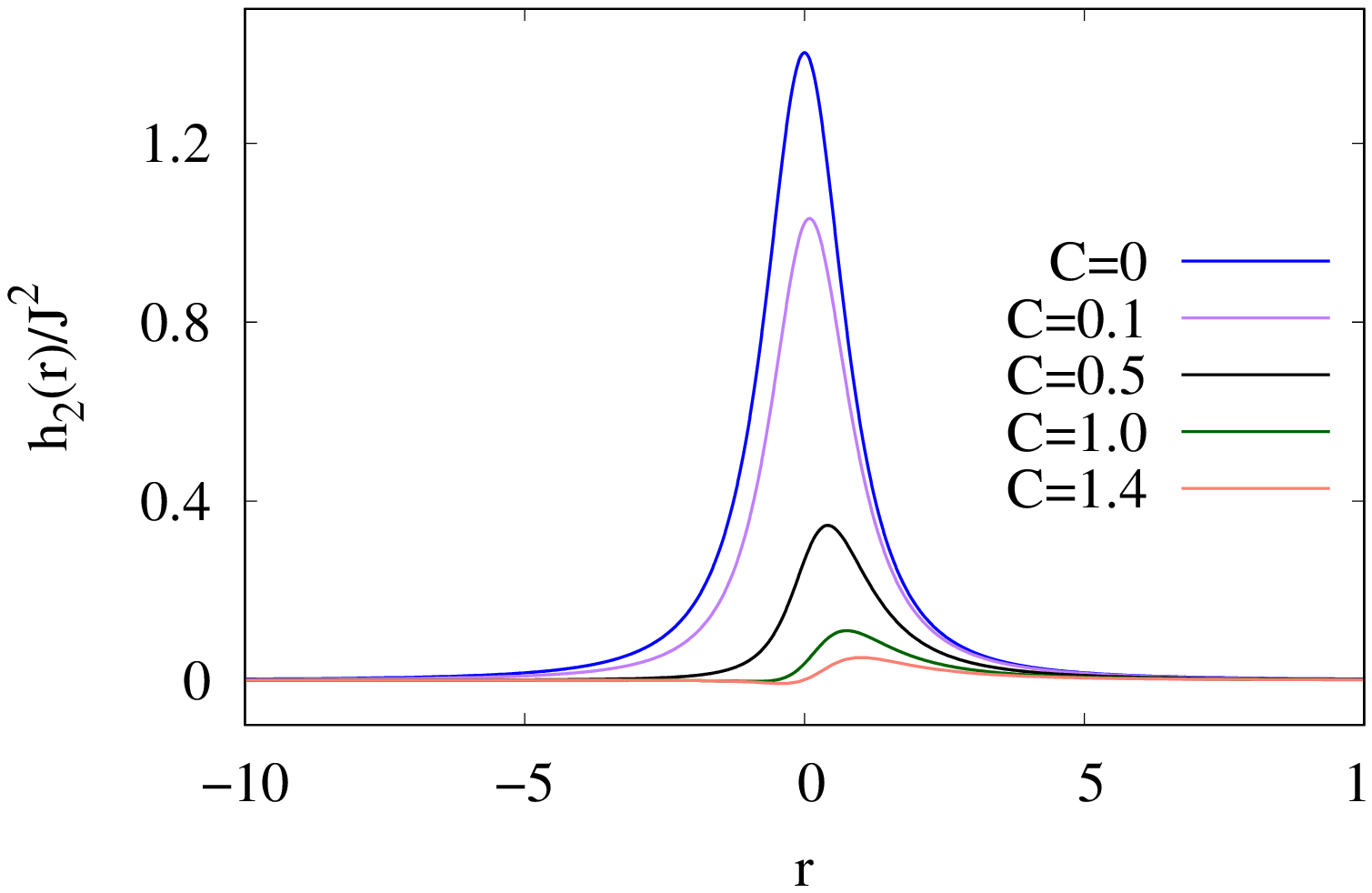}
\includegraphics[width=0.3\textwidth,angle=-90]{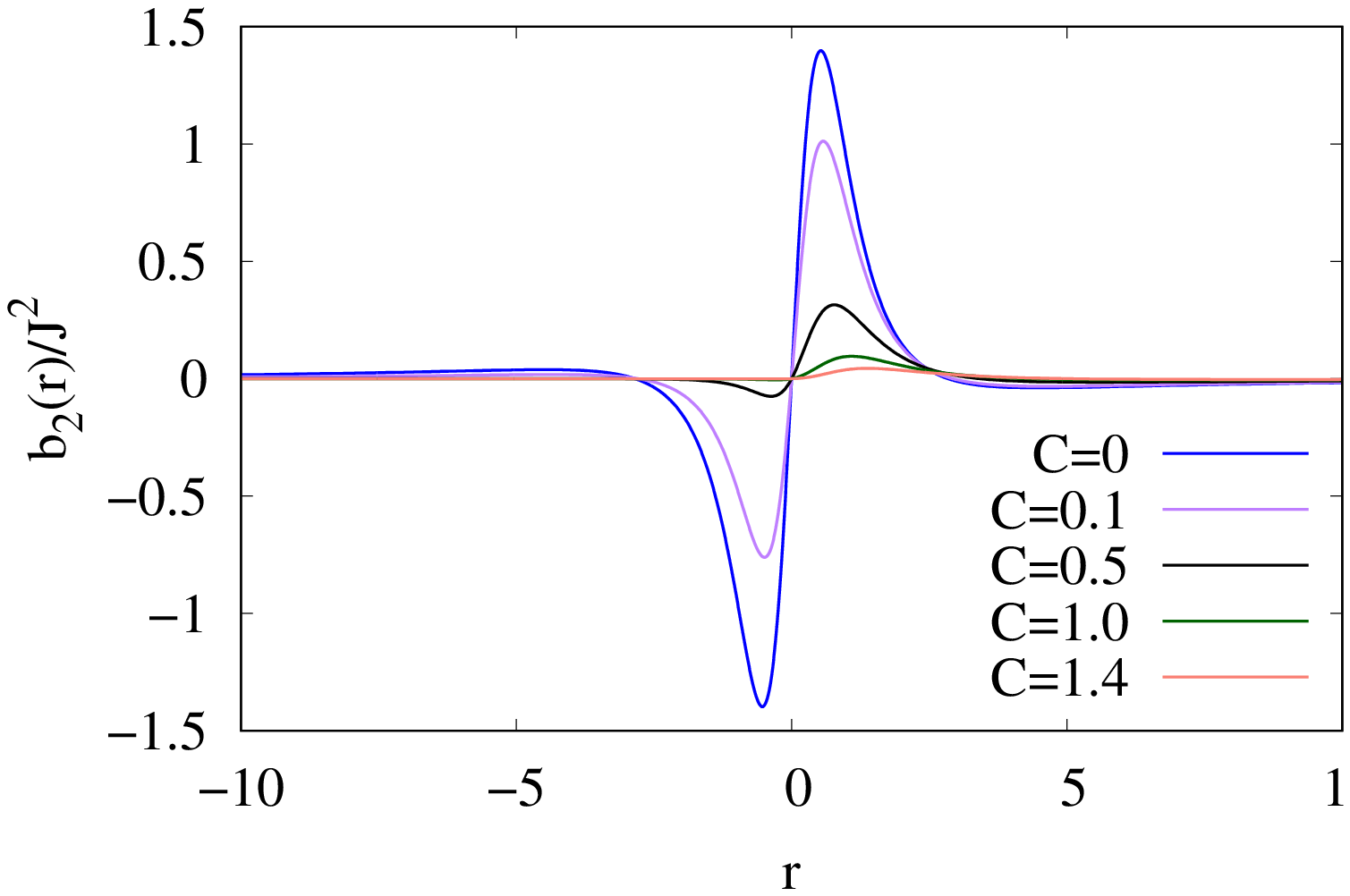}\\
\includegraphics[width=0.3\textwidth,angle=-90]{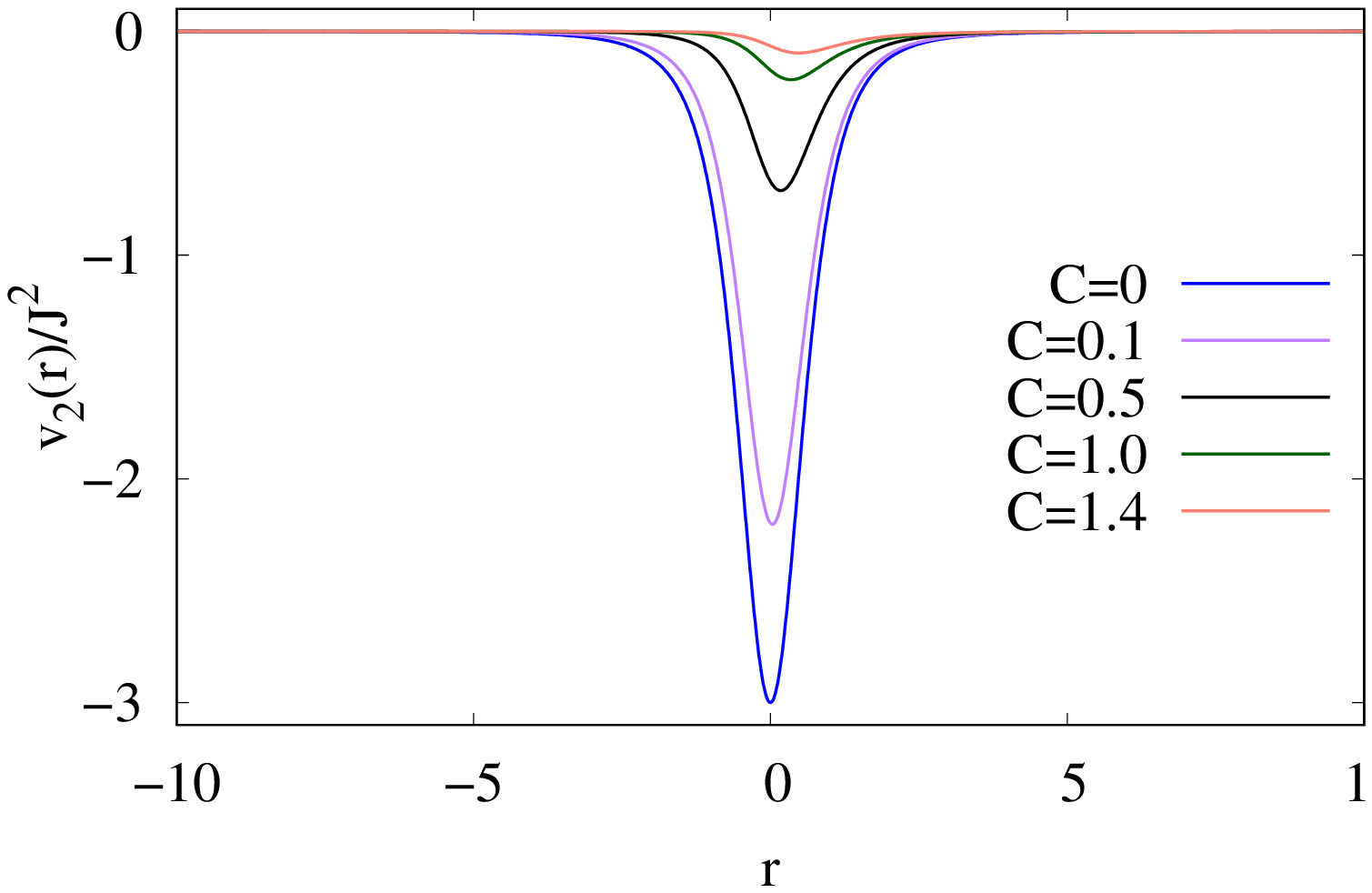}
\includegraphics[width=0.3\textwidth,angle=-90]{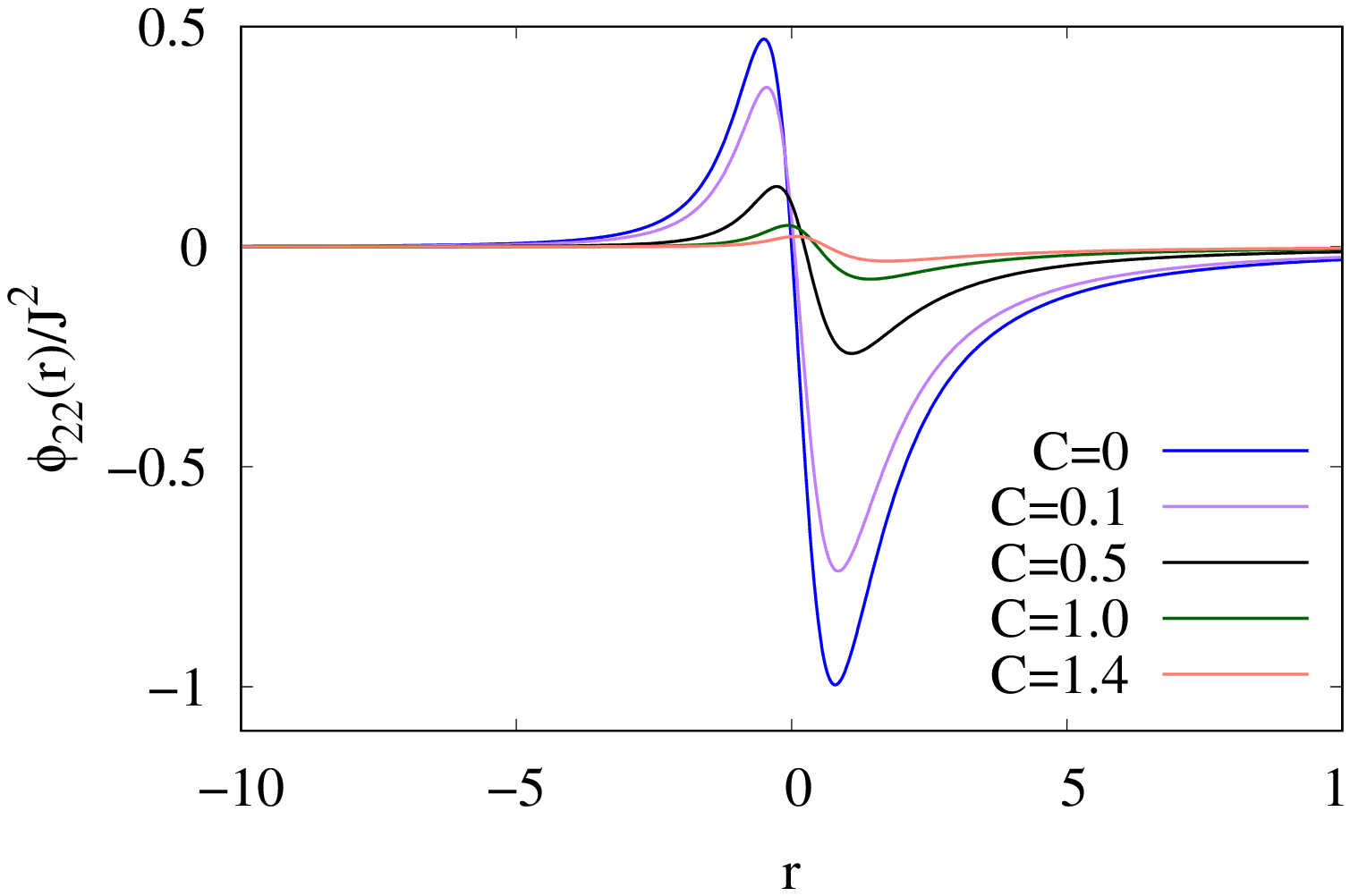}
  \caption{Scaled background perturbation functions vs. the radial coordinate $r$ for several values of the asymmetry parameter $C$ ($\mathit{r_0}=1$). 
  The scaling parameter is the angular momentum $J$.} 
		\label{fig1}
	\end{figure}

For the first order we obtain the function $w(r)$ for the slowly rotating background~\mbox{\cite{Azad:2023iju,Azad:2024axu}},
\begin{eqnarray}
    w(r)= \frac{3J}{2C(C^2+\mathit{r_0}^2)}
    \left[
    1 - \left(
    1+2C \, \frac{C+r}{R^2}
    \right)e^{2f}
    \right] \, ,
\end{eqnarray}
where $J$ denotes the angular momentum of the wormhole.
The remaining seven functions $h_0, h_2, b_0, b_2, \nu_2, \phi_{20}, \phi_{22}$ decouple into two sets of functions, $\mathcal{P}_0 = 
\{h_0,b_0,\phi_{20}\}$ and \linebreak  $\mathcal{P}_2 = \{h_2,\nu_2,b_2,\phi_{22}\}$.
Closed-form expressions for these functions can be found in \cite{Azad:2023iju,Azad:2024axu}.
We exhibit all eight functions in Figure~\ref{fig1} for several values of the asymmetry parameter $C$.

The mass of the wormholes at plus infinity is then given in the second order in rotation by $M=M_0+\Delta M$, where the mass correction $\Delta M$ is extracted from the solution for $b_0(r)$,
\begin{eqnarray}
    \Delta M &=&  
    \frac{3 J^{2} }{\left(\mathit{C}^{2}+4 \mathit{r_0}^{2}\right) \left(\mathit{C }^{2}+\mathit{r_0}^{2}\right)^{2} \left(\left(\mathit{C}^{2}+4 \mathit{r_0}^{2}\right) \mathrm{cot}^{-1}\! \left(\frac{\mathit{C}}{2 \mathit{r_0}}\right)-2 \mathit{C} \mathit{r_0} \right) \mathit{C}^{2}} \times 
    \nn \\ &&
    \Bigg(\left(17 \mathit{C}^{4} \mathit{r_0}^{3}-2 \mathit{C}^{2} \mathit{r_0}^{5}+8 \mathit{r_0}^{7}\right) {\mathrm e}^{-\frac{2 \mathit{C} \,\mathrm{cot}^{-1}\left(\frac{\mathit{C}}{2 \mathit{r_0}}\right)}{\mathit{r_0}}}
    +\left(\mathit{C}^{2}+4 \mathit{r_0}^{2}\right) \times
    \nn \\ &&
    \Big(\left(\mathit{C}^{5}+5 \mathit{C}^{3} \mathit{r_0}^{2}+4 \mathit{C} \,\mathit{r_0}^{4}\right) \mathrm{cot}^{-1}\! \left(\frac{\mathit{C}}{2 \mathit{r_0}}\right)-2 \mathit{C}^{4} \mathit{r_0} -7 \mathit{r_0}^{3} \mathit{C}^{2}-2 \mathit{r_0}^{5}\Big)\Bigg) \, ,
    \label{Delta_M}
\end{eqnarray}
which
reduces for $C=0$ to
\begin{eqnarray}
\Delta M =3J^2(\pi^2-8)/(2\pi \mathit{r_0}^3) \, ,
\label{Delta_M_C0}
\end{eqnarray}
and is read off at plus infinity.
Thus the mass $M$ of the rotating wormhole is non-vanishing for $C=0$,
i.e., $M|_{C=0} \, = \Delta M$
(see also \cite{Kashargin:2008pk}).
Analogously, to the mass $M$, the scalar charge $Q$ also consists of the static charge $Q_0$ and the correction term $\Delta Q$,
\begin{equation}
Q = Q_0 + \epsilon_r^2\Delta Q \, ,
\label{charge}
\end{equation}
where $\Delta Q=- \Delta M$.

\section{Quadrupole Perturbations}
\label{sec4}

Having established the background solutions $\{g^{(sr)}, {\Phi}^{(sr)}\}$ up to second order in rotation, we now perturb up to the first order in $\epsilon_q$ the metric field
\begin{eqnarray}
g_{\mu\nu} &=& g^{(sr)}_{\mu\nu} + \epsilon_q 
\delta h_{\mu\nu}(t,r,\theta,{\varphi}) \nonumber \\
 &=& g^{(sr)}_{\mu\nu} + \epsilon_q 
\left( \delta h^{(A)}_{\mu\nu}(t,r,\theta,{\varphi}) 
    +  \delta h^{(P)}_{\mu\nu}(t,r,\theta,{\varphi}) \right)
\, , 
\label{per_m}
\end{eqnarray}
and the phantom scalar field
\begin{eqnarray}
\Phi &=& {\Phi}^{(sr)} + \epsilon_q  \delta{\phi}^{(P)} (t,r,\theta,{\varphi})
\label{per_p}
\, .
\end{eqnarray}
The metric field perturbations consist of both the axial-led ($A$) and polar-led ($P$) perturbations, while the scalar field contributes solely to the polar perturbations.

In the static limit, the metric perturbations are given by
\begin{eqnarray}
\label{met_pert}
\delta h_{\mu\nu}= 
e^{i\omega t}
    \begin{pmatrix}
-e^{f}NY & -\mathit{H1} \, Y    & -\mathit{h0} \, {\partial_{\varphi}Y}/{s_{\theta}} 
& \mathit{h0} \, s_{\theta}\partial_{\theta}Y  \\
-\mathit{H1} \, Y   & -e^{-f}LY & -\mathit{h1} \, \partial_{\varphi}Y/s_{\theta} & \mathit{h1} \, s_{\theta}\partial_{\theta}Y \\
-\mathit{h0} \, {\partial_{\varphi}Y}/s_{\theta} & 
-\mathit{h1} \, \partial_{\varphi}Y/s_{\theta}
& R^2 TY & 0 \\
\mathit{h0} \, s_{\theta}\partial_{\theta}Y & \mathit{h1} \, s_{\theta}\partial_{\theta}Y
& 0 &R^2 T s_{\theta}^2 Y
\end{pmatrix} \, ,
\label{an_met}
\end{eqnarray}
where 
$Y$ denotes the spherical harmonics, $s_{\theta}=\sin{\theta}$, $R^2=r^2+\mathit{r_0}^{2}$,
and $\omega$ is an eigenvalue.
The axial metric perturbations are given by
$\mathit{h0}(r)$ and $ \mathit{h1}(r)$,
while the polar metric perturbations are 
$T(r),L(r),N(r)$ and $\mathit{H1}(r)$.
The scalar perturbations are decomposed as
\begin{eqnarray}
   \delta{\phi} (t,r,\theta,{\varphi}) =  e^{i\omega t}
   \phi\mathit{1}(r) \, Y(\theta,\varphi) \, .
   \label{an_sc}
\end{eqnarray}
Besides the spherical harmonics $Y(\theta,\varphi)$, all the $r$-dependent perturbation functions also carry two quantum numbers, $l$ and $M_z$.

In the presence of rotation, the perturbations need to be summed over all possible values of $l$ and $M_z$.
While the axial symmetry of the background configurations still leads to a decoupling of the different values of $M_z$, the $l$ values are now coupled and one obtains a tower of equations, when the ansatz is inserted in the field equations
for the metric and phantom scalar components,
\begin{eqnarray}
\mathcal{G}_{\mu\nu} = \mathcal{G}_{\mu\nu}^{(sr)} + \epsilon_q\delta\mathcal{G}_{\mu\nu}e^{-i\omega t}=0 \, , \\
\mathcal{S} = \mathcal{S}^{(sr)} + \epsilon_q\delta\mathcal{S}e^{-i\omega t}=0 \, ,
\end{eqnarray}
where the slowly rotating background solution ensures $\mathcal{G}_{\mu\nu}^{(sr)}=0$, and $\mathcal{S}^{(sr)}=0$.

Here, we follow \cite{Kojima:1992ie,Blazquez-Salcedo:2022eik} and decompose the field equations in terms of spherical harmonics, and then truncate the tower of equations in the slow rotation approximation, mixing different $l$. 
The resulting modes can be identified
by the value of $l$ (as well as by $M_z$), to which the modes reduce in the static limit of the background solutions, referring to the modes then as $l$-led modes.
Moreover, in the perturbative scheme, the equations decouple into two sets, the polar-led and the axial-led perturbations \cite{Blazquez-Salcedo:2022eik}.

The resulting system of equations then has the generic {structure} 
\begin{eqnarray}
\vec{z}_i \, ' = \mathbf{M}_i \vec{z}_i \, ,
\end{eqnarray}
where $\vec{z}_i$ denotes a vector of perturbation functions or their derivatives; $\mathbf{M}_i$ is a matrix containing the background functions, which depend on the radial coordinate $r$, the free parameter $\mathit{r_0}$, the asymmetry parameter $C$, and the eigenvalue $\omega$; and $i=p,a$ denotes the polar-led and axial-led equations, respectively.
In the following, we focus on quadrupole, i.e., $l=2$-led, perturbations, and select $M_z=2$.

The vector $\vec{z}_a$ for the axial-led perturbations reads
\begin{eqnarray}
   \vec{z}_a =  && \Big[ 
   \mathit{h0}_{2,2} \, ,
   \mathit{h1}_{2,2} \, ,
       \mathit{H1}_{3,2} \, ,  
       T_{3,2} \, ,
       L_{3,2} \, ,
       N_{3,2} \, ,            
           \phi\mathit{1}_{3,2} \, ,
           \phi\mathit{1}_{3,2}' \Big]^T \, ,
\end{eqnarray}
where $\mathbf{M}_a$ is a $8 \times 8$ matrix 
whose entries depend on the functions of the slowly rotating background and the eigenvalue $\omega$. 
Analogously, the vector $\vec{z}_p$ for the polar-led perturbations reads
\begin{eqnarray}
   \vec{z}_p =  && \Big[ 
   \mathit{h0}_{3,2} \, ,
   \mathit{h1}_{3,2} \, ,
       \mathit{H1}_{2,2} \, ,  
       T_{2,2} \, ,
       L_{2,2} \, ,
       N_{2,2} \, ,            
           \phi\mathit{1}_{2,2} \, ,
           \phi\mathit{1}_{2,2}' \Big]^T \, .
\end{eqnarray}
Thus, the $l=2, M_z=2$ axial-led perturbation functions ${\cal A}$ consist of the axial $\mathit{h0}_{2,2},\mathit{h1}_{2,2}$ functions and the polar $\mathit{H1}_{3,2},T_{3,2},L_{3,2},N_{3,2},\phi \mathit{1}_{3,2}$ functions, while the $l=2, M_z=2$ polar-led perturbation functions ${\cal P}$ consist of the polar $\mathit{H1}_{2,2},T_{2,2},L_{2,2},N_{2,2},\phi \mathit{1}_{2,2}$ functions and the axial $\mathit{h0}_{3,2},\mathit{h1}_{3,2}$ functions.

Consequently, there are, for each of the axial-led and polar-led $l=M_z=2$ perturbations in total, seven perturbation equations that have to be determined.
We present the explicit expression for these perturbation equations for the symmetric wormholes, $C=0$, in the \highlighting{Appendix \ref{appA}}, since in this case they already lead to rather lengthy expressions.

To obtain the solutions for the modes, one has to numerically solve the resulting system of equations as an eigenvalue problem, subject to the appropriate set of boundary conditions.
For the wormholes, suitable boundary conditions are purely outgoing boundary conditions at both radial infinities.

To obtain the appropriate behavior at the boundaries,
we consider the following parametrization of the perturbation functions at plus infinity,
\begin{eqnarray}
 \mathit{h0}_{l,2} (r) &=& \sqrt{r^2+ \mathit{r_0}^{2}} \, \mathit{h0}(r) \, e^{\mathrm{i}\omega R^*} \, , \\ 
      \mathit{h1}_{l,2} (r) &=& \sqrt{r^2+ \mathit{r_0}^{2}} \, \mathit{h1}(r)\, e^{\mathrm{i}\omega R^*} \, , \\ 
        \mathit{H1}_{l,2} (r) &=& \sqrt{r^2+ \mathit{r_0}^{2}} \, \mathit{H1}(r) \, e^{\mathrm{i}\omega R^*} \, , \\ 
    T_{l,2} (r) &=& T(r) \, e^{\mathrm{i}\omega R^*} \, , \\ 
      \mathit{L}_{l,2} (r) &=& \sqrt{r^2+ \mathit{r_0}^{2}} \, \mathit{L}(r) \, e^{\mathrm{i}\omega R^*} \, , \\ 
      \mathit{N}_{l,2} (r) &=& \sqrt{r^2+ \mathit{r_0}^{2}} \, \mathit{N}(r) \, e^{\mathrm{i}\omega R^*} \, , \\ 
       \phi\mathit{1}_{l,2} (r) &=& \left(1/\sqrt{r^2+ \mathit{r_0}^{2}} \right) \phi\mathit{1}(r) \, e^{\mathrm{i}\omega R^*} \, , 
\end{eqnarray}
{for}
 $l=2,3$.
At minus infinity,
for axial-led perturbations, 
we consider the following parametrization,
\begin{eqnarray}
 \mathit{h0}_{2,2} (r) &=& (r^2+ \mathit{r_0}^{2})^{3/2} \,\mathit{h0}(r) \, e^{-\mathrm{i}\omega R^*} \, , \\ 
      \mathit{h1}_{2,2} (r) &=& (r^2+ \mathit{r_0}^{2})^{3/2} \, \mathit{h1}(r) \, e^{-\mathrm{i}\omega R^*} \, , \\ 
\mathit{H1}_{3,2} (r) &=& (r^2+ \mathit{r_0}^{2}) \, \mathit{H1}(r) \, e^{-\mathrm{i}\omega R^*} \, , \\ 
     T_{3,2} (r) &=& \sqrt{r^2+ \mathit{r_0}^{2}} \, T(r) \, e^{-\mathrm{i}\omega R^*} \, , \\ 
      \mathit{L}_{3,2} (r) &=&  \mathit{L}(r) \, e^{-\mathrm{i}\omega R^*} \, , \\ 
      \mathit{N}_{3,2} (r) &=&  \mathit{N}(r) \, e^{-\mathrm{i}\omega R^*} \, , \\ 
       \phi\mathit{1}_{3,2} (r) &=&  \phi\mathit{1}(r) \, e^{-\mathrm{i}\omega R^*} \, . 
\end{eqnarray}
{For} polar-led perturbations at minus infinity, we have
\begin{eqnarray}
 \mathit{h0}_{3,2} (r) &=& (r^2+ \mathit{r_0}^{2}) \,\mathit{h0}(r) \, e^{-\mathrm{i}\omega R^*} \, , \\ 
      \mathit{h1}_{3,2} (r) &=& (r^2+ \mathit{r_0}^{2}) \, \mathit{h1}(r) \, e^{-\mathrm{i}\omega R^*} \, , \\ 
 \mathit{H1}_{2,2} (r) &=& (r^2+ \mathit{r_0}^{2})^{3/2} \, \mathit{H1}(r) \, e^{-\mathrm{i}\omega R^*} \, , \\ 
     T_{2,2} (r) &=& (r^2+ \mathit{r_0}^{2}) \, T(r) \, e^{-\mathrm{i}\omega R^*} \, , \\ 
      \mathit{L}_{2,2} (r) &=&  \mathit{L}(r) \, e^{-\mathrm{i}\omega R^*} \, , \\ 
      \mathit{N}_{2,2} (r) &=&  \mathit{N}(r) \, e^{-\mathrm{i}\omega R^*} \, , \\ 
       \phi\mathit{1}_{2,2} (r) &=&  \sqrt{r^2+ \mathit{r_0}^{2}} \, \phi\mathit{1}(r) \, e^{-\mathrm{i}\omega R^*} \, . 
\end{eqnarray}
{The} tortoise coordinate $R^*$ is given by
\begin{eqnarray}
    \frac{d R^*}{dr} = e^{-f(r)} - \frac{e^{-2f(r)}}{r} \left(re^{f(r)}h_0(r) - b_0(r)\right) \, ,
\end{eqnarray}
where $f(r),h_0(r), b_0(r)$ are the background functions.

At plus infinity, an outgoing solution implies the following behavior for axial-led perturbations,
\begin{eqnarray}
  \mathit{h0}(r) &\approx& - C_{\mathit{h1}_0} - \frac{2\mathrm{i}}{r\omega} C_{\mathit{h1}_0} \, ,\\
       \mathit{h1}(r) &\approx& C_{\mathit{h1}_0} + \frac{3}{r\pi\omega\mathit{r_0}^3} ((\pi^2-8)J^2\omega + \mathrm{i}\pi\mathit{r_0}^3) C_{\mathit{h1}_0} \, ,\\
  \mathit{H1}(r) &\approx& \mathrm{i} \omega C_{T_0} - \frac{5 C_{T_0}}{r} \, ,\\ 
    T(r) &\approx& C_{T_0} + \frac{C_{T_0}}{2r^2\omega^2} (2\mathrm{i}\omega^2J-(\mathit{r_0}\omega)^2+30) \, , \\
    L(r) &\approx& -\mathrm{i} \omega C_{T_0} + \frac{5 C_{T_0}}{r} \, ,\\
     N(r) &\approx& -\mathrm{i} \omega C_{T_0} + \frac{5 C_{T_0}}{r} \, ,\\ 
       \phi\mathit{1}(r) &\approx& C_{\phi\mathit{1}_0} + \frac{6\mathrm{i}}{r\omega} C_{\phi\mathit{1}_0} \, ,
\end{eqnarray}
and for polar-led perturbations,
\begin{eqnarray}
 \mathit{h0}(r) &\approx& -C_{\mathit{h1}_0} - \frac{5\mathrm{i}}{r\omega}  C_{\mathit{h1}_0} \, ,\\
       \mathit{h1}(r) &\approx& C_{\mathit{h1}_0} + \frac{6\mathrm{i}}{r\omega}  C_{\mathit{h1}_0} \, ,\\ 
 \mathit{H1}(r) &\approx& \mathrm{i} \omega C_{T_0} + \frac{3 C_{T_0}}{r\pi\mathit{r_0}^3} (\mathrm{i}J^2\pi^2\omega - 8\mathrm{i}J^2\omega-2\pi\mathit{r_0}^3/3) \, ,\\ 
    T(r) &\approx& C_{T_0} + \frac{1}{140 \pi  \,\omega^{2} \mathit{r_0}^{3} r^{2}}
   \Big( -80\sqrt{7}\, J^{2} \omega^{3} \pi^{2}  C_{\mathit{h1}_0}  \mathit{r_0}^{2}
   \nn \\ 
   &+& 32 \mathit{C_{T_0}} \Big(\mathrm{i} \Big(\mathit{r_0}^{2} \omega^{2}-\frac{315}{32}\Big) \omega \,J^{2}  \pi^{2}-\frac{35 \pi  \,\mathit{r_0}^{3} (\mathit{r_0}^{2} \omega^{2}-6)}{16}+\frac{315 \,\mathrm{i} J^{2} \omega}{4}\Big) \Big)
    \, , \\
    L(r) &\approx& -\mathrm{i} \omega C_{T_0} 
    -\frac{6 C_{T_0}}{r\pi^2\mathit{r_0}^6} (\mathrm{i}\omega J^2\pi^3 \mathit{r_0}^3 - \mathit{r_0}^6\pi^2/3 - 8\mathrm{i} \omega\mathit{r_0}^3J^2\pi)
    \, ,\\
     N(r) &\approx& -\mathrm{i} \omega C_{T_0} + \frac{2 C_{T_0}}{r} \, ,\\
       \phi\mathit{1}(r) &\approx& C_{\phi\mathit{1}_0} + \frac{1}{2 r\omega\pi\mathit{r_0}^2} (3(\pi^2-8)\omega J^2 C_{T_0} + 6\mathrm{i}\pi\mathit{r_0}^2 C_{\phi\mathit{1}_0}) \, .
\end{eqnarray}
At minus infinity, the axial-led perturbation functions behave as
\begin{eqnarray}
  \mathit{h0}(r) &\approx& -\frac{6\mathrm{i}\sqrt{7}J^2\pi^2}{7\mathit{r_0}^6} {C_{T_1}} - \frac{J\pi}{7r\omega\mathit{r_0}^6} (\sqrt{7}\omega\mathit{r_0}^3 {C_{T_1}})
       \, ,\\ 
       \mathit{h1}(r) &\approx& 
       -\frac{18}{\mathit{r_0}^6} J^2\pi^2 C_{\mathit{h1}_2}
-\frac{54\mathrm{i}}{r\omega\mathit{r_0}^6} J\pi C_{\mathit{h1}_2} (J\pi-\omega\mathit{r_0}^3/9)
       \, ,\\ 
 \mathit{H1}(r) &\approx& \frac{11  J\pi \omega {C_{T_1}}}{2\mathit{r_0}^3} 
+ \frac{{C_{T_1}}}{7r\mathit{r_0}^3} (-7\mathrm{i}\omega\mathit{r_0}^3-175\mathrm{i}J\pi )
\, ,\\ 
    T(r) &\approx& \frac{6 \mathrm{i} J\pi {C_{T_1}}}{\mathit{r_0}^3} + \frac{{C_{T_1}}}{r}  \, , \\
    L(r) &\approx& \frac{6{C_{T_1}}}{\mathit{r_0}^3} J\omega\pi r^2 - \frac{\mathrm{i} {C_{T_1}}}{\mathit{r_0}^3} (\omega\mathit{r_0}^3 + 24J\pi)r - 5{C_{T_1}} + \frac{3{C_{T_1}}J\pi}{\omega\mathit{r_0}^3} (\omega^2\mathit{r_0}^2 - 30) \, ,\\
     N(r) &\approx& \frac{5{C_{T_1}}}{\mathit{r_0}^3} J\omega\pi r^2 - \frac{\mathrm{i} {C_{T_1}}}{7\mathit{r_0}^3} (7\omega\mathit{r_0}^3 + 168 J\pi)r - 5{C_{T_1}} 
     \, ,\\
       \phi\mathit{1}(r) &\approx& 
       \frac{6\mathrm{i}J\pi C_{\phi\mathit{1}_1}}{\mathit{r_0}^3} + \frac{C_{\phi\mathit{1}_1}}{r}
       \, ,
\end{eqnarray}
and for polar-led perturbations, they are
\begin{eqnarray}
   \mathit{h0}(r) &\approx& 
      \frac{\sqrt{7}J\pi C_{T_2}}{7\mathit{r_0}^3} 
      + \frac{6\mathrm{i}J\pi C_{\mathit{h1}_1}}{\mathit{r_0}^3}
       \, ,\\ 
       \mathit{h1}(r) &\approx& 
       \frac{6\mathrm{i}J\pi C_{\mathit{h1}_1}}{\mathit{r_0}^3} + \frac{C_{\mathit{h1}_1}}{r}
       \, ,\\ 
  \mathit{H1}(r) &\approx& \frac{12\mathrm{i}  J^2\pi^2 \omega C_{T_2}}{\mathit{r_0}^6} 
+ \frac{J\pi C_{T_2}}{7r\mathit{r_0}^6}  (35 \omega\mathit{r_0}^3+ 126 J\pi )
\, ,\\ 
    T(r) &\approx& -\frac{18  J^2 \pi^2 {C_{T_2}}}{\mathit{r_0}^6} + \frac{6\mathrm{i}J\pi C_{T_2}}{r\mathit{r_0}^3}  \, , \\
    L(r) &\approx& 
    -\frac{18 \,\mathrm{i} \omega \,J^{2} \pi^{2} C_{T_2} \, r^{3}}{\mathit{r_0}^{6}}-\frac{6 \pi  J  C_{T_2} \,\omega r^{2}}{\mathit{r_0}^{3}}
    \, ,\\
     N(r) &\approx& 
      -\frac{6 \,\mathrm{i} \omega \,J^{2} \pi^{2} C_{T_2} \, r^{3}}{\mathit{r_0}^{6}}-\frac{2 \pi  J r^2  C_{T_2} }{7\mathit{r_0}^{6}} (14\omega \mathit{r_0}^{3}+27J\pi) 
     \, ,\\
       \phi\mathit{1}(r) &\approx& 
       -\frac{18 J^2\pi^2 C_{\phi\mathit{1}_2}}{\mathit{r_0}^6} 
      - \frac{54 \,\mathrm{i} C_{\phi\mathit{1}_2} \left(-\mathit{r_0}^{3} \omega /9 +\pi  J  \right)  J \pi}{\mathit{r_0}^{6} \omega r}
       \, .
\end{eqnarray}
In general, $C_{X_k}$ is a constant obtained from a specific order $k$ of an expansion of the respective perturbation function $X = T,\mathit{H1}, L, N, \mathit{h0}, \mathit{h1}, \phi\mathit{1}$,
in $\mathcal{O}\left(\frac{1}{r^k}\right)$ for $k=0,1,2,..$.
Constants for $k>0$ can all be expressed in terms of the three free constants at the zeroth order, i.e., $C_{T_0}, C_{\mathit{h1}_0}, C_{\phi\mathit{1}_0}$. 
{All these behaviors of the functions ensure an outgoing wave at the infinities.}

Since the modes are determined in perturbation theory, one can also consider the eigenvalue $\omega$ as consisting of the static eigenvalue $\omega_0$ and the higher-order contributions.
Thus, in the second order in rotation, the eigenvalue will read
\begin{equation}
\omega_{2,2} = \omega^{(0)}_{2,2} + \epsilon_r\, \delta \omega^{(1)}_{2,2} + \epsilon_r^2\, \delta \omega^{(2)}_{2,2} \, ,
\end{equation}
where, for clarity, we have omitted further classifications like the harmonic number $n$ or indices for axial/polar.
The modes can be obtained by integrating the perturbation equations, and they should connect smoothly to the static limit. 
In general, the scaled modes display quadratic behavior with respect to the scaled angular momentum, such that
\begin{equation}
    M\omega = M \omega^{(0)} + M \delta\omega^{(1)} \left(\frac{J}{M^2}\right) + M \delta\omega^{(2)} \left(\frac{J}{M^2}\right)^2 \, ,
\end{equation}
as reported in our previous studies in slow rotation up to the second order 
{(see e.g., \cite{Blazquez-Salcedo:2022eik})}.


\section{Further Remarks}

\label{sec5}

By performing a double expansion, we have derived the coupled sets of ordinary differential equations that are needed to obtain the quasinormal modes of slowly rotating Ellis--Bronnikov wormholes.
In particular, we have focused on $l=2, M_z=2$-led  perturbations, which are expected to be accounted for (or at least among) the most prominent modes during a ringdown.
{The detectability of these modes will depend on a high signal-to-noise 
	ratio from observations.}
The numerical implementation of the developed scheme, however, still presents challenges in the case of the quadrupole modes of Ellis--Bronnikov wormholes.

The quasinormal modes of the static Ellis--Bronnikov wormholes exhibit a threefold isospectrality in the symmetric case ($C=0)$ \cite{Azad:2022qqn}.
However, this isospectrality is broken for finite values of the asymmetry parameter $C$.
Likewise, we expect rotation to break isospectrality for the slowly rotating wormholes addressed here. 
Note that for a set of rapidly rotating wormholes, isospectrality is broken, as well \cite{Khoo:2024yeh}.

\begin{figure}[H]
	
\includegraphics[width=0.3\textwidth,angle=-90]{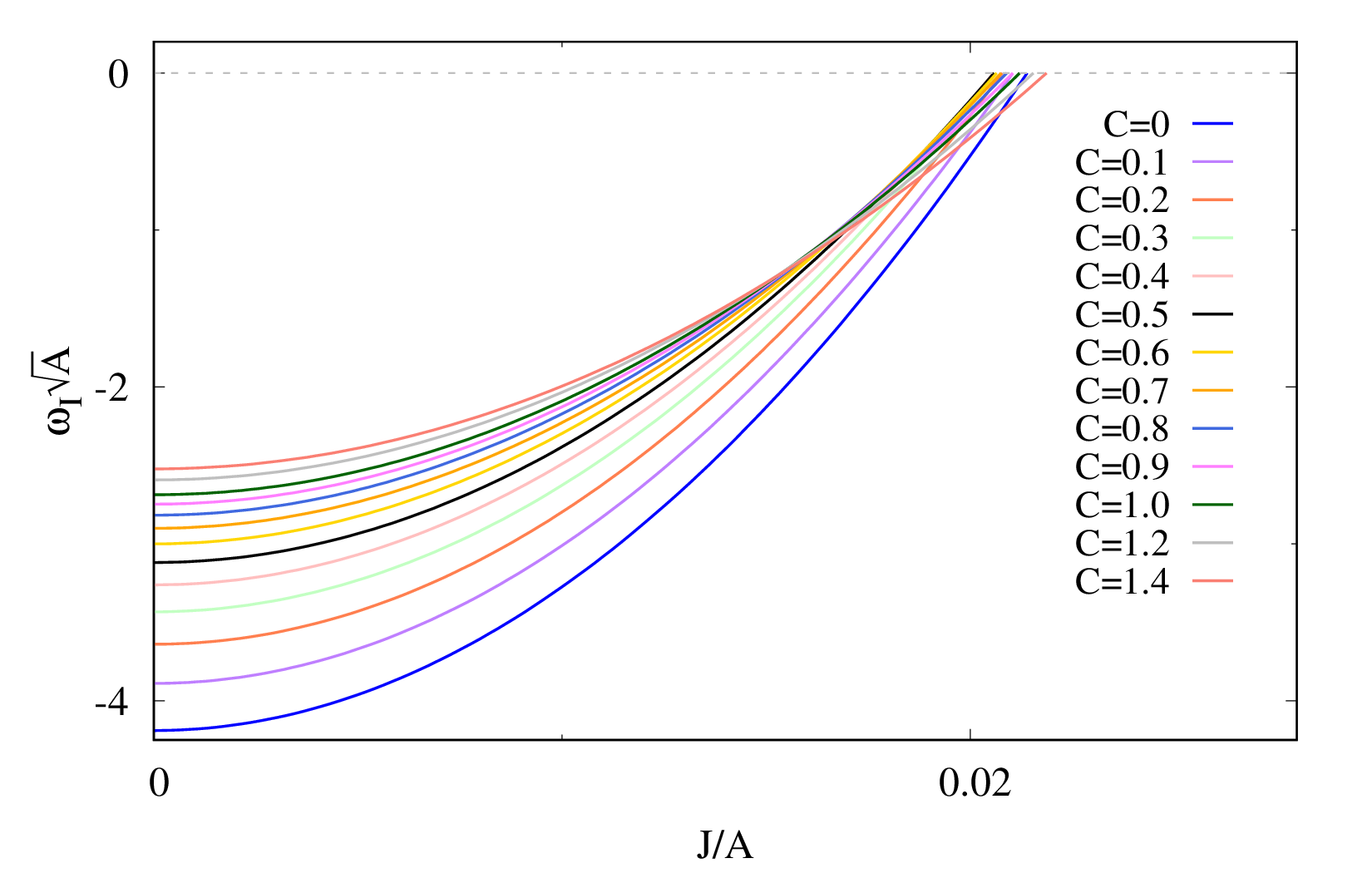}
\includegraphics[width=0.3\textwidth,angle=-90]{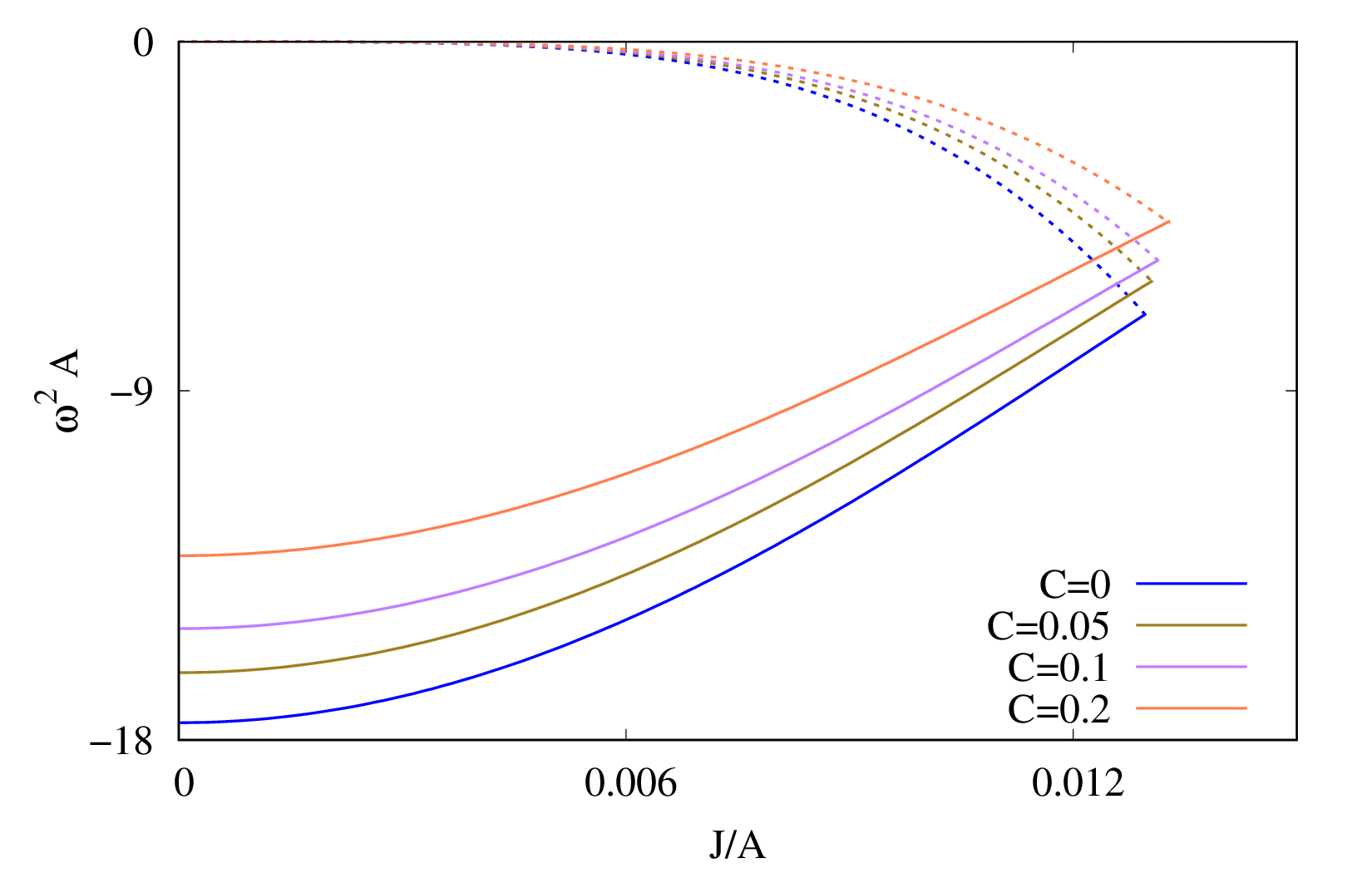}
  \caption{{Unstable}
 radial modes in second-order perturbation theory in rotation: a purely imaginary eigenvalue $\omega_I$ vs. the angular momentum $J$, both scaled with appropriate powers of the corrected throat area (compare \cite{Azad:2023iju,Azad:2024axu}) for several values of the asymmetry parameter $C$.
  The left plot shows the unstable mode already present in the static case, and the right plot includes the second unstable mode emerging from a zero mode, as well.
  Close to the bifurcation of two modes, the second order approximation breaks down.
  }
		\label{fig2}
	\end{figure}

The developed scheme has already been successfully applied in the case of radial modes of the Ellis--Bronnikov wormholes \cite{Azad:2023iju,Azad:2024axu}.
{As shown in the left plot of Figure~\ref{fig2}, the well-known purely imaginary unstable mode of the wormholes \cite{Shinkai:2002gv,Gonzalez:2008wd,Gonzalez:2008xk,Cremona:2018wkj,Blazquez-Salcedo:2018ipc} was seen to become more stable with increasing rotation, as the absolute value of the scaled imaginary eigenvalue is decreasing towards zero.
As the asymmetry parameter $C$ increases up to about $C=0.5$, the critical value of the scaled angular momentum where the imaginary eigenvalue vanishes decreases.
However, for $C>0.5$, the critical angular momentum value grows.
While in the right plot of Figure~\ref{fig2}, a second branch of an unstable mode was seen to emerge from a zero mode in the static limit, this becomes unstable with increasing rotation.
Nonetheless, this branch of the mode merges with the previously discussed unstable mode, and, moreover, at an even smaller value of the scaled angular momentum. Based on this evidence, therefore, the radial instability is conjectured to disappear.
}

Unfortunately, perturbation theory cannot resolve the question of radial stability completely, and (in rotation) non-perturbative calculations along the lines of \cite{Khoo:2024yeh} are needed.
A possible scenario for the further evolution of the modes would be one where both purely imaginary modes merge, develop a real part, and slowly become more stable, until the instability disappears when approaching the final extremal configuration.

Wormholes may also suffer from non-radial instabilities, of course.
Since the study of the quasinormal mode spectrum will reveal instabilities, determination of the spectrum represents an essential aspect concerning the viability of wormhole solutions as potentially observable compact objects.
In recent years, studies in numerical relativity (see e.g.,~\cite{Dent:2020nfa}) are also starting to address potentially observational aspects of wormholes.
{Finally, perturbations of wormholes in alternative theories of gravity, and of other multipolar nature, are left for future studies.}


\vspace{6pt} 


\authorcontributions{{Conceptualization, B.A., J.L.B.-S., F.S.K., J.K. and F.N.-L.; methodology, B.A., J.L.B.-S., F.S.K., J.K. and F.N.-L.; software, B.A., J.L.B.-S., F.S.K., J.K. and F.N.-L.; validation, B.A., J.L.B.-S., F.S.K., J.K. and F.N.-L.; formal analysis, B.A., J.L.B.-S., F.S.K., J.K. and F.N.-L.; investigation, B.A., J.L.B.-S., F.S.K., J.K. and F.N.-L.; resources, B.A., J.L.B.-S., F.S.K., J.K. and F.N.-L.; data curation, B.A., J.L.B.-S., F.S.K., J.K. and F.N.-L.; writing---original draft preparation, B.A., J.L.B.-S., F.S.K., J.K. and F.N.-L.; writing---review and editing, B.A., J.L.B.-S., F.S.K., J.K. and F.N.-L.; visualization, B.A., J.L.B.-S., F.S.K., J.K. and F.N.-L.; supervision, B.A., J.L.B.-S., F.S.K., J.K. and F.N.-L.; project administration, B.A., J.L.B.-S., F.S.K., J.K. and F.N.-L.; funding acquisition, B.A., J.L.B.-S., F.S.K., J.K. and F.N.-L. All authors have read and agreed to the published version of the manuscript.}
}

\funding{{We} 
 gratefully acknowledge the support of DAAD
 and
MICINN project PID2021-125617NB-I00, ``QuasiMode''.
J.L.B.-S. gratefully acknowledges support from MICINN, project CNS2023-144089 ``Quasinormal modes''.
F.S.K. gratefully acknowledges support from ``Atracci\'on de Talento Investigador Cesar Nombela'' of the Comunidad de Madrid, under the grant number \mbox{2024-T1/COM-31385.}
}

\dataavailability{The data are available from the authors upon request.
}

\acknowledgments{{We} 
 would like to thank Luis Manuel Gonz\'alez-Romero and Hendrik Mennenga for their support.
}

\conflictsofinterest{The authors declare no conflicts of interest.
}


\appendixtitles{no} 
\appendixstart
\appendix
\section[\appendixname~\thesection]{}
\label{appA}
We exhibit the final sets of equations for the symmetric case $C=0$ below.
For asymmetric wormholes with $C>0$, the perturbation equations are extremely lengthy.
Therefore, we do not present them here, but provide them in a {supplementary file} 
 (Maple) available upon request.

\appendixtitles{yes} 
\subsection[\appendixname~\thesubsection]{Axial Perturbation Equations}

We first show the axial perturbation equations.
Here, the minimal system of equations is described by five ordinary differential equations (ODEs) for $\phi\mathit{1}_{3,2}, \mathit{H1}_{3,2}, T_{3,2},  \mathit{h0}_{2,2}$, and $ \mathit{h1}_{2,2}$, and by two algebraic equations for $ L_{3,2}, N_{3,2}$.
As this is an axial-led perturbation system,
one finds that only the axial metric perturbation functions $\mathit{h0}, \mathit{h1}$ carry the quantum numbers $l=M_z=2$.
They are coupled with the other polar perturbation functions $\mathit{H1},T,L,N, \phi\mathit{1}$ with $l=3, M_z=2$.

For the perturbations of the phantom scalar field, we {obtain} 
{
\begin{align}
\frac{d^{2}}{d r^{2}}{\phi\mathit{1}}_{3,2}
\! 
\left(r \right) &= 
-\frac{2 r \left(\frac{d}{d r}{\phi \mathit{1}}_{3,2}\! \left(r \right)\right)}{r^{2}+\mathit{r_0}^{2}}-\frac{2 J \mathit{r_0} \omega  T_{3,2}\! \left(r \right)}{\left(r^{2}+\mathit{r_0}^{2}\right)^{2}}
-\frac{12 J \sqrt{7}\, \mathit{r_0}  \mathit{h0}_{2,2}\! \left(r \right)}{7 \left(r^{2}+\mathit{r_0}^{2}\right)^{3}}
\nn
\\ &
+\frac{48 \,\mathrm{i} J \sqrt{7}\, r \mathit{r_0}  \mathit{h1}_{2,2}\! \left(r \right)}{7 \left(r^{2}+\mathit{r_0}^{2}\right)^{4} \omega} + 
\Big(-\frac{12 \omega J \left(\left(r^{2}+\mathit{r_0}^{2}\right) \phi \! \left(r \right)+r \mathit{r_0} \right) }{\left(r^{2}+\mathit{r_0}^{2}\right) \mathit{r_0}^{3}}
\nn
\\ &
+\frac{-r^{4} \omega^{2}+\left(-2 \mathit{r_0}^{2} \omega^{2}+12\right) r^{2}
-\mathit{r_0}^{4} \omega^{2}+8 \mathit{r_0}^{2}}{\left(r^{2}+\mathit{r_0}^{2}\right)^{2}}\Big) {\phi \mathit{1}}_{3,2}\! \left(r \right) \, .
\label{sc_ax}
\end{align}
}


Concerning the metric perturbations, there are six relevant perturbation functions.
First, we show the equation for the $\mathit{H1}$ function,
{\small
\begin{align}
\frac{d}{d r}\mathit{H1}_{3,2}(r) 
&= 
\frac{\mathrm{i} }{25 \mathit{r_0}^{2} (r^{2}+\mathit{r_0}^{2})^{2}}
\Big( - (r^{2}+\mathit{r_0}^{2}) 
\big( (r^{2} \omega^{2}+50) \mathit{r_0}^{2} + r^{4} \omega^{2} + 40 r^{2} \big) 
J \phi(r) 
\nn\\
& 
- \mathit{r_0} \big( 
10 \omega \mathit{r_0}^{6} + 20 r^{2} \omega \mathit{r_0}^{4} + r ( J r^{2} \omega^{2}  + 10 r^{3} \omega + 50 J  ) \mathit{r_0}^{2} 
\nn\\
&
+ J r^{3}  (r^{2} \omega^{2} + 40) 
\big) 
\Big) 
\frac{d}{d r}\phi\mathit{1}_{3,2}(r)
\nn\\
&+ \Big( 
\frac{ J\mathit{r_0}}{50 \mathit{r_0}^{3} (r^{2}+\mathit{r_0}^{2})^{2} \omega} 
\Big(
(r^{2}+\mathit{r_0}^{2}) 
\big( r^{4} \omega^{4} + r^{2} \mathit{r_0}^{2} \omega^{4} + 130 r^{2} \omega^{2} 
\nn\\
&
+ 140 \mathit{r_0}^{2} \omega^{2}
- 300 \big) 
r \phi(r) 
\nn\\
&
+  r^{6} \omega^{4} + r^{4} \mathit{r_0}^{2} \omega^{4} + 130 r^{4} \omega^{2} + 200 r^{2} \mathit{r_0}^{2} \omega^{2} + 60 \mathit{r_0}^{4} \omega^{2} - 300 r^{2} 
\Big)
\nn\\
& 
+ \frac{\omega^{2} r}{5} \Big) \mathit{H1}_{3,2}(r)
\nn\\&
+ \frac{\mathrm{i} }{50 (r^{2}+\mathit{r_0}^{2}) \mathit{r_0}^{3}} 
\Big(
\big( - (r^{2}+\mathit{r_0}^{2}) J 
\big( (r^{2} \omega^{4} + 145 \omega^{2}) \mathit{r_0}^{2}
\nn\\
&
+ r^{4} \omega^{4} 
+ 125 r^{2} \omega^{2} - 250 \big) \phi(r)  
- \big( 10 \omega^{3} \mathit{r_0}^{6} + 20 (r^{2} \omega^{3} - 5 \omega) \mathit{r_0}^{4} 
\nn\\
&
+ r \omega ( J r^{2} \omega^{3}  + 10 r^{3} \omega^{2} + 155 J \omega  - 100 r ) \mathit{r_0}^{2} 
\nn\\
&
+ J r  ( r^{4} \omega^{4} + 125 r^{2} \omega^{2} - 250 ) \big) \mathit{r_0} \big)
\Big) T_{3,2}(r)
\nn\\
&+ \frac{6 \mathrm{i} \omega J \sqrt{7}\,  \mathit{h0}_{2,2}(r)}{7 \mathit{r_0}^{3} (r^{2}+\mathit{r_0}^{2})^{2}}
\Big( (r^{2}+\mathit{r_0}^{2})^{2} \phi(r) 
+ r^{3} \mathit{r_0} + \frac{13 \mathit{r_0}^{3} r}{15} \Big) 
\nn\\
&+ \frac{6 J  \sqrt{7}\,  \mathit{h1}_{2,2}(r)}{35 \mathit{r_0}^{3} (r^{2}+\mathit{r_0}^{2})^{3}} \Big( r \omega^{2} (r^{2}+\mathit{r_0}^{2})^{3} \phi(r) 
+ \big( (r^{2} \omega^{2} + \frac{40}{3}) \mathit{r_0}^{4} 
\nn\\ &
+ (2 r^{4} \omega^{2} + \frac{32}{3} r^{2}) \mathit{r_0}^{2} 
+ r^{6} \omega^{2} \big) \mathit{r_0} \Big)
\nn\\
&+ \frac{3 \mathrm{i} }{25 \mathit{r_0}^{2} (r^{2}+\mathit{r_0}^{2})^{3}} 
\Big(
\big(  (r^{2}+\mathit{r_0}^{2}) (r^{2}+\frac{5 \mathit{r_0}^{2}}{3}) J (r^{2} \omega^{2} + \mathit{r_0}^{2} \omega^{2} - 10) \phi(r) \big) 
\nn\\
&
+ \big( -\frac{20 \omega \mathit{r_0}^{6}}{3} + \frac{5 r \omega (J \omega  - 8 r) \mathit{r_0}^{4}}{3} 
+ \frac{8 r ( J r^{2} \omega^{2}  - \frac{5}{2} r^{3} \omega - \frac{25}{4} J  ) \mathit{r_0}^{2}}{3} 
\nn\\
&
+ J r^{3}  (r^{2} \omega^{2} - 10) \big) \mathit{r_0}
\Big) r \phi\mathit{1}_{3,2}(r) \, .
\end{align}}


For the $T$ function, the equation is given by
{\small\begin{align}
\frac{d}{d r} T_{3,2}(r) & = 
\Big( 
-\frac{12 J   }{5 \mathit{r_0}^{2} \omega (r^{2} + \mathit{r_0}^{2})^{2}} 
\Big( \frac{r (r^{2} + \mathit{r_0}^{2}) (r^{2} \omega^{2} + 30) \phi(r)}{60} 
\nn\\
&
+ \mathit{r_0} \big( \frac{r^{4} \omega^{2}}{60} + \mathit{r_0}^{2} + \frac{r^{2}}{2} \big) \Big)
- \frac{2 r \mathit{r_0}}{5 (r^{2} + \mathit{r_0}^{2})} \Big) \frac{d}{d r} \phi\mathit{1}_{3,2}(r) 
\nn\\
&
+ \frac{\mathrm{i} }
{50 \mathit{r_0}^{3} \omega^{2} (r^{2} + \mathit{r_0}^{2})^{2}} 
\Big( - \left( r^{4} \omega^{4} + 120 r^{2} \omega^{2} - 900 \right)  (r^{2} + \mathit{r_0}^{2}) J \phi(r) 
\nn\\
&
-  \mathit{r_0} \big( 10 (r^{2} \omega^{3} + 30 \omega) \mathit{r_0}^{4} 
\nn\\
&
+ 10 (r^{4} \omega^{3} 
+ 12 J r \omega^{2}  
+ 30 r^{2} \omega) \mathit{r_0}^{2} 
+ J r  ( r^{4} \omega^{4} + 120 r^{2} \omega^{2} 
\nn\\
&
- 900 ) \big) 
\Big) 
\mathit{H1}_{3,2}(r) 
+ \Big( -\frac{J  \mathit{r_0} }{50 \mathit{r_0}^{3} \omega (r^{2} + \mathit{r_0}^{2})^{2}} 
\Big( (r^{2} + \mathit{r_0}^{2}) ( r^{4} \omega^{4} + r^{2} \mathit{r_0}^{2} \omega^{4} 
\nn\\
&
+ 140 r^{2} \omega^{2} + 150 \mathit{r_0}^{2} \omega^{2} 
- 300 ) r \phi(r) 
\nn\\
&
+ 
 r^{6} \omega^{4} + r^{4} \mathit{r_0}^{2} \omega^{4} + 140 r^{4} \omega^{2} + 220 r^{2} \mathit{r_0}^{2} \omega^{2} + 60 \mathit{r_0}^{4} \omega^{2} - 300 r^{2} 
- 300 \mathit{r_0}^{2} \Big)  
\nn\\
&
- \frac{\omega^{2} r}{5} \Big) T_{3,2}(r) 
- \frac{4 r^{2} J \sqrt{7}  \mathit{h0}_{2,2}(r)}{35 (r^{2} + \mathit{r_0}^{2})^{3}} 
- \frac{\mathcal{T}_1}{35 \mathit{r_0}^{3} (r^{2} + \mathit{r_0}^{2})^{4} \omega} 
\nn\\
&
+ \left( \frac{\mathcal{T}_2}{5 \mathit{r_0}^{2} \omega (r^{2} + \mathit{r_0}^{2})^{3}} + \frac{2 (3 r^{2} + 5 \mathit{r_0}^{2}) \mathit{r_0}}{5 (r^{2} + \mathit{r_0}^{2})^{2}} \right) \phi\mathit{1}_{3,2}(r) \, ,
\end{align}}
where
\vspace{6pt}
\begin{align}
\mathcal{T}_1 & = \, 6 \sqrt{7}  \, \mathrm{i} 
J \,  \mathit{h1}_{2,2}(r)
\Big( 
(r^{2} + \mathit{r_0}^{2})^{3} (r^{2} \omega^{2} + 30) \, \phi(r) 
\nn\\
&+ \Big( 
(r^{2} \omega^{2} + 50) \mathit{r_0}^{4} 
+ \Big( 2 r^{4} \omega^{2} + \frac{232}{3} r^{2} \Big) \mathit{r_0}^{2} 
+ r^{6} \omega^{2} + 30 r^{4} 
\Big) r \mathit{r_0} 
\Big) \, ,
\end{align}

\noindent
and

\vspace{-18pt}
\begin{align}
\mathcal{T}_2 &=  \, J \Big( 
(r^{2} \omega^{2} + 30) (r^{2} + \mathit{r_0}^{2}) \Big( \mathit{r_0}^{2} + \frac{3 r^{2}}{5} \Big)  \phi(r) 
\nonumber
\\
& 
+ \Big( 
\frac{3 r^{4} \omega^{2}}{5} 
+ (\mathit{r_0}^{2} \omega^{2} + 18) r^{2} 
+ 6 \mathit{r_0}^{2} 
\Big) \mathit{r_0}  r 
\Big) \, .
\end{align}


For the two algebraic equations, $L$ is given by
\begin{align}
L_{3,2}(r) & =  \left( \frac{\omega J r^{2} (\mathit{r_0}^{2} \phi(r) + \phi(r) r^{2} + r \mathit{r_0}) }{25 (r^{2} + \mathit{r_0}^{2}) \mathit{r_0}^{2}} + \frac{2 \mathit{r_0}}{5} \right) \frac{d}{d r} \phi\mathit{1}_{3,2}(r) 
\nn\\
&+ \frac{\mathrm{i} }{50 (r^{2} + \mathit{r_0}^{2}) \mathit{r_0}^{3}} 
\Big(  (r^{2} \omega^{2} + 90) (r^{2} + \mathit{r_0}^{2}) r J \phi(r) + \mathit{r_0} (10 r \omega \mathit{r_0}^{4} 
\nn\\
&
+ 10 (r^{3} \omega + 6 J ) \mathit{r_0}^{2} 
+ J r^{2}  (r^{2} \omega^{2} + 90)) \Big) \mathit{H1}_{3,2}(r) 
\nn\\
&+ 
\Big( 
\frac{\omega J }{50 (r^{2} + \mathit{r_0}^{2}) \mathit{r_0}^{3}}
( (r^{2} + \mathit{r_0}^{2})(r^{4} \omega^{2} + (\mathit{r_0}^{2} \omega^{2} + 110) r^{2} + 120 \mathit{r_0}^{2}) \phi(r) 
\nn\\
&+ \mathit{r_0} r (r^{4} \omega^{2} + (\mathit{r_0}^{2} \omega^{2} + 110) r^{2} + 130 \mathit{r_0}^{2}) )  
- 1 + \frac{(r^{2} + \mathit{r_0}^{2}) \omega^{2}}{5} 
\Big) 
T_{3,2}(r) 
\nn\\&
+ \frac{4 \sqrt{7} J  r  \, \mathit{h0}_{2,2}(r)}{35 (r^{2} + \mathit{r_0}^{2})^{2}} 
+ \frac{6 \sqrt{7}  \, \mathrm{i} J 
}{35 \mathit{r_0}^{3} \omega (r^{2} + \mathit{r_0}^{2})^{3}} 
\Big( r \omega^{2} (r^{2} + \mathit{r_0}^{2})^{3} \phi(r)
\nn\\
&
+ \Big( (r^{2} \omega^{2} + 20) \mathit{r_0}^{4} + (2 r^{4} \omega^{2} + \frac{52}{3} r^{2}) \mathit{r_0}^{2} + r^{6} \omega^{2} \Big) \mathit{r_0} \Big) 
 \mathit{h1}_{2,2}(r) 
\nn\\
&+ \frac{r }{5}
\Big( -\frac{3  \omega J ((r^{2} + \mathit{r_0}^{2}) \phi(r) + r \mathit{r_0}) }{5 \mathit{r_0}^{2} (r^{2} + \mathit{r_0}^{2})^{2}} \Big( r^{2} + \frac{5 \mathit{r_0}^{2}}{3} \Big) 
+ \frac{4 \mathit{r_0}}{r^{2} + \mathit{r_0}^{2}} \Big) \phi\mathit{1}_{3,2}(r) \, ,
\end{align}
and
$N$ is given by
\begin{align}
N_{3,2}(r) & =  
\left( 
\frac{\omega J r^{2} \left( \mathit{r_0}^{2} \phi(r) + \phi(r) r^{2} + r \mathit{r_0} \right) }{25 (r^{2} + \mathit{r_0}^{2}) \mathit{r_0}^{2}} 
+ \frac{2 \mathit{r_0}}{5} 
\right) \frac{d}{d r} \phi\mathit{1}_{3,2}(r) 
\nn\\
&+ \frac{ \mathrm{i}}{50 (r^{2} + \mathit{r_0}^{2}) \mathit{r_0}^{3}}
\Big( 
 (r^{2} \omega^{2} + 90) (r^{2} + \mathit{r_0}^{2}) r J \phi(r) 
+ \mathit{r_0} ( 
10 r \omega \mathit{r_0}^{4} 
\nn\\
&
+ 10 (r^{3} \omega + 6 J ) \mathit{r_0}^{2} 
+ J r^{2}  (r^{2} \omega^{2} + 90) 
) 
\Big) \mathit{H1}_{3,2}(r)  
\nn\\
&+ \Big( 
\frac{\omega J  }{50 (r^{2} + \mathit{r_0}^{2}) \mathit{r_0}^{3}}
( 
(r^{4} \omega^{2} + (\mathit{r_0}^{2} \omega^{2} + 60) r^{2} + 70 \mathit{r_0}^{2}) (r^{2} + \mathit{r_0}^{2}) \phi(r) 
\nn\\ &
+ (r^{4} \omega^{2} + (\mathit{r_0}^{2} \omega^{2} + 60) r^{2} + 80 \mathit{r_0}^{2}) \mathit{r_0} r 
)  
-1  + \frac{(r^{2} + \mathit{r_0}^{2}) \omega^{2}}{5} 
\Big) 
T_{3,2}(r) 
\nn \\
&
- \frac{12 J \sqrt{7}   }{7 \mathit{r_0}^{3} (r^{2} + \mathit{r_0}^{2})^{2}} 
\left( (r^{2} + \mathit{r_0}^{2})^{2} \phi(r) + r^{3} \mathit{r_0} + \frac{14 \mathit{r_0}^{3} r}{15} \right)
\mathit{h0}_{2,2}(r)
\nn\\
&+ \frac{6 \sqrt{7} \, \mathrm{i} J  \, \mathit{h1}_{2,2}(r)}{35 \mathit{r_0}^{3} \omega (r^{2} + \mathit{r_0}^{2})^{3}} 
\Big( 
r \omega^{2} (r^{2} + \mathit{r_0}^{2})^{3} \phi(r) 
\nn\\
&
+ \Big( 
(r^{2} \omega^{2} + \frac{20}{3}) \mathit{r_0}^{4} 
+ (2 r^{4} \omega^{2} + \frac{52}{3} r^{2}) \mathit{r_0}^{2} 
+ r^{6} \omega^{2} 
\Big) \mathit{r_0} 
\Big) 
\nn\\
&+ \frac{ r }{5}
\left( 
-\frac{3  \omega J \left( (r^{2} + \mathit{r_0}^{2}) \phi(r) + r \mathit{r_0} \right)  }{5 \mathit{r_0}^{2} (r^{2} + \mathit{r_0}^{2})^{2}} \left( r^{2} + \frac{5 \mathit{r_0}^{2}}{3} \right)
+ \frac{4 \mathit{r_0}}{r^{2} + \mathit{r_0}^{2}} 
\right) \phi\mathit{1}_{3,2}(r) \, .
\end{align}


Lastly, we show the perturbation equations for the two axial metric functions,  $\mathit{h0}$ and  $\mathit{h1}$.
For $\mathit{h0}$, the equation is given by
\begin{align}
\frac{d}{d r}\mathit{h0}_{2,2}\! \left(r \right) &= 
-\frac{\mathfrak{h}_1}{175 \omega \,\mathit{r_0}^{5} \left(r^{2}+\mathit{r_0}^{2}\right)^{3}}
-\frac{\mathfrak{h}_2}{350 \mathit{r_0}^{6} \omega^{2} \left(r^{2}+\mathit{r_0}^{2}\right)^{3}} 
-\frac{\mathfrak{h}_3}{350 \mathit{r_0}^{6} \omega \left(r^{2}+\mathit{r_0}^{2}\right)^{2}}
\nn\\
&
+ \left(-\frac{\mathfrak{h}_4}{7 \mathit{r_0}^{6} \omega^{2} \left(r^{2}+\mathit{r_0}^{2}\right)^{3}}+ 
\frac{4 J }{\omega \left(r^{2}+\mathit{r_0}^{2}\right)^{2}}+\frac{2 r}{r^{2}+\mathit{r_0}^{2}}\right) \mathit{h0}_{2,2}\! \left(r \right) 
\nn\\
&
+\frac{\mathfrak{h}_5}{35 \mathit{r_0}^{6} \left(r^{2}+\mathit{r_0}^{2}\right)^{4} \pi  \,\omega^{3}}+\frac{\mathfrak{h}_6}{175 \mathit{r_0}^{5} \omega \left(r^{2}+\mathit{r_0}^{2}\right)^{4}}-\frac{\mathfrak{h}_7}{35 \mathit{r_0}^{6} \omega \left(r^{2}+\mathit{r_0}^{2}\right)^{2}} \, ,
\end{align}

\noindent
where
\vspace{-12pt}

\begin{align}
\mathfrak{h}_1 & =  \sqrt{7}  J \frac{d}{d r} \phi\mathit{1}_{3,2}(r)
\Big( (r^{2}+\mathit{r_0}^{2})^{2}  \big( (r^{2} \omega^{2}+270) \mathit{r_0}^{2} + r^{4} \omega^{2} + 30 r^{2} \big) r J \phi(r)^{2} 
\nn\\&
+ 2 (r^{2}+\mathit{r_0}^{2})^{2} \big( 5 r \omega \mathit{r_0}^{4} + 5 (r^{3} \omega + 54 J ) \mathit{r_0}^{2} + 120 J r  \pi \mathit{r_0} +  J r^{2} (r^{2} \omega^{2} + 30) \big) \mathit{r_0} \phi(r) 
\nn\\ &
+ \big( 10 r^{2} \omega \mathit{r_0}^{6} + 240 J  \pi \mathit{r_0}^{5} + 20 (r^{4} \omega - 3 J r ) \mathit{r_0}^{4} + 480 J r^{2} \pi  \mathit{r_0}^{3} 
\nn\\ &
+ r^{3} (J r^{2} \omega^{2}  + 10 r^{3} \omega + 30 J ) \mathit{r_0}^{2} + 240 J r^{4}  \pi \mathit{r_0} + J r^{5}  (r^{2} \omega^{2} + 30) \big) \mathit{r_0}^{2}
\Big) \, ,
\end{align}\vspace{-16pt}
{\small\begin{align}
\mathfrak{h}_2 & =  \, \mathrm{i} \sqrt{7}  J \mathit{H1}_{3,2}(r) \Big( 
J  (r^{2}+\mathit{r_0}^{2})^{3} (r^{4} \omega^{4} - 1500 r^{2} \omega^{2} - 660 \mathit{r_0}^{2} \omega^{2} - 900) \phi(r)^{2} 
\nn \\ &
+ 2 \Big( 
5 (r^{2} \omega^{3} + 30 \omega) \mathit{r_0}^{5} 
- 330 J \omega^{2}  \pi \mathit{r_0}^{4} 
+ 5 (r^{4} \omega^{3} - 336 J r \omega^{2}  + 30 r^{2} \omega) \mathit{r_0}^{3} 
\nn\\ &
- 1140 J r^{2} \omega^{2}  \pi \mathit{r_0}^{2} 
+ J r  (r^{4} \omega^{4} - 1500 r^{2} \omega^{2} - 900) \mathit{r_0} 
- 810 J r^{4} \pi \omega^{2}  
\Big) (r^{2}+\mathit{r_0}^{2})^{2} \phi(r) 
\nn\\ &
+ \mathit{r_0} \Big( 
10 (r^{3} \omega^{3} - 198 J \omega^{2}  + 30 r \omega) \mathit{r_0}^{7} 
- 1740 J r \omega^{2}  \pi \mathit{r_0}^{6} 
\nn\\ &
+ 20 (r^{5} \omega^{3} - 231 J r^{2} \omega^{2}  + 30 r^{3} \omega + 90 J ) \mathit{r_0}^{5} 
- 5100 J r^{3} \omega^{2}  \pi \mathit{r_0}^{4} 
\nn \\ &
+ r^{2} \left( J r^{4} \omega^{4}  + 10 r^{5} \omega^{3} - 4200 J r^{2} \omega^{2}  + 300 r^{3} \omega - 900 J  \right) \mathit{r_0}^{3} 
\nn \\ &
- 4980 J r^{5} \pi \omega^{2}  \mathit{r_0}^{2} 
+ J r^{4}  (r^{4} \omega^{4} - 1500 r^{2} \omega^{2} - 900) \mathit{r_0} 
- 1620 J r^{7} \omega^{2}  \pi 
\Big) 
\Big) \, ,
\end{align}}
\vspace{-16pt}
{\small\begin{align}
\mathfrak{h}_3 & = \sqrt{7}\,  J T_{3,2}(r) \Big(
(r^{2}+\mathit{r_0}^{2})^{2}  \left( (r^{2} \omega^{4} + 390 \omega^{2}) \mathit{r_0}^{2} + r^{4} \omega^{4} + 140 r^{2} \omega^{2} - 2100 \right) r J \phi(r)^{2} 
\nn \\ &
+ 2 (r^{2}+\mathit{r_0}^{2}) \Big(
5 r \omega^{3} \mathit{r_0}^{7} 
+ (10 r^{3} \omega^{3} + 295 J \omega^{2} ) \mathit{r_0}^{5}
+ 120 J r \omega^{2}  \pi \mathit{r_0}^{4}
\nn \\ &
+ (J r^{4} \omega^{4}  + 5 r^{5} \omega^{3} + 450 J r^{2} \omega^{2}  - 1350 J ) \mathit{r_0}^{3}
+ 120 \pi  r J \left( r^{2} \omega^{2} - \tfrac{15}{2} \right) \mathit{r_0}^{2} 
\nn \\ &
+ J r^{2}  (r^{4} \omega^{4} + 140 r^{2} \omega^{2} - 2100) \mathit{r_0}
- 900 J r^{3} \pi 
\Big) \phi(r) 
\nn \\ &
+ \Big(
(10 r^{2} \omega^{3} + 100 \omega) \mathit{r_0}^{7}
+ 240 J \omega^{2}  \pi \mathit{r_0}^{6}
+ (20 r^{4} \omega^{3} + 290 J r \omega^{2}  + 100 r^{2} \omega) \mathit{r_0}^{5} 
\nn \\ &
+ 480 \pi  \left( r^{2} \omega^{2} - \tfrac{5}{2} \right) J \mathit{r_0}^{4}
+ r \left( J r^{4} \omega^{4}  + 10 r^{5} \omega^{3} + 510 J r^{2} \omega^{2}  - 2700 J  \right) \mathit{r_0}^{3} 
\nn \\ &
+ 240 \pi \left( r^{2} \omega^{2} - \tfrac{25}{2} \right)  r^{2} J \mathit{r_0}^{2}
+ J r^{3}  (r^{4} \omega^{4} + 140 r^{2} \omega^{2} - 2100) \mathit{r_0}
- 1800 J r^{4}  \pi
\Big) \mathit{r_0}
\Big) \, ,
\end{align}}\vspace{-16pt}
\begin{align}
\mathfrak{h}_4 & = 36 J^{2} \Big(
r \omega^{2} (r^{2}+\mathit{r_0}^{2})^{3} \phi(r)^{2} 
+ (r^{2}+\mathit{r_0}^{2}) \Big(
\tfrac{4 \mathit{r_0}^{5} \omega^{2}}{3}
+ \pi r \mathit{r_0}^{4} \omega^{2}
\nn \\ &
+ \tfrac{(151 r^{2} \omega^{2} + 210) \mathit{r_0}^{3}}{45}
+ 2 \pi r^{3} \mathit{r_0}^{2} \omega^{2}
+ 2 \mathit{r_0} r^{4} \omega^{2}
+ \pi r^{5} \omega^{2}
\Big) \phi(r) 
\nn \\ &
+ \Big(
\tfrac{2 \pi \mathit{r_0}^{6} \omega^{2}}{3}
+ 2 \omega^{2} r \mathit{r_0}^{5}
+ \tfrac{7 \pi r^{2} \mathit{r_0}^{4} \omega^{2}}{3}
+ \tfrac{2 r (53 r^{2} \omega^{2} + 105) \mathit{r_0}^{3}}{45}
+ \tfrac{8 r^{4} \omega^{2} \pi \mathit{r_0}^{2}}{3}
\nn \\ &
+ r^{5} \omega^{2} \mathit{r_0}
+ r^{6} \pi \omega^{2}
\Big) \mathit{r_0}
\Big) ^{2} \, ,
\end{align}
\begin{align}
\mathfrak{h}_5 & = 6 \mathrm{i} \Big( 
-\pi \Big( 
\left( r^{2} \omega^{4} - 150 \omega^{2} \right) \mathit{r_0}^{2} 
+ r^{4} \omega^{4} - 90 r^{2} \omega^{2} - 840 
\Big) (r^{2} + \mathit{r_0}^{2})^{3}  J^{2} \phi(r)^{2} 
\nn\\ &
- 2 (r^{2} + \mathit{r_0}^{2}) \Big( 
\tfrac{35 \omega^{3} \pi \mathit{r_0}^{9}}{2} 
+ 35 \left( \tfrac{3}{2} r^{2} \omega^{3} + 2 \omega \right) \pi \mathit{r_0}^{7}
- 55 \omega^{2} \left( \pi^{2} - \tfrac{56}{11} \right)  J \mathit{r_0}^{6} 
\nn\\ &
+ r \omega \pi \left( J r^{2} \omega^{3}  + \tfrac{105}{2} r^{3} \omega^{2} - 80 J \omega  + 140 r \right) \mathit{r_0}^{5} 
- 170 \omega^{2} \left( \pi^{2} - \tfrac{56}{17} \right)  r^{2} J \mathit{r_0}^{4} 
\nn\\ &
+ 2 \pi \left( J r^{4} \omega^{4}  + \tfrac{35}{4} r^{5} \omega^{3} - \tfrac{257}{3} J r^{2} \omega^{2}  + 35 r^{3} \omega - 420 J  \right) r \mathit{r_0}^{3} 
\nn\\ &
- 175 \left( \pi^{2} - \tfrac{8}{5} \right) \omega^{2}  r^{4} J \mathit{r_0}^{2}
+ J r^{3} \pi  \left( r^{4} \omega^{4} - 90 r^{2} \omega^{2} - 840 \right) \mathit{r_0}
- 60 J r^{6} \omega^{2}  \pi^{2} 
\Big)  J \phi(r) 
\nn\\ &
- \Big( 
\tfrac{35 \omega^{4} \mathit{r_0}^{13}}{6} 
+ \tfrac{70 (r^{2} \omega^{4} - \omega^{2}) \mathit{r_0}^{11}}{3} 
+ 35 \omega^{2} r \left( r^{3} \omega^{2} + J \omega  - 2 r \right) \mathit{r_0}^{9} 
\nn\\ &
+ 5 \left( 21 J r^{3} \omega^{3}  + 28 J r \omega  + \tfrac{14}{3} r^{6} \omega^{4} - 14 r^{4} \omega^{2} + \tfrac{128}{3} J^{2} \omega^{2}  \right) \mathit{r_0}^{7} 
- 120 J^{2} r \omega^{2}  \pi \mathit{r_0}^{6} 
\nn\\ &
+ r^{2} \omega \Big( 
J^{2} r^{2} \omega^{3} + 105 J r^{3} \omega^{2}  + \tfrac{610}{3} J^{2} \omega  + 280 J r  + \tfrac{35}{6} r^{6} \omega^{3} - \tfrac{70}{3} r^{4} \omega 
\Big) \mathit{r_0}^{5} 
\nn\\ &
- 360 J^{2} r^{3} \omega^{2} \pi \mathit{r_0}^{4}
+ 2  \Big( J r^{4} \omega^{4}  + \tfrac{35}{2} r^{5} \omega^{3} - \tfrac{154}{3} J r^{2} \omega^{2}  + 70 r^{3} \omega - 420 J  \Big) r^{2} J \mathit{r_0}^{3} 
\nn\\ &
- 360 J^{2} r^{5} \omega^{2}  \pi \mathit{r_0}^{2} 
+ J^{2} r^{4} \left( r^{4} \omega^{4} - 90 r^{2} \omega^{2} - 840 \right) \mathit{r_0} 
- 120 J^{2} r^{7} \omega^{2}  \pi 
\Big) \pi \mathit{r_0}
\Big) \mathit{h1}_{2,2}(r) \, ,
\end{align}
\vspace{-16pt}
\begin{align}
\mathfrak{h}_6 & = 3 \sqrt{7}\,  \Big( 
(r^{2} + \mathit{r_0}^{2})^{2}  J \Big( 
\Big( 90 + \tfrac{5 r^{2} \omega^{2}}{3} \Big) \mathit{r_0}^{4}
+ \Big( \tfrac{8}{3} r^{4} \omega^{2} - 80 r^{2} \Big) \mathit{r_0}^{2}
+ r^{6} \omega^{2} 
\nn\\ &
- 330 r^{4}
\Big) \phi(r)^{2} 
+ 2 (r^{2} + \mathit{r_0}^{2})^{2} \Big( 
\tfrac{25 \mathit{r_0}^{7} \omega}{3}
+ \tfrac{40 r^{2} \mathit{r_0}^{5} \omega}{3}
+ 20 J  \pi \mathit{r_0}^{4}
\nn\\ &
+ \tfrac{5 r (J r^{2} \omega^{2}  + 3 r^{3} \omega - 6 J ) \mathit{r_0}^{3}}{3} 
- 80 J r^{2}  \pi \mathit{r_0}^{2}
+ J r^{3}  (r^{2} \omega^{2} - 330) \mathit{r_0}
\nn\\ &
- 180 J r^{4}  \pi
\Big) \phi(r) 
+ \Big( 
\tfrac{50 r \mathit{r_0}^{9} \omega}{3}
+ \left( 120 J  + \tfrac{130 r^{3} \omega}{3} \right) \mathit{r_0}^{7}
- 40 J \pi r \mathit{r_0}^{6} 
\nn\\ &
+ \tfrac{5 r^{2} (J r^{2} \omega^{2}  + 22 r^{3} \omega + 18 J ) \mathit{r_0}^{5}}{3} 
\nn\\ &
- 440 J \pi r^{3} \mathit{r_0}^{4}  
+ \left( \tfrac{8}{3} J r^{6} \omega^{2}  - 300 J r^{4}  + 10 r^{7} \omega \right) \mathit{r_0}^{3} 
\nn\\ &
- 760 J \pi r^{5} \mathit{r_0}^{2} 
+ J r^{6}  (r^{2} \omega^{2} - 330) \mathit{r_0}
- 360 J \pi r^{7}  
\Big) \mathit{r_0}
\Big) J \phi\mathit{1}_{3,2}(r) \, ,
\end{align}
\noindent
and 
\begin{align}
\mathfrak{h}_7 & = 18 \mathrm{i} \sqrt{7} \Big(
\Big( r^{2} + \tfrac{\mathit{r_0}^{2}}{3} \Big) (r^{2} + \mathit{r_0}^{2})^{2} \phi(r)^{2} 
+ (r^{2} + \mathit{r_0}^{2})^{2} \Big( r^{2} \pi + \tfrac{1}{3} \pi \mathit{r_0}^{2} + 2 r \mathit{r_0} \Big) \phi(r) 
\nn\\ &
+ \mathit{r_0} \Big( 
r^{5} \pi + 2 r^{3} \pi \mathit{r_0}^{2} + r \pi \mathit{r_0}^{4}
+ \mathit{r_0} r^{4} + \tfrac{5}{3} \mathit{r_0}^{3} r^{2} + \mathit{r_0}^{5}
\Big)
\Big) J^{2} 
\nn\\ &
\quad \times \Big(
2 \mathrm{i} \frac{d^{2}}{d r^{2}} \phi\mathit{1}_{3,2}(r)  \mathit{r_0} 
+ \left( r^{2} \omega^{2} + \mathit{r_0}^{2} \omega^{2} - 5 \right) \mathrm{i} \frac{d}{d r} T_{3,2}(r) 
- \omega r \frac{d}{d r} H1_{3,2}(r)
\Big) \, .
\end{align}


For the $\mathit{h1}$ function, the equation is given by
\begin{align}
\frac{d}{d r}\mathit{h1}_{2,2}\! \left(r \right) & = 
\frac{\gamma_1}{35 \mathit{r_0}^{5} \left(r^{2}+\mathit{r_0}^{2}\right)^{2}}-\frac{\gamma_2}{35 \mathit{r_0}^{6} \omega \left(r^{2}+\mathit{r_0}^{2}\right)^{2}}+\frac{\gamma_3}{35 \left(r^{2}+\mathit{r_0}^{2}\right)^{2} \mathit{r_0}^{6}} 
\nn\\
&
+\frac{\gamma_4}{7 r^{2} \omega \,\mathit{r_0}^{6} \pi  \left(r^{2}+\mathit{r_0}^{2}\right)^{3}}-\frac{\gamma_5}{7 \omega^{2} r^{3} \mathit{r_0}^{6} \pi  \left(r^{2}+\mathit{r_0}^{2}\right)^{3}}+\frac{\gamma_6}{35 \mathit{r_0}^{5} \left(r^{2}+\mathit{r_0}^{2}\right)^{3}} \, ,
\end{align}

\noindent
where
\begin{align}
\gamma_1 & = 54 \mathrm{i} J^{2} 
\sqrt{7} \frac{d}{d r} \phi\mathit{1}_{3,2}(r)
\Big(
\Big( r^{2} + \tfrac{\mathit{r_0}^{2}}{3} \Big) (r^{2} + \mathit{r_0}^{2})^{2} \phi(r)^{2} 
\nn \\ &
+ \Big( \pi r^{4} + \tfrac{4}{3} \pi r^{2} \mathit{r_0}^{2} + \tfrac{1}{3} \pi \mathit{r_0}^{4} + 2 r^{3} \mathit{r_0} + \tfrac{53}{27} \mathit{r_0}^{3} r \Big) (r^{2} + \mathit{r_0}^{2}) \phi(r) 
\nn \\ &
+ \mathit{r_0} \Big( r^{5} \pi + 2 r^{3} \pi \mathit{r_0}^{2} + r \pi \mathit{r_0}^{4} + \mathit{r_0} r^{4} + \tfrac{44}{27} \mathit{r_0}^{3} r^{2} + \mathit{r_0}^{5} \Big)
\Big) \, ,
\end{align}
\begin{align}
\gamma_2 & = 27 J^{2} \sqrt{7}\, \Big(
\Big( r^{2} + \tfrac{\mathit{r_0}^{2}}{3} \Big) \omega^{2} r (r^{2} + \mathit{r_0}^{2})^{2} \phi(r)^{2} 
\nn\\
&
+ \Big( 
\tfrac{\pi r \mathit{r_0}^{4} \omega^{2}}{3}
+ \Big( \tfrac{53 r^{2} \omega^{2}}{27} - \tfrac{10}{9} \Big) \mathit{r_0}^{3}
+ \tfrac{4 \pi r^{3} \mathit{r_0}^{2} \omega^{2}}{3}
+ 2 \mathit{r_0} r^{4} \omega^{2}
\nn\\
&
+ \pi r^{5} \omega^{2}
\Big) (r^{2} + \mathit{r_0}^{2}) \phi(r) 
+ r \Big(
\mathit{r_0}^{5} \omega^{2}
+ \pi r \mathit{r_0}^{4} \omega^{2}
+ \Big( \tfrac{44 r^{2} \omega^{2}}{27} - \tfrac{10}{9} \Big) \mathit{r_0}^{3}
\nn\\
&
+ 2 \pi r^{3} \mathit{r_0}^{2} \omega^{2}
+ \mathit{r_0} r^{4} \omega^{2}
+ \pi r^{5} \omega^{2}
\Big) \mathit{r_0}
\Big)  \mathit{H1}_{3,2}(r) \, ,
\end{align}\vspace{-18pt}
{\small\begin{align}
\gamma_3 & = 27 \mathrm{i} \sqrt{7}\, \Big(
(r^{2}+\mathit{r_0}^{2})^{2} \Big( 
\tfrac{\mathit{r_0}^{4} \omega^{2}}{3}
+\Big(-\tfrac{5}{3} + 4 r^{2} \omega^{2} \Big) \frac{\mathit{r_0}^{2}}{3}
+ r^{4} \omega^{2} - \tfrac{35 r^{2}}{9}
\Big)  J \phi(r)^{2} 
\nn \\ &
+ (r^{2}+\mathit{r_0}^{2})^{2} \Big(
\tfrac{5 \mathit{r_0}^{5} \omega}{27}
+ \tfrac{J \omega^{2}  \pi \mathit{r_0}^{4}}{3}
+ 53 \omega r (J \omega  + \tfrac{5 r}{53}) \tfrac{ \mathit{r_0}^{3}}{27}
+ \tfrac{4 }{3} \pi \Big(r^{2} \omega^{2} - \tfrac{5}{4}\Big)  J \mathit{r_0}^{2}
\nn \\ &
+ 2  \Big( r^{2} \omega^{2} - \tfrac{35}{9} \Big) r J \mathit{r_0}
+ J r^{2}  \pi \left( r^{2} \omega^{2} - 5 \right)
\Big) \phi(r) 
\nn \\ &
+ \mathit{r_0} \Big(
\omega \Big( \tfrac{5 r}{27} + J \omega  \Big) \mathit{r_0}^{7}
+ J r \omega^{2}  \pi \mathit{r_0}^{6}
+ \tfrac{ \left(  \left(71 r^{2} \omega^{2} - 70 \right) J + 10 r^{3} \omega \right) \mathit{r_0}^{5} }{27} 
\nn \\ &
+ 3 \pi  \Big( r^{2} \omega^{2} - \tfrac{5}{3} \Big) r J \mathit{r_0}^{4}
+ \tfrac{71 r^{2} \mathit{r_0}^{3} }{27} \Big(  \Big( r^{2} \omega^{2} - \tfrac{195}{71} \Big) J + \tfrac{5 r^{3} \omega}{71} \Big) 
\nn \\ &
+ 3 \pi  \Big( r^{2} \omega^{2} - \tfrac{10}{3} \Big) r^{3} J \mathit{r_0}^{2}
+ J r^{4}  \Big( r^{2} \omega^{2} - \tfrac{35}{9} \Big) \mathit{r_0}
+ J \pi r^{5}  \left( r^{2} \omega^{2} - 5 \right)
\Big)
\Big)  J T_{3,2}(r) \, ,
\end{align}}\vspace{-16pt}
\begin{align}
\gamma_4 & = 72 \mathrm{i} \Big(
-\omega^{2} \pi (r^{2}+\mathit{r_0}^{2})^{3}  J^{2} \Big( r^{4} - \tfrac{1}{4} r^{2} \mathit{r_0}^{2} - \tfrac{7}{12} \mathit{r_0}^{4} \Big) \phi(r)^{2} 
\nn \\ &
- \omega (r^{2}+\mathit{r_0}^{2})^{3}  J \Big(
-\tfrac{7 J \omega  (\pi^{2}-8) \mathit{r_0}^{4}}{24}
+ \tfrac{7 \pi r \mathit{r_0}^{3}}{6}  (J \omega  + \tfrac{r}{2})
\nn \\ &
+ \tfrac{J r^{2} \omega  (\pi^{2}+56) \mathit{r_0}^{2}}{24} 
+ 2 J r^{3} \omega  \pi \mathit{r_0}
+ J r^{4} \omega  \pi^{2}
\Big) \phi(r) 
\nn \\ &
- \Big(
\tfrac{7 r^{2} \omega^{2} \pi \mathit{r_0}^{11}}{72}
+ \Big( J^{2}  + \tfrac{r^{4}}{4} \Big)\tfrac{7 \omega^{2} \pi  \mathit{r_0}^{9}}{6}
+ \tfrac{7 J^{2} r \omega^{2}  (\pi^{2}-8) \mathit{r_0}^{8}}{24} 
\nn \\ &
+ \Big( \Big( J^{2}  + \tfrac{7 r^{4}}{122} \Big) \omega + \tfrac{7 J r }{61} \Big)\tfrac{61 \omega \pi r^{2}  \mathit{r_0}^{7} }{12}
+ \Big( \pi^{2} - \tfrac{56}{15} \Big) \tfrac{15 \omega^{2}   r^{3} J^{2} \mathit{r_0}^{6} }{8} 
\nn \\ &
+ \Big( \Big( J^{2} r^{2}  + \tfrac{7}{528} r^{6} \Big) \omega^{2} + \tfrac{7 J r^{3} \omega }{44} + \tfrac{7 J^{2} }{66} \Big)
\tfrac{22 \pi r^{2}  \mathit{r_0}^{5} }{3}
\nn\\ &
+ \Big( \pi^{2} - \tfrac{56}{31} \Big) \tfrac{31 \omega^{2}   r^{5} J^{2} \mathit{r_0}^{4} }{8} 
+ \Big( J \omega  + \tfrac{7 r}{53} \Big) \tfrac{53 \omega \pi   r^{6} J \mathit{r_0}^{3}}{12}
\nn\\ &
+ \Big( \pi^{2} - \tfrac{56}{79} \Big)\tfrac{79 \omega^{2}   r^{7} J^{2} \mathit{r_0}^{2} }{24} 
+ J^{2} r^{8} \omega^{2} \pi \mathit{r_0}
+ J^{2} r^{9} \omega^{2}  \pi^{2}
\Big) \mathit{r_0}
\Big) \mathit{h0}_{2,2}(r) \, ,
\end{align}\vspace{-16pt}
\begin{align}
\gamma_5 & = 36 \Big(
\omega^{2} \pi (r^{2} + \mathit{r_0}^{2})^{3} \Big( r^{4} - \tfrac{7 \mathit{r_0}^{4}}{6} \Big)  J \phi(r)^{2} 
\nn \\ &
+ (r^{2} + \mathit{r_0}^{2})  J \Big(
-\tfrac{7 \omega^{2} (\pi^{2}-8) \mathit{r_0}^{8}}{12}
+ \tfrac{7 r \pi \mathit{r_0}^{7} \omega^{2}}{3}
- \tfrac{7 r^{2} \omega^{2} (\pi^{2}-8) \mathit{r_0}^{6}}{6}
+ 6 r^{3} \pi \omega^{2} \mathit{r_0}^{5} 
\nn \\ &
+ \tfrac{5}{12} \omega^{2} \Big(\pi^{2} + \tfrac{56}{5}\Big) r^{4} \mathit{r_0}^{4}
+ \tfrac{17 }{3} \Big(r^{2} \omega^{2} + \tfrac{28}{17} \Big) r^{3} \pi \mathit{r_0}^{3}
+ 2 r^{6} \pi^{2} \omega^{2} \mathit{r_0}^{2}
+ 2 r^{7} \pi \omega^{2} \mathit{r_0}
\nn \\ &
+ \pi^{2} r^{8} \omega^{2}
\Big) \phi(r) 
+ \Big(
\tfrac{7 J \pi \mathit{r_0}^{9} \omega^{2} }{3}
+ \tfrac{7 J \omega^{2} r  (\pi^{2}-8) \mathit{r_0}^{8}}{12}
\nn \\ &
+ \tfrac{35}{6} \omega \pi \Big( J \omega  - \tfrac{4 r}{15} \Big) r^{2} \mathit{r_0}^{7}
+ \tfrac{29}{12} \omega^{2}  \Big( \pi^{2} - \tfrac{168}{29} \Big) r^{3} J \mathit{r_0}^{6} 
\nn \\ &
+ \tfrac{21}{2} \omega \pi \Big( J \omega  - \tfrac{4 r}{27} \Big) r^{4} \mathit{r_0}^{5}
+ \tfrac{49}{12} \Big(\pi^{2} - \tfrac{24}{7} \Big) \omega^{2}  r^{5} J \mathit{r_0}^{4}
+ \tfrac{14}{3}  J r^{4}  \pi (r^{2} \omega^{2} + 2) \mathit{r_0}^{3}
\nn \\ &
+ \tfrac{13}{4} \omega^{2}  \Big(\pi^{2} - \tfrac{56}{39}\Big) r^{7} J \mathit{r_0}^{2}
+ J \pi r^{8} \mathit{r_0} \omega^{2} 
+ J \pi^{2} r^{9} \omega^{2} 
\Big) \mathit{r_0}
\Big)  J \mathit{h1}_{2,2}(r) \, ,
\end{align}

\noindent
and 
\begin{align}
\gamma_6 & = 108 \mathrm{i} J^{2} \Big(
\Big( r^{2} + \tfrac{\mathit{r_0}^{2}}{3} \Big) (r^{2} + \mathit{r_0}^{2})^{2} r \phi(r)^{2} 
\nn \\ &
+ (r^{2} + \mathit{r_0}^{2}) \Big(
r^{5} \pi
+ \tfrac{4}{3} r^{3} \pi \mathit{r_0}^{2}
+ \tfrac{1}{3} r \pi \mathit{r_0}^{4}
+ 2 \mathit{r_0} r^{4}
+ \tfrac{37}{18} \mathit{r_0}^{3} r^{2}
+ \tfrac{5}{54} \mathit{r_0}^{5}
\Big) \phi(r) 
\nn \\ &
+ r \mathit{r_0} \Big(
r^{5} \pi
+ 2 r^{3} \pi \mathit{r_0}^{2}
+ r \pi \mathit{r_0}^{4}
+ \mathit{r_0} r^{4}
+ \tfrac{23}{9} \mathit{r_0}^{3} r^{2}
+ \tfrac{52}{27} \mathit{r_0}^{5}
\Big)
\Big) \sqrt{7}\, \phi\mathit{1}_{3,2}(r) \, .
\end{align}

Depending on the perturbation function, its differential equation in the minimal system may contain only up to first-order contributions from the rotation.
In fact, in this sector of axial perturbations, this is true for all the ODEs of the polar metric functions  $\mathit{H1}(r),T(r),L(r),N(r)$ and the scalar function $\phi\mathit{1}(r)$,
whereas for the axial metric functions $\mathit{h0}(r)$
and $\mathit{h1}(r)$, they contain terms up to the second order in rotation.

\subsection[\appendixname~\thesubsection]{Polar Perturbation Equations}

Next, we show the polar perturbation equations.
Here, the mixing of $l$ is reversed. 
The minimal system of equations is described by five ODEs for $\phi\mathit{1}_{2,2}, \mathit{H1}_{2,2}, T_{2,2},  \mathit{h0}_{3,2}$, and $ \mathit{h1}_{3,2}$, and two algebraic equations for $ L_{2,2}, N_{2,2}$.
Due to the nature of the polar perturbations,
the polar metric perturbation functions $\mathit{H1}, T, L, N$ and the phantom scalar perturbation function $\phi\mathit{1}$ are the ones carrying the quantum numbers $l=M_z=2$.
They are coupled with the axial metric perturbation functions $\mathit{h0}, \mathit{h1}$ with $l=3, M_z=2$.
Note that here the polar perturbation equations are significantly more involved compared to the axial pe\mbox{rturbation equati}ons.

First we obtain the equation for the polar-led perturbations of the phantom scalar field
\begin{align}
\frac{d^{2}}{d r^{2}}{\phi \mathit{1}}_{2,2}\! \left(r \right) & = 
\frac{\chi_1}{7 \mathit{r_0}^{3} r^{3} \pi  \left(r^{2}+\mathit{r_0}^{2}\right)^{4}}+\frac{\chi_2}{7 \mathit{r_0}^{5} \omega \pi  \left(r^{2}+\mathit{r_0}^{2}\right)^{4}}-\frac{\chi_3}{7 \mathit{r_0}^{5} \pi  \left(r^{2}+\mathit{r_0}^{2}\right)^{4}} 
\nn \\ &
+ 
\frac{\chi_4}{\mathit{r_0}^{5} \omega \left(r^{2}+\mathit{r_0}^{2}\right)^{5}}-\frac{\chi_5}{7 \mathit{r_0}^{5} \omega^{2} \left(r^{2}+\mathit{r_0}^{2}\right)^{5}}+\frac{\chi_6}{7 r^{2} \pi  \left(r^{2}+\mathit{r_0}^{2}\right)^{5} \mathit{r_0}^{6}} \, ,
\end{align}

\noindent
where
\begin{align}
\chi_1 & =  \frac{d}{d r} {\phi \mathit{1}}_{2,2}(r)  \Big(
-6 J^{2} \pi \, \mathit{r_0} \left( r^{4} - 14 r^{2} \mathit{r_0}^{2} - 7 \mathit{r_0}^{4} \right) \left( r^{2} + \mathit{r_0}^{2} \right)^{2} \phi(r)^{2} 
\nn \\ &
-84  \left( r^{2} + \mathit{r_0}^{2} \right) J^{2} \Big( 
\Big( -\frac{\pi^{2}}{4} + 2 \Big) \mathit{r_0}^{7} 
+ r \pi \mathit{r_0}^{6} 
- \frac{3 r^{2} \left( \pi^{2} - 8 \right) \mathit{r_0}^{5}}{4} 
+ \frac{59 r^{3} \pi \mathit{r_0}^{4}}{21} 
\nn \\ &
+ r^{4} \Big( -\frac{5 \pi^{2}}{28} + 6 \Big) \mathit{r_0}^{3}
+ \frac{73 \pi r^{5} \mathit{r_0}^{2}}{21}
+ r^{6} \Big( \frac{9 \pi^{2}}{28} + 2 \Big) \mathit{r_0}
+ r^{7} \pi 
\Big) \phi(r) 
\nn \\ &
+ 14 \pi \left( -6 J^{2}  - r^{4} \right) \mathit{r_0}^{9} 
- 21 J^{2} r  \left( \pi^{2} - 8 \right) \mathit{r_0}^{8} 
+ 42 \pi \left( -7 J^{2} r^{2}  - r^{6} \right) \mathit{r_0}^{7} 
\nn \\ &
- 111  r^{3} J^{2} \Big( \pi^{2} - \frac{168}{37} \Big) \mathit{r_0}^{6}
+ 2 \pi \left( -58 J^{2} r^{4}  - 21 r^{8} \right) \mathit{r_0}^{5}
\nn \\ &
- 180  r^{5} J^{2} \Big( \pi^{2} - \frac{56}{15} \Big) \mathit{r_0}^{4} 
+ 2 \pi \left( -89 J^{2} r^{6}  - 7 r^{10} \right) \mathit{r_0}^{3}
\nn \\ &
- 111  r^{7} J^{2} \Big( \pi^{2} - \frac{168}{37} \Big) \mathit{r_0}^{2}
- 84 J^{2} \pi r^{8} \mathit{r_0}
- 21 J^{2} r^{9} \left( \pi^{2} - 8 \right)
\Big) \, ,
\end{align}\vspace{-16pt}
\begin{align}
\chi_2 & = 72 \,\mathrm{i} \Big(
\left( r^{2}+\mathit{r_0}^{2} \right)^{2} \pi \Big( \frac{\mathit{r_0}^{4} \omega^{2}}{3} + \Big( r^{2} \omega^{2} - \frac{1}{2} \Big) \mathit{r_0}^{2} + r^{4} \omega^{2} - \frac{3 r^{2}}{2} \Big) \phi(r)^{2} 
\nn \\ &
+ \left( r^{2} + \mathit{r_0}^{2} \right) \pi \Big( \left( r^{2} + \mathit{r_0}^{2} \right) \Big( \frac{\mathit{r_0}^{4} \omega^{2}}{3} + \Big( r^{2} \omega^{2} - \frac{1}{2} \Big) \mathit{r_0}^{2} + r^{4} \omega^{2} - \frac{3 r^{2}}{2} \Big) \pi 
\nn \\ &
+ 2 r \mathit{r_0} \Big( \frac{19 \mathit{r_0}^{4} \omega^{2}}{18} + \Big( \frac{67 r^{2} \omega^{2}}{36} - \frac{8}{3} \Big) \mathit{r_0}^{2} + r^{4} \omega^{2} - \frac{3 r^{2}}{2} \Big) \Big) \phi(r) 
\nn \\ &
+ \Big( \left( r^{2} + \mathit{r_0}^{2} \right)^{2} r \Big( r^{2} \omega^{2} + \frac{13}{16} \mathit{r_0}^{2} \omega^{2} - \frac{3}{2} \Big) \pi^{2} 
\nn \\ &
+ \Big( \mathit{r_0}^{6} \omega^{2} + \frac{77 \mathit{r_0}^{4} r^{2} \omega^{2}}{18} + \Big( \frac{107}{36} r^{4} \omega^{2} - \frac{29}{6} r^{2} \Big) \mathit{r_0}^{2} + r^{6} \omega^{2} - \frac{3 r^{4}}{2} \Big) 
\mathit{r_0} \pi 
\nn \\ &
- \frac{7 \omega^{2} r \mathit{r_0}^{2} \left( r^{2} + \mathit{r_0}^{2} \right)^{2} }{6} \Big) \mathit{r_0}
\Big) J^{2}  \mathit{H1}_{2,2}(r) \, ,
\end{align}
\vspace{-16pt}
\begin{align}
\chi_3 & = 24  J \Big(
\pi \left( r^{2} + \mathit{r_0}^{2} \right)^{2}  \Big( \mathit{r_0}^{4} \omega^{2} + \Big( r^{2} \omega^{2} - \frac{3}{2} \Big) \mathit{r_0}^{2} + \frac{3 r^{2}}{2} \Big) r J \phi(r)^{2} 
\nn \\ &
+ \pi \left( r^{2} + \mathit{r_0}^{2} \right)^{2}  \Big( -\frac{\mathit{r_0}^{5} \omega^{2}}{3} + \pi r \mathit{r_0}^{4} \omega^{2} + \frac{ \left( 5 r^{2} \omega^{2} + 31 \right) \mathit{r_0}^{3} }{6} 
+ r \pi \Big( r^{2} \omega^{2} - \frac{3}{2} \Big) \mathit{r_0}^{2}
\nn \\ &
+ 3 r^{2} \mathit{r_0} + \frac{3 r^{3} \pi}{2} \Big) J \phi(r) 
+ \frac{9}{16} \Big( \frac{56 \pi \mathit{r_0}^{9} \omega}{27} + J \omega^{2}  \Big( \pi^{2} + \frac{56}{9} \Big) \mathit{r_0}^{8} 
\nn \\ &
- \frac{232  \omega \pi r \mathit{r_0}^{7}}{27} \Big( J \omega  - \frac{14 r}{29} \Big)
+ 3   J \mathit{r_0}^{6} \Big( r^{2} \omega^{2} - \frac{2}{3} \Big) \Big( \pi^{2} + \frac{56}{9} \Big)  
\nn \\ &
- \frac{92 }{9} \left(  \Big( r^{2} \omega^{2} - \frac{110}{69} \Big) J - \frac{14 r^{3} \omega}{69} \right) \pi r \mathit{r_0}^{5}
\nn \\ &
+ 3  r^{2} J \left( \Big( r^{2} \omega^{2} - \frac{4}{9} \Big) \pi^{2} + \frac{56 r^{2} \omega^{2}}{9} - \frac{224}{27} \right) \mathit{r_0}^{4} 
\nn \\ &
- \frac{44 \pi  r^{3} J  \mathit{r_0}^{3}}{27}  \Big( r^{2} \omega^{2} - \frac{116}{11} \Big)
+  r^{4} J \left( \Big( r^{2} \omega^{2} + \frac{10}{3} \Big) \pi^{2} + \frac{56 r^{2} \omega^{2}}{9} - \frac{112}{9} \right) \mathit{r_0}^{2} 
\nn \\ &
+ \frac{8 J \pi r^{5} \mathit{r_0} }{3} + \frac{8 J \pi^{2} r^{6} }{3} \Big) \mathit{r_0}
\Big) T_{2,2}(r) \, ,
\end{align}\vspace{-16pt}
\begin{align}
\chi_4 & = 36 \Big(
J \omega^{2} r  \left( r^{2} + \mathit{r_0}^{2} \right)^{4} \phi(r)^{2}
+ \omega^{2} \left( r^{2} + \mathit{r_0}^{2} \right)^{3}  \Big( r^{3} \pi + \pi r \mathit{r_0}^{2} + 2 r^{2} \mathit{r_0} 
\nn \\ &
+ \frac{4}{3} \mathit{r_0}^{3} \Big) J \phi(r) 
\nn \\ &
+ \Big(
\frac{5 \mathit{r_0}^{9} \omega}{21}
+ \frac{2 J \pi \mathit{r_0}^{8} \omega^{2} }{3}
+ \frac{4}{7} \omega r \Big( J \omega  + \frac{5 r}{6} \Big) \mathit{r_0}^{7}
+ 3 J \pi r^{2} \mathit{r_0}^{6} \omega^{2}  
\nn \\ &
+ \frac{61}{21} \Big( J r^{2} \omega^{2}  + \frac{5}{61} r^{3} \omega - \frac{2}{61} J  \Big) r \mathit{r_0}^{5}
+ 5 J \pi r^{4} \mathit{r_0}^{4} \omega^{2} 
+ \frac{10 J r^{5} \mathit{r_0}^{3} \omega^{2} }{3} 
\nn \\ &
+ \frac{11 J \pi r^{6} \mathit{r_0}^{2} \omega^{2} }{3}
+ J r^{7} \mathit{r_0} \omega^{2} 
+ J \pi r^{8} \omega^{2} 
\Big) \mathit{r_0}
\Big) \sqrt{7}\,  J \mathit{h0}_{3,2}(r) \, ,
\end{align}
\begin{adjustwidth}{-\extralength}{0cm}
\vspace{-16pt}\begin{align}
\chi_5 & = 90 \,\mathrm{i} \sqrt{7}  \Bigg[
\left( \frac{\mathit{r_0}^{4} \omega^{2}}{3} + \left( \frac{4 r^{2} \omega^{2}}{3} - \frac{14}{5} \right) \mathit{r_0}^{2} + r^{4} \omega^{2} - \frac{34 r^{2}}{5} \right) \omega^{2} \left( r^{2} + \mathit{r_0}^{2} \right)^{3}  J \phi(r)^{2} 
\nn \\ &
+ \Big( \frac{\omega^{4} \pi \,\mathit{r_0}^{8}}{3} + 2 r \,\mathit{r_0}^{7} \omega^{4} + 2 \pi \omega^{2} \Big( r^{2} \omega^{2} - \frac{7}{5} \Big) \mathit{r_0}^{6}
+ \Big( 6 r^{3} \omega^{4} - \frac{44}{3} \omega^{2} r \Big) \mathit{r_0}^{5} 
\nn \\ &
+ 4 r^{2} \pi \Big( r^{2} \omega^{2} - \frac{31}{10} \Big) \omega^{2} \mathit{r_0}^{4}
+ \Big( 6 r^{5} \omega^{4} - \frac{424}{15} r^{3} \omega^{2} - 8 r \Big) \mathit{r_0}^{3} 
\nn \\ &
+ \frac{10}{3} \Big( r^{2} \omega^{2} - \frac{123}{25} \Big) r^{4} \pi \omega^{2} \mathit{r_0}^{2}
+ \Big( 2 r^{7} \omega^{4} - \frac{68}{5} r^{5} \omega^{2} \Big) \mathit{r_0}
\nn \\ &
+ \pi r^{6} \omega^{2} \Big( r^{2} \omega^{2} - \frac{34}{5} \Big)
\Big) 
\left( r^{2} + \mathit{r_0}^{2} \right)  J \phi(r) \nn \\ &
+ \Big(  \Big( 
\mathit{r_0}^{9} \omega^{4} + \pi r \mathit{r_0}^{8} \omega^{4}
+ \Big( \frac{11}{3} r^{2} \omega^{4} - \frac{44}{5} \omega^{2} \Big) \mathit{r_0}^{7}
+ 4 \Big( r^{2} \omega^{2} - \frac{11}{6} \Big) r \pi \omega^{2} \mathit{r_0}^{6} 
\nn \\ &
+ \Big( \frac{16}{3} r^{4} \omega^{4} - 28 r^{2} \omega^{2} \Big) \mathit{r_0}^{5}
+ 6 \Big( r^{2} \omega^{2} - \frac{161}{45} \Big) r^{3} \pi \omega^{2} \mathit{r_0}^{4} 
\nn \\ &
+ \Big( -8 r^{2} + \frac{11}{3} r^{6} \omega^{4} - \frac{262}{15} r^{4} \omega^{2} \Big) \mathit{r_0}^{3}
+ 4 r^{5} \pi \omega^{2} \Big( r^{2} \omega^{2} - \frac{157}{30} \Big) \mathit{r_0}^{2} 
\nn \\ &
+ \Big( r^{8} \omega^{4} - \frac{34}{5} r^{6} \omega^{2} \Big) \mathit{r_0}
+ \pi r^{7} \omega^{2} \Big( r^{2} \omega^{2} - \frac{34}{5} \Big)
\Big) J + \frac{4 r \omega \mathit{r_0}^{5} \Big( r^{2} + \mathit{r_0}^{2} \Big)}{3} \Big) \mathit{r_0}
\Bigg] J \mathit{h1}_{3,2}(r) \, ,
\end{align}
\end{adjustwidth}

\noindent
and
\vspace{6pt}
{\small
\begin{align}
 \chi_6  &= 72 \, {\phi \mathit{1}}_{2,2}(r) \Big(
\pi \left(r^{2}+\mathit{r_0}^{2}\right)^{2}  
\Big(
\frac{7 \mathit{r_0}^{10} \omega^{2}}{12}
+ \frac{ \mathit{r_0}^{8}}{3} \left(-\frac{7}{2}+8 r^{2} \omega^{2}\right)
\nn \\ &
+ \frac{ \mathit{r_0}^{6}}{2} \left(11 r^{4} \omega^{2}-\frac{37}{3} r^{2}\right)
+ \frac{\left(19 r^{6} \omega^{2}-43 r^{4}\right) \mathit{r_0}^{4}}{3}
+ \frac{ \mathit{r_0}^{2}}{2} \left( \frac{47}{6} r^{8} \omega^{2} - 23 r^{6} \right)
\nn \\ &
+ r^{8} \left(r^{2} \omega^{2} - \frac{7}{2}\right)
\Big) 
J^{2} \phi(r)^{2} 
\nn \\ &
+ \left(r^{2}+\mathit{r_0}^{2}\right) \Big(
\frac{7 J \omega^{2}  (\pi^{2}-8) \mathit{r_0}^{12}}{24}
- \frac{7 r \omega \pi (J \omega  + r) \mathit{r_0}^{11}}{6}
\nn \\ &
+ \frac{43}{24}  J \left(\omega^{2} \left(\pi^{2} - \frac{280}{43}\right) r^{2} - \frac{14 \pi^{2}}{43} + \frac{112}{43}\right) \mathit{r_0}^{10} 
\nn \\ &
- \frac{8  \pi r \mathit{r_0}^{9}}{3} \left(J r^{2} \omega^{2}  + \frac{7}{4} r^{3} \omega - \frac{7}{8} J  \right)
+ \frac{21   r^{2} J \mathit{r_0}^{8}}{4} \left( \omega^{2} \left( \pi^{2} - \frac{40}{9} \right) r^{2} - \frac{25 \pi^{2}}{63} + \frac{8}{3} \right)
\nn \\ &
+ r^{3} \pi \left( J r^{2} \omega^{2}  - 7 r^{3} \omega + \frac{56}{9} J  \right) \mathit{r_0}^{7} 
\nn \\ &
+ \frac{107}{12} \left( \omega^{2} \left( \pi^{2} - \frac{280}{107} \right) r^{2} - \frac{57 \pi^{2}}{107} + \frac{168}{107}\right)  r^{4} J \mathit{r_0}^{6}
\nn \\ &
+ \frac{22}{3}  \pi r^{5} \left(J r^{2} \omega^{2}  - \frac{7}{11} r^{3} \omega - \frac{10}{11} J  \right) \mathit{r_0}^{5}
\nn \\ &
+ \frac{211}{24}  r^{6} J \left(\omega^{2} \left( \pi^{2} - \frac{280}{211} \right) r^{2} - \frac{102 \pi^{2}}{211} + \frac{112}{211} \right) \mathit{r_0}^{4}
\nn \\ &
+ \frac{41}{6}  \pi \left(J r^{2} \omega^{2}  - \frac{7}{41} r^{3} \omega - \frac{96}{41} J  \right) r^{7} \mathit{r_0}^{3}
\nn \\ &
+ \frac{37}{8} \left(\omega^{2} \left( \pi^{2} - \frac{56}{111} \right) r^{2} - \frac{8 \pi^{2}}{37}\right)  r^{8} J \mathit{r_0}^{2}
\nn \\ &
+ 2 \pi  r^{9} J \left(r^{2} \omega^{2} - \frac{7}{2}\right) \mathit{r_0}
+ J r^{12} \omega^{2}  \pi^{2}
\Big) \phi(r)  J 
\nn \\ &
+ \Big(
-\frac{7 r^{2} \omega^{2} \pi \mathit{r_0}^{15}}{72}
- \frac{7}{6} \left( J^{2} \omega^{2}  + \frac{5}{12} r^{4} \omega^{2} 
- \frac{1}{6} r^{2}\right) \pi \mathit{r_0}^{13}
- \frac{7 J^{2} r \omega^{2}  (\pi^{2}-8) \mathit{r_0}^{12}}{24}
\nn \\ &
-  \frac{65}{12}  \pi \left( J^{2} r^{2} \omega^{2}  + \frac{7}{39} r^{6} \omega^{2} + \frac{14}{65} J r^{3} \omega  - \frac{28}{65} J^{2}  - \frac{14}{65} r^{4}\right) \mathit{r_0}^{11}
\nn \\ &
- \frac{11 }{24}  r J^{2} \left( \omega^{2} \left( \pi^{2} - \frac{280}{11} \right) r^{2} - \frac{14 \pi^{2}}{11} + \frac{112}{11} \right) \mathit{r_0}^{10}
\nn \\ &
-  \frac{28}{3} \pi r^{2} \left( J^{2} r^{2} \omega^{2}  + \frac{5}{48} r^{6} \omega^{2} + \frac{1}{2} J r^{3} \omega  - \frac{43}{56} J^{2}  - \frac{1}{4} r^{4} \right) \mathit{r_0}^{9} 
\nn \\ &
+ \frac{25}{12} \left( \omega^{2} \left( \pi^{2} + \frac{56}{5} \right) r^{2} + \frac{21 \pi^{2}}{25} - \frac{392}{25} \right)  r^{3} J^{2} \mathit{r_0}^{8}
\nn \\ &
-  \frac{37}{6}  \pi r^{4} \left( J^{2} r^{2} \omega^{2}  + \frac{35}{444} r^{6} \omega^{2} + \frac{42}{37} J r^{3} \omega  - \frac{391}{111} J^{2}  - \frac{35}{111} r^{4} \right) \mathit{r_0}^{7}
\nn \\ &
+ \frac{85}{12}  \left( \omega^{2} \left( \pi^{2} + \frac{56}{17} \right) r^{2} + \frac{6 \pi^{2}}{17} - \frac{784}{85} \right) r^{5} J^{2} \mathit{r_0}^{6}
\nn \\ &
+  \frac{5}{6} \pi r^{6} \left( J^{2} r^{2} \omega^{2}  - \frac{7}{60} r^{6} \omega^{2} - \frac{28}{5} J r^{3} \omega  + \frac{64}{5} J^{2}  + \frac{7}{10} r^{4} \right) \mathit{r_0}^{5} 
\nn \\ &
+ \frac{205 }{24} \left( \omega^{2} \left( \pi^{2} + \frac{56}{41} \right) r^{2} + \frac{10 \pi^{2}}{41} - \frac{1232}{205} \right)  r^{7} J^{2} \mathit{r_0}^{4}
\nn \\ &
+ \frac{35}{12} \pi  r^{8} \left( J r^{2} \omega^{2}  - \frac{2}{5} r^{3} \omega - \frac{54}{35} J  \right) J \mathit{r_0}^{3} \nn \\ &
+ \frac{113}{24} \left( \omega^{2} \left( \pi^{2} + \frac{56}{113} \right) r^{2} + \frac{18 \pi^{2}}{113} - \frac{336}{113} \right) r^{9} J^{2} \mathit{r_0}^{2}
\nn \\ &
+ J^{2} r^{10} \pi \left( r^{2} \omega^{2} - \frac{7}{2} \right) \mathit{r_0}
+ J^{2} r^{13} \omega^{2}  \pi^{2}
\Big) \mathit{r_0}
\Big)  \, .
\end{align}}


For the metric perturbation function $\mathit{H1}$, we obtain the equation
{
\footnotesize
\begin{align}
\frac{d}{d r}\mathit{H1}_{2,2}\! \left(r \right) & = 
\frac{\alpha_1}{4 r^{2} \mathit{r_0}^{5} \omega \pi  \left(r^{2}+\mathit{r_0}^{2}\right)^{3}}
\nn\\ &
+\left(\frac{\alpha_2}{8 \pi  \,\omega^{2} \left(r^{2}+\mathit{r_0}^{2}\right)^{3} r^{3} \mathit{r_0}^{6}}+\frac{\alpha_3}{4 \omega \,\mathit{r_0}^{3} \left(r^{2}+\mathit{r_0}^{2}\right)^{2}}+\frac{\omega^{2} r}{2}\right) \mathit{H1}_{2,2}\! \left(r \right)
\nn\\ &
+  
\frac{\alpha_4}{8 r^{2} \omega \pi  \left(r^{2}+\mathit{r_0}^{2}\right)^{3} \mathit{r_0}^{6}}+\frac{\alpha_5}{7 \mathit{r_0}^{6} \left(r^{2}+\mathit{r_0}^{2}\right)^{4}}-\frac{\alpha_6}{7 \omega \,\mathit{r_0}^{6} \left(r^{2}+\mathit{r_0}^{2}\right)^{5}}
\nn\\ &
+\frac{\alpha_7}{14 r \,\mathit{r_0}^{5} \omega \pi  \left(r^{2}+\mathit{r_0}^{2}\right)^{4}} \, ,
\end{align}}

\noindent
{where}
{\footnotesize
\begin{align}
\alpha_1 & = \mathrm{i} \frac{d}{dr} \phi\mathit{1}_{2,2}(r) \Big(
-\Big(
-12 \mathit{r_0}^{6} \omega^{2} + \tfrac{60}{7} \mathit{r_0}^{4} r^{2} \omega^{2} 
+ \left( r^{6} \omega^{4} + \tfrac{782}{7} r^{4} \omega^{2} \right) \mathit{r_0}^{2} 
\nn\\ &
+ r^{4} ( r^{4} \omega^{4} + \tfrac{442}{7} r^{2} \omega^{2} - 24 )
\Big) 
 \pi (r^{2} + \mathit{r_0}^{2})^{2}  J^{2} \phi(r)^{2} 
\nn\\ &
- 2 (r^{2} + \mathit{r_0}^{2})  \Big(
-3 J \omega^{2}  (\pi^{2} - 8) \mathit{r_0}^{8} 
+ 12 \omega \pi r \left( \tfrac{1}{12} r^{3} \omega^{2} + J \omega  + \tfrac{2}{3} r \right) \mathit{r_0}^{7} 
\nn\\ &
+ \tfrac{51}{7} \omega^{2} \left(\pi^{2} + \tfrac{168}{17}\right)  r^{2} J \mathit{r_0}^{6} 
+ \tfrac{94}{47} \left( r^{3} \omega^{2} + J \omega  + \tfrac{6}{47} r \right) \omega \pi r^{3} \mathit{r_0}^{5} 
\nn\\ &
+ \tfrac{393}{7} \left(\pi^{2} + \tfrac{168}{131}\right) \omega^{2}  r^{4} J \mathit{r_0}^{4} 
+ \tfrac{465}{7} \omega^{2}  \left(\pi^{2} + \tfrac{56}{155}\right) r^{6} J \mathit{r_0}^{2}
\nn\\ &
+ r^{3} \pi \left( J r^{4} \omega^{4}  + r^{5} \omega^{3} + \tfrac{1072}{7} J r^{2} \omega^{2}  + 4 r^{3} \omega - 24 J  \right) \mathit{r_0}^{3} 
\nn\\ &
+ J r^{5}  \pi \left( r^{4} \omega^{4} + \tfrac{442}{7} r^{2} \omega^{2} - 24 \right) \mathit{r_0} 
+ \tfrac{144}{7} J r^{8} \omega^{2}  \pi^{2}
\Big) J \phi(r) 
\nn\\ &
- \mathit{r_0} \Big(
4 r^{2} \omega^{2} \pi \mathit{r_0}^{11} 
+ 24 \omega^{2} \pi \left( J^{2} + \tfrac{r^{4}}{2} \right) \mathit{r_0}^{9} 
+ 6 J^{2} r \omega^{2}  (\pi^{2} - 8) \mathit{r_0}^{8} 
\nn\\ &
+ \tfrac{1272}{7} \left( \tfrac{7}{636} J r^{3} \omega^{2}  + J^{2} \omega  + \tfrac{7}{106} r^{4} \omega + \tfrac{14}{159} J r  \right) \omega \pi r^{2} \mathit{r_0}^{7} 
\nn\\ &
+ 93 \omega^{2} \left( \pi^{2} - \tfrac{24}{31} \right)  r^{3} J^{2} \mathit{r_0}^{6}
+ \tfrac{1916}{7} \omega \pi \Biggl( \tfrac{7}{479} J r^{3} \omega^{2}  + J^{2} \omega  + \tfrac{7}{479} r^{4} \omega 
\nn\\ &
+ \tfrac{42}{479} J r  \Biggr) r^{4} \mathit{r_0}^{5} 
+ \tfrac{1464}{7} J^{2} r^{5} \omega^{2}  \pi^{2} \mathit{r_0}^{4}
\nn\\ &
+ J r^{4}  \pi \left( J r^{4} \omega^{4}  + 2 r^{5} \omega^{3} + \tfrac{1410}{7} J r^{2} \omega^{2}  + 8 r^{3} \omega - 48 J  \right) \mathit{r_0}^{3} 
\nn\\ &
+ \tfrac{1143}{7} \omega^{2}  r^{7} J^{2} \left( \pi^{2} + \tfrac{56}{381} \right) \mathit{r_0}^{2}
+ J^{2} r^{6} \pi \left( r^{4} \omega^{4} + \tfrac{442}{7} r^{2} \omega^{2} - 24 \right) \mathit{r_0} 
\nn\\ &
+ \tfrac{288}{7} J^{2} r^{9} \omega^{2}  \pi^{2}
\Big)
\Big) \, ,
\end{align}
\begin{align}
\alpha_2 & = \Big(
\pi 
\Big( 
(r^{10} + r^{8} \mathit{r_0}^{2}) \omega^{6} 
- \tfrac{156}{7} r^{4} \left( \mathit{r_0}^{4} - \tfrac{62}{39} r^{2} \mathit{r_0}^{2} - \tfrac{4}{3} r^{4} \right) \omega^{4} 
\nn\\ &
+ \left( -\tfrac{60}{7} r^{6} + 108 r^{4} \mathit{r_0}^{2} + 48 r^{2} \mathit{r_0}^{4} + 48 \mathit{r_0}^{6} \right) \omega^{2}
+ 144 r^{4}
\Big) 
(r^{2} + \mathit{r_0}^{2})^{2} \phi(r)^{2} 
\nn\\ &
+ 2 \Big( 
\pi \mathit{r_0} r^{9} (r^{2} + \mathit{r_0}^{2}) \omega^{6} 
- \tfrac{36}{7} \Big( 
\tfrac{8 \pi \mathit{r_0}^{7}}{3} 
+ r (\pi^{2} - \tfrac{28}{3}) \mathit{r_0}^{6} 
+ \tfrac{5 r^{2} \pi \mathit{r_0}^{5}}{9} 
\nn\\ &
+ \tfrac{2 r^{3} (\pi^{2} - 28) \mathit{r_0}^{4}}{3} 
- \tfrac{85 r^{4} \pi \mathit{r_0}^{3}}{9} 
+ \tfrac{2 r^{5} (\pi^{2} - 14) \mathit{r_0}^{2}}{3} 
- \tfrac{52 r^{6} \pi \mathit{r_0}}{9} 
+ r^{7} \pi^{2}
\Big) r^{3} \omega^{4} 
\nn\\ &
+ \Big( 
(12 \pi^{2} - 96) \mathit{r_0}^{8} 
- 48 \pi r \mathit{r_0}^{7} 
+ 24 r^{2} (\pi^{2} - 8) \mathit{r_0}^{6} 
- \tfrac{960}{7} \pi r^{3} \mathit{r_0}^{5} 
- 24 r^{4} (\pi^{2} + 4) \mathit{r_0}^{4} 
\nn\\ &
- \tfrac{180}{7} r^{5} \pi \mathit{r_0}^{3} 
- \tfrac{576}{7} \pi^{2} r^{6} \mathit{r_0}^{2} 
- \tfrac{60}{7} \pi r^{7} \mathit{r_0} 
- \tfrac{324}{7} \pi^{2} r^{8}
\Big) \omega^{2} 
+ 144 \mathit{r_0} r^{5} \pi 
\Big) (r^{2} + \mathit{r_0}^{2}) \phi(r) 
\nn\\ &
+ \Big(
\pi \mathit{r_0} r^{10} (r^{2} + \mathit{r_0}^{2}) \omega^{6} 
- \tfrac{96}{7} r^{3} \Big( 
\pi^{2} \mathit{r_0}^{8} 
- \tfrac{35}{8} \pi r \mathit{r_0}^{7} 
+ r^{2} \left( \tfrac{41 \pi^{2}}{32} - \tfrac{21}{4} \right) \mathit{r_0}^{6} 
\nn\\ &
- \tfrac{61}{12} \pi r^{3} \mathit{r_0}^{5} 
+ \tfrac{5}{16} r^{4} \left( \pi^{2} - \tfrac{168}{5} \right) \mathit{r_0}^{4} 
- \tfrac{9}{2} r^{5} \pi \mathit{r_0}^{3} 
+ r^{6} \left( \tfrac{25 \pi^{2}}{32} - \tfrac{21}{4} \right) \mathit{r_0}^{2} 
\nn\\ &
- \tfrac{13}{6} \pi r^{7} \mathit{r_0} 
+ \tfrac{3}{4} \pi^{2} r^{8}
\Big) \omega^{4} 
\nn\\ &
+ \Big( 
-96 \pi \mathit{r_0}^{9} 
+ (-24 \pi^{2} + 192) r \mathit{r_0}^{8} 
- 240 r^{2} \pi \mathit{r_0}^{7} 
+ r^{3} \left( 720 - \tfrac{918 \pi^{2}}{7} \right) \mathit{r_0}^{6} 
\nn\\ &
- \tfrac{2208}{7} r^{4} \pi \mathit{r_0}^{5} 
+ r^{5} \left( 864 - \tfrac{1980 \pi^{2}}{7} \right) \mathit{r_0}^{4} 
- \tfrac{1116}{7} r^{6} \pi \mathit{r_0}^{3} 
+ r^{7} \left( 336 - \tfrac{1878 \pi^{2}}{7} \right) \mathit{r_0}^{2} 
\nn\\ &
- \tfrac{60}{7} \pi r^{8} \mathit{r_0} 
- \tfrac{648}{7} \pi^{2} r^{9}
\Big) \omega^{2} 
+ 144 r^{6} \mathit{r_0} \pi 
\Big) \mathit{r_0}
\Big) J^{2} \, ,
\end{align}
}
\vspace{6pt}
\begin{align}
\alpha_3 & = \Big(
r \left( 
r^{4} \omega^{4}
+ \left( \mathit{r_0}^{2} \omega^{4} + 22 \omega^{2} \right) r^{2}
+ 26 \mathit{r_0}^{2} \omega^{2}
- 24 
\right) (r^{2} + \mathit{r_0}^{2}) \phi(r) 
\nn\\ &
+ \mathit{r_0} \Big(
r^{6} \omega^{4}
+ \left( \mathit{r_0}^{2} \omega^{4} + 22 \omega^{2} \right) r^{4}
+ \left( 38 \mathit{r_0}^{2} \omega^{2} - 24 \right) r^{2}
+ 12 \mathit{r_0}^{4} \omega^{2}
\Big)
\Big) J \, ,
\end{align}\vspace{-16pt}
\begin{align}
\alpha_4 & = \mathrm{i} \Big(
-\Big(
-24 \omega^{4} \mathit{r_0}^{8}
+ 12 \Big( -\tfrac{27}{7} r^{2} \omega^{4} + 8 \omega^{2} \Big) \mathit{r_0}^{6}
\nn\\ &
+ r^{2} \omega^{2} \left( r^{4} \omega^{4} + 74 r^{2} \omega^{2} + \tfrac{1800}{7} \right) \mathit{r_0}^{4} 
+ 2 \left( r^{8} \omega^{6} + \tfrac{556}{7} r^{6} \omega^{4} - \tfrac{160}{7} r^{4} \omega^{2} \right) \mathit{r_0}^{2}
\nn\\ &
+ r^{4} \Big( r^{6} \omega^{6} + \tfrac{438}{7} r^{4} \omega^{4} - \tfrac{1048}{7} r^{2} \omega^{2} + 96 \Big)
\Big) 
\pi (r^{2} + \mathit{r_0}^{2})^{2}  J^{2} \phi(r)^{2} 
\nn\\ &
- 2 \Big(
-6  \omega^{4} J (\pi^{2} - 8) \mathit{r_0}^{10}
+ 24 \omega^{3} \pi \left( \tfrac{1}{24} r^{3} \omega^{2} + J \omega  + \tfrac{7}{6} r \right) r \mathit{r_0}^{9}
\nn\\ &
- \tfrac{78}{7} \Big( \omega^{2} \left( \pi^{2} - \tfrac{224}{13} \right) r^{2} - \tfrac{28 \pi^{2}}{13} + \tfrac{224}{13} \Big) \omega^{2}  J \mathit{r_0}^{8} 
\nn\\ &
+ \tfrac{878}{7} \omega \pi \left( \tfrac{21}{878} r^{5} \omega^{4} + J r^{2} \omega^{3}  + \tfrac{266}{439} r^{3} \omega^{2} - \tfrac{336}{439} J \omega  - \tfrac{56}{439} r \right) r \mathit{r_0}^{7} 
\nn\\ &
+ \tfrac{174}{7} \omega^{2}  \left( \omega^{2} \left( \pi^{2} + \tfrac{336}{29} \right) r^{2} + \tfrac{2 \pi^{2}}{29} - \tfrac{448}{29} \right) r^{2} J \mathit{r_0}^{6} 
\nn\\ &
+ r^{3} \omega \pi \left( \tfrac{1788}{7} J r^{2} \omega^{3}  + J r^{4} \omega^{5}  + 3 r^{5} \omega^{4} + 68 r^{3} \omega^{2} - 32 r - \tfrac{1916}{7} J \omega  \right) \mathit{r_0}^{5} 
\nn\\ &
+ \tfrac{486}{7} \omega^{2} \left( \omega^{2} \left( \pi^{2} + \tfrac{224}{81} \right) r^{2} - \tfrac{218 \pi^{2}}{81} - \tfrac{224}{81} \right)  r^{4} J \mathit{r_0}^{4} 
\nn\\ &
+ 2 \Big( J r^{6} \omega^{6}  + \tfrac{1}{2} r^{7} \omega^{5} + \tfrac{758}{7} J r^{4} \omega^{4}  + 10 r^{5} \omega^{3} 
\nn\\ &
- \tfrac{1230}{7} J r^{2} \omega^{2}  - 8 r^{3} \omega + 24 J  \Big) \pi r^{3} \mathit{r_0}^{3} 
+ \tfrac{348}{7} \left( \omega^{2} \left( \pi^{2} + \tfrac{28}{29} \right) r^{2} - \tfrac{153 \pi^{2}}{29} \right) \omega^{2}  r^{6} J \mathit{r_0}^{2} 
\nn\\ &
+ J r^{5}  \pi \left( r^{6} \omega^{6} + \tfrac{438}{7} r^{4} \omega^{4} - \tfrac{1048}{7} r^{2} \omega^{2} + 96 \right) \mathit{r_0} 
\nn\\ &
+ \tfrac{72}{7} \omega^{2} \pi^{2} \left( r^{2} \omega^{2} - \tfrac{19}{2} \right)  r^{8} J 
\Big) (r^{2} + \mathit{r_0}^{2})  J \phi(r) 
\nn\\ &
- \mathit{r_0} \Big(
4 r^{2} \omega^{4} \pi \mathit{r_0}^{13}
+ 48 \omega^{2} \pi \left( J^{2} \omega^{2}  + \tfrac{1}{3} r^{4} \omega^{2} - \tfrac{1}{3} r^{2} \right) \mathit{r_0}^{11}
+ 12 J^{2} r \omega^{4}  (\pi^{2} - 8) \mathit{r_0}^{10} 
\nn\\ &
+ \tfrac{1980}{7} \omega^{2} \pi \Big( \tfrac{7}{990} J r^{5} \omega^{3}  + J^{2} r^{2} \omega^{2}  + \tfrac{14}{165} r^{6} \omega^{2} + \tfrac{112}{495} J r^{3} \omega  - \tfrac{112}{165} J^{2}  - \tfrac{28}{165} r^{4} \Big) \mathit{r_0}^{9} 
\nn\\ &
+ \tfrac{717}{7} \omega^{2}  \left( \omega^{2} \left( \pi^{2} - \tfrac{728}{239} \right) r^{2} - \tfrac{112 \pi^{2}}{239} + \tfrac{896}{239} \right) r J^{2} \mathit{r_0}^{8} 
\nn\\ &
+ \tfrac{3784}{7} \omega \pi \Big( \tfrac{21}{1892} J r^{5} \omega^{4}  + J^{2} r^{2} \omega^{3}  + \tfrac{14}{473} r^{6} \omega^{3} + \tfrac{147}{473} J r^{3} \omega^{2}  - \tfrac{670}{473} J^{2} \omega  
\nn\\ &
- \tfrac{42}{473} r^{4} \omega - \tfrac{28}{473} J r  \Big) r^{2} \mathit{r_0}^{7} 
+ \tfrac{1791}{7} \left( \omega^{2} \left( \pi^{2} - \tfrac{280}{199} \right) r^{2} - \tfrac{308 \pi^{2}}{199} + \tfrac{2464}{597} \right) \omega^{2}  r^{3} J^{2} \mathit{r_0}^{6} 
\nn\\ &
+ \omega \pi r^{4} \Big( 4 r^{6} \omega^{3} + 6 J r^{5} \omega^{4}  + (J^{2} \omega^{5}  - 16 \omega) r^{4} + 144 J r^{3} \omega^{2}  + \tfrac{3622}{7} J^{2} r^{2} \omega^{3} 
\nn\\ &
- 64 J r  - \tfrac{7312}{7} J^{2} \omega  \Big) \mathit{r_0}^{5} 
+ 273 \omega^{2} \left( \omega^{2} \left( \pi^{2} - \tfrac{8}{13} \right) r^{2} - \tfrac{1968 \pi^{2}}{637} + \tfrac{320}{91} \right) r^{5} J^{2} \mathit{r_0}^{4} 
\nn\\ &
+ 2 \pi  r^{4} J \Big( J r^{6} \omega^{6}  + r^{7} \omega^{5} + \tfrac{960}{7} J r^{4} \omega^{4}  + 20 r^{5} \omega^{3} - \tfrac{2300}{7} J r^{2} \omega^{2}  - 16 r^{3} \omega + 48 J  \Big) \mathit{r_0}^{3} 
\nn\\ &
+ \tfrac{897}{7} \left( \omega^{2} \left( \pi^{2} - \tfrac{56}{299} \right) r^{2} - \tfrac{124 \pi^{2}}{23} + \tfrac{672}{299} \right) \omega^{2}  r^{7} J^{2} \mathit{r_0}^{2} 
\nn\\ &
+ J^{2} r^{6}  \pi \left( r^{6} \omega^{6} + \tfrac{438}{7} r^{4} \omega^{4} - \tfrac{1048}{7} r^{2} \omega^{2} + 96 \right) \mathit{r_0} 
\nn\\ &
+ \tfrac{144}{7} \omega^{2} \pi^{2} \left( r^{2} \omega^{2} - \tfrac{19}{2} \right)  r^{9} J^{2}
\Big)
\Big) T_{2,2}(r) \, ,
\end{align}
\vspace{6pt}
\begin{align}
\alpha_5 & = 198 \,\mathrm{i} \Big(
\left(r^{2}+\mathit{r_0}^{2}\right)^{3}  J \left( \frac{5 \mathit{r_0}^{4} \omega^{2}}{11} + \left( \frac{58 r^{2} \omega^{2}}{33} - \frac{6}{11} \right) \mathit{r_0}^{2} + r^{4} \omega^{2} - \frac{14 r^{2}}{11} \right) \phi(r)^{2} 
\nn\\ &
+ \left(r^{2}+\mathit{r_0}^{2}\right) \Big(
-\frac{2 \mathit{r_0}^{9} \omega}{33} 
+ \frac{5 J \pi \,\mathit{r_0}^{8} \omega^{2} }{11}
+ \frac{94}{33} \omega \left( J \omega  - \frac{3 r}{47} \right) r \mathit{r_0}^{7}
\nn\\ &
+ \frac{8}{3}  \left( r^{2} \omega^{2} - \frac{3}{44} \right) \pi  J \mathit{r_0}^{6}
+ \left( \frac{767}{99} J r^{3} \omega^{2}  - \frac{2}{11} r^{4} \omega - \frac{248}{99} J r  \right) \mathit{r_0}^{5}
\nn\\ &
+ \frac{164 \pi   r^{2} J \mathit{r_0}^{4}}{33} \left( r^{2} \omega^{2} - \frac{21}{82} \right) 
+ \left( \frac{683}{99} J r^{5} \omega^{2}  - \frac{2}{33} r^{6} \omega - \frac{520}{99} J r^{3}  \right) \mathit{r_0}^{3}
\nn\\ &
+ \frac{124}{33}  \left( r^{2} \omega^{2} - \frac{33}{62} \right) \pi  r^{4} J \mathit{r_0}^{2}
+ 2  \left( r^{2} \omega^{2} - \frac{14}{11} \right) r^{5} J \mathit{r_0}
+ J r^{6}  \pi \left( r^{2} \omega^{2} - \frac{10}{11} \right)
\Big) \phi(r) 
\nn\\ &
+ \Big(
\left( \frac{4}{99} r \omega + \frac{15}{11} J \omega^{2}  \right) \mathit{r_0}^{9}
+ \frac{47 J \pi r \mathit{r_0}^{8} \omega^{2} }{33}
+ \left( -\frac{104}{99} J  + \frac{2}{99} r^{3} \omega + \frac{52}{11} J r^{2} \omega^{2}  \right) \mathit{r_0}^{7} 
\nn\\ &
+ \frac{58}{11} \pi  \left( r^{2} \omega^{2} - \frac{13}{87} \right) r J \mathit{r_0}^{6}
+ \frac{644}{99} \left( J r^{2} \omega^{2}  - \frac{2}{161} r^{3} \omega - \frac{3}{7} J  \right) r^{2} \mathit{r_0}^{5} 
\nn\\ &
+ \frac{80}{11} \left( r^{2} \omega^{2} - \frac{41}{120} \right) \pi  r^{3} J \mathit{r_0}^{4}
+ \frac{410 }{99} \left( J r^{2} \omega^{2}  - \frac{3}{205} r^{3} \omega - \frac{34}{41} J  \right) r^{4} \mathit{r_0}^{3}
\nn\\ &
+ \frac{146}{33} \pi  r^{5} J \left( r^{2} \omega^{2} - \frac{43}{73} \right) \mathit{r_0}^{2}
+ J r^{6}  \left( r^{2} \omega^{2} - \frac{14}{11} \right) \mathit{r_0}
\nn\\ &
+ J r^{7}  \pi \Big( r^{2} \omega^{2} - \frac{10}{11} \Big)
\Big) \mathit{r_0}
\Big) \sqrt{7}\,  J \mathit{h0}_{3,2}(r) \, ,
\end{align}\vspace{-16pt}
\begin{align}
\alpha_6 & = 3 \Big( 
\left( \left( \left( r^{2} \omega^{4} + 60 \omega^{2} \right) \mathit{r_0}^{2} + r^{4} \omega^{4} + 16 r^{2} \omega^{2} - 24 \right) \left( r^{2} + \mathit{r_0}^{2} \right)^{4}  r J \phi(r)^{2} \right) 
\nn\\ &
+ 2 \left( r^{2} + \mathit{r_0}^{2} \right) \Big( 
r \omega^{3} \mathit{r_0}^{11} 
+ \left( 4 r^{3} \omega^{3} + 38 J \omega^{2}  \right) \mathit{r_0}^{9} 
+ 17 J \pi r \mathit{r_0}^{8} \omega^{2}  
\nn\\ &
+ \left( J r^{4} \omega^{4}  + 6 r^{5} \omega^{3} + 148 J r^{2} \omega^{2}  - 120 J  \right) \mathit{r_0}^{7}
+ 48 J \pi r^{3} \mathit{r_0}^{6} \omega^{2}  
\nn\\ &
+ \left( 3 J r^{6} \omega^{4}  + 4 r^{7} \omega^{3} + \frac{589}{3} J r^{4} \omega^{2}  - \frac{772}{3} J r^{2}  \right) \mathit{r_0}^{5}
+ 42 J \pi r^{5} \mathit{r_0}^{4} \omega^{2}  
\nn\\ &
+ \left( 3 J r^{8} \omega^{4}  + r^{9} \omega^{3} + \frac{307}{3} J r^{6} \omega^{2}  - \frac{464}{3} J r^{4}  \right) \mathit{r_0}^{3}
+ 8 J r^{7} \omega^{2}  \pi \mathit{r_0}^{2} 
\nn\\ &
+ J r^{6}  \left( r^{4} \omega^{4} + 16 r^{2} \omega^{2} - 24 \right) \mathit{r_0}
- 3 J r^{9} \omega^{2}  \pi 
\Big) \phi(r) 
\nn\\ &
+ \Big( \left( \frac{136}{3} \omega + 2 r^{2} \omega^{3} \right) \mathit{r_0}^{11}
+ 32 J \pi \mathit{r_0}^{10} \omega^{2} 
+ \left( -\frac{70}{3} J r \omega^{2}  + \frac{388}{3} r^{2} \omega + 8 r^{4} \omega^{3} \right) \mathit{r_0}^{9} 
\nn\\ &
+ 126 J \pi r^{2} \mathit{r_0}^{8} \omega^{2} 
+ r \left( J r^{4} \omega^{4}  + \frac{208}{3} J r^{2} \omega^{2}  - \frac{320}{3} J  + \frac{368}{3} r^{3} \omega + 12 r^{5} \omega^{3} \right) \mathit{r_0}^{7} 
\nn\\ &
+ 180 J \pi r^{4} \mathit{r_0}^{6} \omega^{2}  
+ 3 \left( J r^{4} \omega^{4}  + \frac{8}{3} r^{5} \omega^{3} + \frac{568}{9} J r^{2} \omega^{2}  + \frac{116}{9} r^{3} \omega - \frac{1072}{9} J  \right) r^{3} \mathit{r_0}^{5} 
\nn\\ &
+ 104 J \pi r^{6} \mathit{r_0}^{4} \omega^{2}  
+ \left( 3 J r^{9} \omega^{4}  + \frac{338}{3} J r^{7} \omega^{2}  - \frac{784}{3} J r^{5}  + 2 r^{10} \omega^{3} \right) \mathit{r_0}^{3} 
\nn\\ &
+ 12 J \pi r^{8} \mathit{r_0}^{2} \omega^{2}  
+ J r^{7}  \left( r^{4} \omega^{4} + 16 r^{2} \omega^{2} - 24 \right) \mathit{r_0} 
- 6 J \pi r^{10} \omega^{2}  
\Big) \mathit{r_0} 
\Big) \sqrt{7}\,  J \mathit{h1}_{3,2}(r) \, ,
\end{align}

\noindent
and
\vspace{6pt}
{\footnotesize
\begin{align}
\alpha_7 & = \mathrm{i} \Big( 
7 \pi \left( r^{2} + \mathit{r_0}^{2} \right)^{2} 
\Big( 
12 \mathit{r_0}^{6} \omega^{2} 
+ \Big( r^{4} \omega^{4} + \frac{206}{7} r^{2} \omega^{2} \Big) \mathit{r_0}^{4} 
\nn\\ &
+ r^{2} \left( r^{4} \omega^{4} - \frac{254}{7} r^{2} \omega^{2} - 24 \right) \mathit{r_0}^{2} 
- \frac{180 r^{6} \omega^{2}}{7}
\Big)  J^{2} \phi(r)^{2} 
\nn\\ &
- 180 \omega \left( r^{2} + \mathit{r_0}^{2} \right) \Big( 
-\frac{7 \mathit{r_0}^{9} r^{2} \omega^{2} \pi}{90} 
- \frac{7 J \omega  \left( \pi^{2} - 8 \right) \mathit{r_0}^{8}}{30} 
\nn\\ &
+ \frac{46 \pi r \mathit{r_0}^{7}}{45}  \left( -\frac{7}{46} r^{3} \omega^{2} + J \omega  + \frac{7}{23} r \right)
+ \frac{J r^{2} \omega  \left( \pi^{2} + 56 \right) \mathit{r_0}^{6}}{10} 
\nn\\ &
- \frac{7 }{90} \pi r^{3} \left( J r^{2} \omega^{3}  + r^{3} \omega^{2} - \frac{356}{7} J \omega  - 4 r \right) \mathit{r_0}^{5}
\nn\\ &
+ \frac{17}{6} \omega \left( \pi^{2} + \frac{168}{85} \right)  r^{4} J \mathit{r_0}^{4} 
- \frac{7}{90} \left( r^{2} \omega^{2} - \frac{500}{7} \right) \omega \pi  r^{5} J \mathit{r_0}^{3} 
\nn\\ &
+ \frac{7}{2} \omega  \left( \pi^{2} + \frac{8}{15} \right) r^{6} J \mathit{r_0}^{2} 
+ 2 J r^{7} \omega  \pi \mathit{r_0} 
+ J r^{8} \omega  \pi^{2} 
\Big)  J \phi(r) 
\nn\\ &
- 180 \Big( 
\frac{7 r^{2} \omega^{2} \pi \mathit{r_0}^{11}}{45} 
+ \frac{14}{15}  \omega^{2} \pi \left( -\frac{1}{12} J r^{3} \omega  + J^{2}  + \frac{1}{2} r^{4} \right) \mathit{r_0}^{9}
\nn\\ &
+ \frac{\omega^{2}   r J^{2} \mathit{r_0}^{8}}{12} \left( \pi^{2} - \frac{168}{5} \right)
+ \frac{77 \omega^{2}  r^{3} J^{2}  \mathit{r_0}^{6}}{30} \left( \pi^{2} - \frac{24}{11} \right)
\nn\\ &
+ \frac{346 \omega \pi  r^{2} \mathit{r_0}^{7}}{45} 
\left( -\frac{7}{346} J r^{3} \omega^{2}  + J^{2} \omega  + \frac{21}{346} r^{4} \omega + \frac{7}{173} J r  \right)
\nn\\ &
- \frac{7  \omega \pi r^{4} \mathit{r_0}^{5}}{180} 
\left( J^{2} r^{2} \omega^{3}  + 2 J r^{3} \omega^{2}  - \frac{1818}{7} J^{2} \omega  - 4 r^{4} \omega - 8 J r  \right)
\nn\\ &
+ \frac{353  \omega^{2}  r^{5} J^{2} \mathit{r_0}^{4}}{60} \left( \pi^{2} - \frac{168}{353} \right)
- \frac{7 J^{2} r^{4}  \pi \mathit{r_0}^{3}}{180} \left( r^{4} \omega^{4} - 134 r^{2} \omega^{2} + 24 \right) 
\nn\\ &
+ \frac{22 J^{2} r^{7} \omega^{2}  \pi^{2} \mathit{r_0}^{2}}{5} 
+ J^{2} r^{8} \omega^{2}  \pi \mathit{r_0} 
+ J^{2} r^{9} \omega^{2}  \pi^{2} 
\Big) \mathit{r_0}
\Big) {\phi \mathit{1}}_{2,2}(r) \, .
\end{align}}


Next, the equation for $T$ is given by
{\footnotesize
\begin{align}
\frac{d}{d r}T_{2,2}\! \left(r \right) & = 
\left(-\frac{\Pi_1}{4 \pi  r \,\mathit{r_0}^{5} \omega^{2} \left(r^{2}+\mathit{r_0}^{2}\right)^{4}}-\frac{\Pi_2}{2 \mathit{r_0}^{2} \omega \left(r^{2}+\mathit{r_0}^{2}\right)^{2}}-\frac{r \mathit{r_0}}{r^{2}+\mathit{r_0}^{2}}\right) \frac{d}{d r}{\phi \mathit{1}}_{2,2}\! \left(r \right)
\nn\\ &
+ 
\frac{\Pi_3}{8 \pi  \,\mathit{r_0}^{6} \omega^{3} \left(r^{2}+\mathit{r_0}^{2}\right)^{4}}
+\Big(-\frac{\Pi_4}{8 \pi  \,\mathit{r_0}^{6} \omega^{2} \left(r^{2}+\mathit{r_0}^{2}\right)^{4}}-\frac{\Pi_5}{4 \omega \,\mathit{r_0}^{3} \left(r^{2}+\mathit{r_0}^{2}\right)^{2}}
\nn\\ &
-\frac{\omega^{2} r}{2}\Big) T_{2,2}\! \left(r \right)
+ \frac{\Pi_6}{7 \omega \,\mathit{r_0}^{6} \left(r^{2}+\mathit{r_0}^{2}\right)^{5}}+\frac{\Pi_7}{7 \mathit{r_0}^{6} \omega^{2} \left(r^{2}+\mathit{r_0}^{2}\right)^{5}}
\nn\\ &
+\left(\frac{\Pi_8}{2 \pi  \,r^{2} \mathit{r_0}^{5} \omega^{2} \left(r^{2}+\mathit{r_0}^{2}\right)^{5}}+\frac{\Pi_9}{\omega \left(r^{2}+\mathit{r_0}^{2}\right)^{3}}+\frac{2 \mathit{r_0}^{3}}{\left(r^{2}+\mathit{r_0}^{2}\right)^{2}}\right) {\phi \mathit{1}}_{2,2}\! \left(r \right) \, ,   
\end{align}}

\noindent
where

{
{\footnotesize
\begin{align}
\Pi_1 & = \Big(
\left(r^{2}+\mathit{r_0}^{2}\right)^{2} \pi \Big(
12 \mathit{r_0}^{6} \omega^{2} + \frac{60}{7} \mathit{r_0}^{4} r^{2} \omega^{2} 
+ \mathit{r_0}^{2} r^{6} \omega^{4} + \frac{384}{7} \mathit{r_0}^{2} r^{4} \omega^{2} 
- 36 r^{2} \mathit{r_0}^{2} 
\nn\\ &
+ r^{8} \omega^{4}
+ \frac{240}{7} r^{6} \omega^{2} - 36 r^{4}
\Big) \phi(r)^{2} 
\nn\\ &
+ 2 \left(r^{2}+\mathit{r_0}^{2}\right) \Big(
3 \omega^{2} (\pi^{2}-8) \mathit{r_0}^{8} 
- \frac{180 r \pi \mathit{r_0}^{7} \omega^{2}}{7}
+ \frac{51}{7} r^{2} \left( \pi^{2} - \frac{56}{17} \right) \omega^{2} \mathit{r_0}^{6}
\nn\\ &
+ 36 \pi \left( -\frac{2}{7} r^{3} \omega^{2} - r \right) \mathit{r_0}^{5} 
+ \frac{159}{7} \left( \pi^{2} + \frac{56}{53} \right) r^{4} \omega^{2} \mathit{r_0}^{4}
\nn\\ &
+ r^{3} \pi \left( r^{4} \omega^{4} + \frac{376}{7} r^{2} \omega^{2} - 72 \right) \mathit{r_0}^{3}
+ \frac{165  r^{6} \omega^{2} \mathit{r_0}^{2}}{7} \left( \pi^{2} + \frac{56}{55} \right)
\nn\\ &
+ r^{5} \pi \left( r^{4} \omega^{4} + \frac{240}{7} r^{2} \omega^{2} - 36 \right) \mathit{r_0}
+ \frac{36 \pi^{2} r^{8} \omega^{2}}{7}
\Big) \phi(r) 
\nn\\ &
+ \Big(
- 24 \mathit{r_0}^{9} \pi \omega^{2}
- \frac{138}{7} r \omega^{2} \left( \pi^{2} - \frac{56}{23} \right) \mathit{r_0}^{8}
+ \frac{48 \mathit{r_0}^{7} r^{2} \pi \omega^{2}}{7}
\nn\\ &
- 15 r^{3} \left( \pi^{2} - \frac{72}{5} \right) \omega^{2} \mathit{r_0}^{6} 
+ 24 \pi \left( -3 r^{2} + \frac{11}{7} r^{4} \omega^{2} \right) \mathit{r_0}^{5}
\nn\\ &
+ \frac{276 r^{5}  \omega^{2} \mathit{r_0}^{4}}{7} \left( \pi^{2} + \frac{168}{23} \right)
+ \pi r^{4} \left( -108 + r^{4} \omega^{4} + \frac{416}{7} r^{2} \omega^{2} \right) \mathit{r_0}^{3} 
\nn\\ &
+ 45 r^{7} \omega^{2} \left( \pi^{2} + \frac{8}{3} \right) \mathit{r_0}^{2}
+ r^{6} \pi \left( r^{4} \omega^{4} + \frac{240}{7} r^{2} \omega^{2} - 36 \right) \mathit{r_0}
+ \frac{72 r^{9} \omega^{2} \pi^{2}}{7}
\Big) \mathit{r_0}
\Big) J^{2} \, ,
\end{align}}
\begin{equation}
\Pi_2 = \Big(
r \left(r^{2}+\mathit{r_0}^{2}\right) \left(r^{2} \omega^{2} + 6\right) \phi(r)
+ \mathit{r_0} \left( r^{4} \omega^{2} + 6 r^{2} + 12 \mathit{r_0}^{2} \right)
\Big) J \, ,
\end{equation}
\small{\begin{align}
\Pi_3 & = \mathrm{i} \mathit{H1}_{2,2}(r) \Big(
- \pi \left(r^{2}+\mathit{r_0}^{2}\right)^{2} 
\Big(
24 \omega^{4} \mathit{r_0}^{6}
+ \frac{12\left(-13 r^{2} \omega^{4}-24 \omega^{2}\right) \mathit{r_0}^{4}}{7}
\nn\\ &
+ \left(r^{2} \omega^{2}+6\right)\Big(r^{4} \omega^{4}-\frac{192}{7} r^{2} \omega^{2}+36\Big) \mathit{r_0}^{2} 
\nn\\ &
+ r^{2}\Big(r^{6} \omega^{6}+\frac{6}{7} r^{4} \omega^{4}-108 r^{2} \omega^{2}+216\Big)
\Big)
 J^{2} \phi(r)^{2} 
\nn\\ &
- 2\left(r^{2}+\mathit{r_0}^{2}\right)  
\Big(
6  \omega^{4} J (\pi^{2}-8) \mathit{r_0}^{8}
- \frac{360 \omega \pi  \mathit{r_0}^{7}}{7} \Big(-\frac{7}{360} r^{4} \omega^{4} + J r \omega^{3}  
\nn\\ &
-\frac{7}{15} r^{2} \omega^{2} + \frac{7}{10}\Big)
- \frac{36}{7}  \omega^{2} \left(\omega^{2}\left(\pi^{2}+\frac{28}{3}\right) r^{2} + 4 \pi^{2}\right)  J \mathit{r_0}^{6}
\nn\\ &
- \frac{750 \omega \pi  r \mathit{r_0}^{5}}{7} \left(\omega  \left(r^{2} \omega^{2}+\frac{6}{5}\right) J - \frac{7 r \left(r^{4} \omega^{4}+24 r^{2} \omega^{2}-36\right)}{375}\right)
\nn\\ &
- \frac{258}{7}  \omega^{2}  \left(\omega^{2}\left(\pi^{2}-\frac{56}{43}\right) r^{2} + \frac{162 \pi^{2}}{43}\right) r^{2} J \mathit{r_0}^{4}
\nn\\ &
+ \pi \Big(  \left(r^{6} \omega^{6} - \frac{356}{7} r^{4} \omega^{4} - \frac{1488}{7} r^{2} \omega^{2} + 216\right) J 
+ r^{3} \omega \left(r^{4} \omega^{4} + 24 r^{2} \omega^{2} - 36\right) \Big) r \mathit{r_0}^{3} 
\nn\\ &
- \frac{324}{7} \omega^{2} \left(\omega^{2}\left(\pi^{2}-\frac{28}{27}\right) r^{2} + \frac{44 \pi^{2}}{9}\right)  r^{4} J \mathit{r_0}^{2} 
\nn\\ &
+ J r^{3}  \pi \left(r^{6} \omega^{6} + \frac{6}{7} r^{4} \omega^{4} - 108 r^{2} \omega^{2} + 216\right) \mathit{r_0} 
\nn\\ &
- \frac{144}{7} \omega^{2} \pi^{2}  r^{6} J \left(r^{2} \omega^{2} + \frac{21}{4}\right)
\Big) J \phi(r) 
\nn\\ &
- \Big(
4 \omega^{2} \pi \left(r^{2} \omega^{2} + 6\right) \mathit{r_0}^{11}
- 48 \omega^{2} \pi \left(J^{2} \omega^{2}  - \frac{1}{4} r^{4} \omega^{2} - J r \omega  - \frac{3}{2} r^{2}\right) \mathit{r_0}^{9} 
\nn\\ &
- \frac{276 \omega^{4}   r J^{2} \mathit{r_0}^{8}}{7} \left(\pi^{2} - \frac{56}{23}\right)
- \frac{804 \omega \pi \mathit{r_0}^{7}}{7}  \Big(\omega  \left(r^{2} \omega^{2} - \frac{24}{67}\right) J^{2} 
\nn\\ &
- \frac{7}{402} r  \left(r^{4} \omega^{4} + 72 r^{2} \omega^{2} - 36\right) J 
- \frac{7 r^{4} \omega \left(r^{2} \omega^{2} + 6\right)}{67}\Big)
\nn\\ &
- \frac{879 \omega^{2}  r J^{2} \mathit{r_0}^{6}}{7} \left(\omega^{2}\left(\pi^{2} - \frac{840}{293}\right) r^{2} + \frac{426 \pi^{2}}{293} - \frac{336}{293}\right) 
\nn\\ &
- \frac{1164 \omega \pi  r^{2} \mathit{r_0}^{5}}{7} \Big(\omega  \left(r^{2} \omega^{2} + \frac{114}{97}\right) J^{2} - \frac{7 r  \left(r^{4} \omega^{4} + 36 r^{2} \omega^{2} - 36\right) J}{291} 
\nn\\ &
- \frac{7 r^{4} \omega \left(r^{2} \omega^{2} + 6\right)}{291}\Big)
- 174 \omega^{2}  r^{3} J^{2} \left(\omega^{2}\left(\pi^{2} - \frac{72}{29}\right) r^{2} + \frac{678 \pi^{2}}{203} - \frac{48}{29}\right) \mathit{r_0}^{4} 
\nn\\ &
+ \pi  \Big( \left(216 + r^{6} \omega^{6} - \frac{562}{7} r^{4} \omega^{4} - \frac{2076}{7} r^{2} \omega^{2}\right) J + 2 r^{7} \omega^{5} + 48 r^{5} \omega^{3} 
\nn\\ &
- 72 r^{3} \omega \Big) r^{2} J \mathit{r_0}^{3} 
- 129 \left(\omega^{2}\left(\pi^{2} - \frac{56}{43}\right) r^{2} + \frac{1434 \pi^{2}}{301} - \frac{48}{43}\right) \omega^{2}  r^{5} J^{2} \mathit{r_0}^{2} 
\nn\\ &
+ J^{2} r^{4}  \pi \left(r^{6} \omega^{6} + \frac{6}{7} r^{4} \omega^{4} - 108 r^{2} \omega^{2} + 216\right) \mathit{r_0}
\nn\\ &
- \frac{288}{7} \omega^{2} \pi^{2}  r^{7} J^{2} \Big(r^{2} \omega^{2} + \frac{21}{4}\Big)
\Big) \mathit{r_0}
\Big) \, ,
\end{align}}\vspace{6pt}
\begin{align}
\Pi_4 & = \Big(
\left(r^{2}+\mathit{r_0}^{2}\right)^{2} \pi \Big(
-\frac{156}{7} \omega^{4} \mathit{r_0}^{6} 
+ \mathit{r_0}^{4} \omega^{6} r^{4} 
+ \frac{120}{7} \mathit{r_0}^{4} r^{2} \omega^{4} 
+ \frac{1548}{7} \mathit{r_0}^{4} \omega^{2} 
\nn\\ &
+ 2 \mathit{r_0}^{2} r^{6} \omega^{6} 
+ \frac{512}{7} r^{4} \mathit{r_0}^{2} \omega^{4} 
+ \frac{1704}{7} r^{2} \mathit{r_0}^{2} \omega^{2} 
+ 144 \mathit{r_0}^{2} 
+ r^{8} \omega^{6} 
+ \frac{236}{7} r^{6} \omega^{4} 
\nn\\ &
+ \frac{444}{7} r^{4} \omega^{2} 
+ 144 r^{2} 
\Big)
r \, \phi(r)^{2} 
\nn\\ &
+ 2 \left(r^{2}+\mathit{r_0}^{2}\right)^{2} \Big(
-\frac{96 \omega^{4} \pi \mathit{r_0}^{7}}{7}
- \frac{36 }{7} r \left(\pi^{2} - \frac{28}{3}\right) \omega^{4} \mathit{r_0}^{6}
\nn\\ &
- \frac{20}{7}  \pi \omega^{2} \left(r^{2} \omega^{2} - \frac{153}{5}\right) \mathit{r_0}^{5}
- \frac{24}{7} r \omega^{2} \left(\omega^{2} \left(\pi^{2} - 28\right) r^{2} + \frac{9 \pi^{2}}{2}\right) \mathit{r_0}^{4}
\nn\\ &
+ \pi \left(r^{6} \omega^{6} + \frac{382}{7} r^{4} \omega^{4} + \frac{1056}{7} r^{2} \omega^{2} + 72\right) \mathit{r_0}^{3} 
\nn\\ &
- \frac{24}{7} r^{3} \left(\omega^{2} \left(\pi^{2} - 14\right) r^{2} + 24 \pi^{2}\right) \omega^{2} \mathit{r_0}^{2}
\nn\\ &
+ \pi r^{2} \left(r^{6} \omega^{6} + \frac{236}{7} r^{4} \omega^{4} + \frac{444}{7} r^{2} \omega^{2} + 144\right) \mathit{r_0} 
- \frac{36 r^{5} \omega^{2} \pi^{2} \left(r^{2} \omega^{2} + 9\right)}{7}
\Big) \phi(r) 
\nn\\ &
+ \mathit{r_0} \Big(
-\frac{96 \pi^{2} \mathit{r_0}^{10} \omega^{4}}{7}
+ 60 r \omega^{4} \pi \mathit{r_0}^{9}
+ 3 \Big(\omega^{4} \left(24 - \frac{73 \pi^{2}}{7}\right) r^{2} 
\nn\\ &
+ 4 \omega^{2} \Big(8 + \frac{9 \pi^{2}}{7}\Big)\Big) \mathit{r_0}^{8} 
+ \frac{908}{7} \left(r^{2} \omega^{2} + \frac{186}{227}\right) r \pi \omega^{2} \mathit{r_0}^{7}
\nn\\ &
- \frac{153}{7} r^{2} \omega^{2} \left(\omega^{2} \left(\pi^{2} - \frac{168}{17}\right) r^{2} + \frac{80 \pi^{2}}{17} - \frac{448}{51}\right) \mathit{r_0}^{6} 
\nn\\ &
+ r \pi \left( \frac{1628}{7} r^{2} \omega^{2} + \frac{976}{7} r^{4} \omega^{4} + r^{6} \omega^{6} + 144 \right) \mathit{r_0}^{5}
\nn\\ &
- 15 \left(\omega^{2} \left(\pi^{2} - \frac{72}{5}\right) r^{2} + \frac{804 \pi^{2}}{35} - \frac{32}{5}\right) r^{4} \omega^{2} \mathit{r_0}^{4} 
\nn\\ &
+ 2 r^{3} \pi \left( r^{6} \omega^{6} + \frac{362}{7} r^{4} \omega^{4} + \frac{648}{7} r^{2} \omega^{2} + 144 \right) \mathit{r_0}^{3} 
\nn\\ &
+ 3 \left( \left( -7 \pi^{2} + 24 \right) \omega^{4} r^{8} - \frac{744 r^{6} \omega^{2} \pi^{2}}{7} \right) \mathit{r_0}^{2}
\nn\\ &
+ r^{5} \pi \Big( r^{6} \omega^{6} + \frac{236}{7} r^{4} \omega^{4} + \frac{444}{7} r^{2} \omega^{2} 
+ 144 \Big) \mathit{r_0} 
- \frac{72}{7} r^{8} \omega^{2} \pi^{2} \left(r^{2} \omega^{2} + 9\right)
\Big)
\Big) J^{2} \, ,
\end{align}
\vspace{-16pt}
\begin{align}
\Pi_5 & =  \Big(
r \left( 
r^{4} \omega^{4} 
+ \left( \mathit{r_0}^{2} \omega^{4} + 26 \omega^{2} \right) r^{2} 
+ 30 \mathit{r_0}^{2} \omega^{2} 
- 24 
\right) \left( r^{2} + \mathit{r_0}^{2} \right) \phi(r) 
\nn\\ &
+ \mathit{r_0} \left( 
r^{6} \omega^{4} 
+ \left( \mathit{r_0}^{2} \omega^{4} + 26 \omega^{2} \right) r^{4} 
+ \left( 46 \mathit{r_0}^{2} \omega^{2} - 24 \right) r^{2} 
+ 12 \mathit{r_0}^{4} \omega^{2} 
- 24 \mathit{r_0}^{2} 
\right)
\Big) J \, ,
\end{align}
\vspace{-16pt}
\begin{align}
\Pi_6 & = 198 \Big( 
\left( r^{2} + \mathit{r_0}^{2} \right)^{3}  \left( \frac{5 \mathit{r_0}^{4} \omega^{2}}{11} + \left( \frac{58 r^{2} \omega^{2}}{33} - \frac{4}{11} \right) \mathit{r_0}^{2} + r^{4} \omega^{2} \right) r J \, \phi(r)^{2} 
\nn\\ &
+ \left( r^{2} + \mathit{r_0}^{2} \right)^{2} \Big( 
\frac{5 r \pi \omega^{2} \mathit{r_0}^{6}}{11} 
+ \left( \frac{94 r^{2} \omega^{2}}{33} - \frac{8}{11} \right) \mathit{r_0}^{5} 
+ \frac{73}{33} \left( r^{2} \omega^{2} - \frac{12}{73} \right) r \pi \mathit{r_0}^{4} 
\nn\\ &
+ \left( \frac{485}{99} r^{4} \omega^{2} - \frac{14}{33} r^{2} \right) \mathit{r_0}^{3}
+ \frac{91}{33} \left( r^{2} \omega^{2} - \frac{12}{91} \right) r^{3} \pi \mathit{r_0}^{2} 
+ 2 r^{6} \mathit{r_0} \omega^{2} 
+ r^{7} \pi \omega^{2} 
\Big)  J \, \phi(r) 
\nn\\ &
+ \Big( 
\frac{15 \omega r  \mathit{r_0}^{9}}{11} \left( J \omega  + \frac{2 r}{27} \right)
+ \frac{47 \pi  J  \mathit{r_0}^{8}}{33} \left( r^{2} \omega^{2} - \frac{12}{47} \right)
\nn\\ &
+ \frac{52 r \mathit{r_0}^{7}}{11}  \left( J r^{2} \omega^{2}  + \frac{5}{117} r^{3} \omega + \frac{3}{26} J  \right)
+ \frac{80 \pi   r^{4} J \mathit{r_0}^{4}}{11} \left( r^{2} \omega^{2} - \frac{3}{20} \right)
\nn\\ &
+ \frac{58 \pi   r^{2} J \mathit{r_0}^{6}}{11} \left( r^{2} \omega^{2} - \frac{6}{29} \right)
+ \frac{644  r^{3} \mathit{r_0}^{5}}{99} \left( J r^{2} \omega^{2}  + \frac{5}{322} r^{3} \omega + \frac{1}{14} J  \right)
\nn\\ &
+ \frac{410  r^{5} J \mathit{r_0}^{3}}{99} \left( r^{2} \omega^{2} - \frac{3}{205} \right) 
+ \frac{146 \pi   r^{6} J \mathit{r_0}^{2}}{33} \left( r^{2} \omega^{2} - \frac{6}{73} \right)
+ J r^{9} \mathit{r_0} \omega^{2}  
\nn\\ &
+ J \pi r^{10} \omega^{2}  
\Big) \mathit{r_0} 
\Big) \sqrt{7} \,  J \, \mathit{h0}_{3,2}(r) \, ,
\end{align}
\begin{align}
\Pi_7 &= 3 \,\mathrm{i} \sqrt{7} \Big(
\left( r^{2} + \mathit{r_0}^{2} \right)^{3}  J \Big( \left( r^{4} \omega^{4} + 46 r^{2} \omega^{2} + 24 \right) \mathit{r_0}^{2} 
\nn\\ &
+ r^{2} \left( r^{4} \omega^{4} + 6 r^{2} \omega^{2} + 144 \right) \Big) \phi(r)^{2} 
\nn\\ &
+ 2 \left( r^{2} + \mathit{r_0}^{2} \right) \Big(
\left( r^{2} \omega^{3} + 6 \omega \right) \mathit{r_0}^{9} 
+ \left( 3 r^{4} \omega^{3} + 36 J r \omega^{2}  + 18 r^{2} \omega \right) \mathit{r_0}^{7} 
\nn\\ &
+ 11 \pi  J \left( r^{2} \omega^{2} + \frac{30}{11} \right) \mathit{r_0}^{6} 
+ 13 \pi  r^{2} J \left( r^{2} \omega^{2} + \frac{150}{13} \right) \mathit{r_0}^{4} 
\nn\\ &
+ r \left( J r^{4} \omega^{4}  + 3 r^{5} \omega^{3} + 92 J r^{2} \omega^{2}  + 18 r^{3} \omega + 84 J  \right) \mathit{r_0}^{5}
\nn\\ &
+ \Big( 238 J r^{3}  + 2 J r^{7} \omega^{4}  + \frac{181}{3} J r^{5} \omega^{2}  + r^{8} \omega^{3} + 6 r^{6} \omega \Big) \mathit{r_0}^{3}
- 7 J r^{4}  \pi \left( r^{2} \omega^{2} - 30 \right) \mathit{r_0}^{2} 
\nn\\ &
+ J r^{5}  \left( r^{4} \omega^{4} + 6 r^{2} \omega^{2} + 144 \right) \mathit{r_0}
- 9 J r^{6}  \pi \left( r^{2} \omega^{2} - 10 \right)
\Big) \phi(r) 
\nn\\ &
+ \Big(
\left( 2 r^{3} \omega^{3} + 52 r \omega \right) \mathit{r_0}^{9}
+ 24 J \pi r \mathit{r_0}^{8} \omega^{2}  
\nn\\ &
+ \Big( 380 J  + \frac{328}{3} r^{3} \omega + 6 r^{5} \omega^{3} - \frac{106}{3} J r^{2} \omega^{2}  \Big) \mathit{r_0}^{7} 
\nn\\ &
+ 58 \pi  \left( r^{2} \omega^{2} + \frac{90}{29} \right) r J \mathit{r_0}^{6}
+ 26 \pi \left( r^{2} \omega^{2} + \frac{270}{13} \right)  r^{3} J \mathit{r_0}^{4}
\nn\\ &
+ r^{2} \Big( 484 J  + \frac{208}{3} r^{3} \omega + 6 r^{5} \omega^{3} + J r^{4} \omega^{4}  + \frac{188}{3} J r^{2} \omega^{2}  \Big) \mathit{r_0}^{5} 
\nn\\ &
+ 2 r^{4} \Big( J r^{4} \omega^{4}  + r^{5} \omega^{3} + \frac{103}{3} J r^{2} \omega^{2}  + 6 r^{3} \omega + 154 J  \Big) \mathit{r_0}^{3} 
\nn\\ &
- 26 \pi \left( r^{2} \omega^{2} - \frac{270}{13} \right)  r^{5} J \mathit{r_0}^{2}
+ J r^{6}  \left( r^{4} \omega^{4} + 6 r^{2} \omega^{2} + 144 \right) \mathit{r_0}
\nn\\ &
- 18 J r^{7}  \pi \left( r^{2} \omega^{2} - 10 \right)
\Big) \mathit{r_0}
\Big) J \, \mathit{h1}_{3,2}(r) \, ,
\end{align}
\vspace{-16pt}
{\small
\begin{align}
\Pi_8 &= 
\Big(
\left( r^{2} + \mathit{r_0}^{2} \right)^{2} \pi \Big( 
-12 \omega^{2} \mathit{r_0}^{8}
- \frac{384 \omega^{2} r^{2} \mathit{r_0}^{6}}{7}
+ r^{2} \Big( r^{4} \omega^{4} - \frac{300}{7} r^{2} \omega^{2} - 36 \Big) \mathit{r_0}^{4} 
\nn\\ &
+ r^{4} \Big( r^{4} \omega^{4} - \frac{276}{7} r^{2} \omega^{2} - 36 \Big) \mathit{r_0}^{2}
- \frac{108 r^{8} \omega^{2}}{7}
\Big) \phi(r)^{2} 
\nn\\ &
+ 2 
\Big( 
\left( -3 \pi^{2} + 24 \right) \mathit{r_0}^{10} 
+ 12 r \pi \mathit{r_0}^{9} 
+ \mathit{r_0}^{8} \left( -\frac{129 \pi^{2}}{7} + 120 \right) r^{2}
+ 42 r^{3} \pi \mathit{r_0}^{7} 
\nn\\ &
+ \mathit{r_0}^{6} \left( -\frac{279 \pi^{2}}{7} + 168 \right) r^{4} 
+ \left( r^{2} \omega^{2} + \frac{208}{7} \right) r^{5} \pi \mathit{r_0}^{5}
+ \mathit{r_0}^{4} \left( -\frac{393 \pi^{2}}{7} + 72 \right) r^{6}
\nn\\ &
+ r^{7} \pi \left( r^{2} \omega^{2} - \frac{138}{7} \right) \mathit{r_0}^{3} 
- \frac{276 r^{8} \pi^{2} \mathit{r_0}^{2}}{7}
- \frac{108 r^{9} \pi \mathit{r_0}}{7}
- \frac{54 r^{10} \pi^{2}}{7}
\Big) 
\left( r^{2} + \mathit{r_0}^{2} \right) \omega^{2} \phi(r) 
\nn\\ &
+ \Big(
24 \omega^{2} \pi \mathit{r_0}^{11}
+ 6 \omega^{2} r \left( \pi^{2} - 8 \right) \mathit{r_0}^{10}
+ \frac{852 \mathit{r_0}^{9} r^{2} \omega^{2} \pi}{7}
+ 27 r^{3} \omega^{2} \left( \pi^{2} - 8 \right) \mathit{r_0}^{8} 
\nn\\ &
+ \frac{792 r^{4} \pi \mathit{r_0}^{7} \omega^{2}}{7}
- \frac{30}{7} r^{5} \omega^{2} \left( \pi^{2} + \frac{504}{5} \right) \mathit{r_0}^{6}
+ r^{4} \pi \left( r^{4} \omega^{4} + \frac{200}{7} r^{2} \omega^{2} + 36 \right) \mathit{r_0}^{5} 
\nn\\ &
- 81 \left( \pi^{2} + \frac{136}{27} \right) r^{7} \omega^{2} \mathit{r_0}^{4}
+ \left( r^{4} \omega^{4} - \frac{144}{7} r^{2} \omega^{2} + 36 \right) r^{6} \pi \mathit{r_0}^{3} 
\nn\\ &
- \frac{498}{7} r^{9} \left( \pi^{2} + \frac{168}{83} \right) \omega^{2} \mathit{r_0}^{2}
- \frac{108 r^{10} \pi \mathit{r_0} \omega^{2}}{7}
- \frac{108 r^{11} \omega^{2} \pi^{2}}{7}
\Big) \mathit{r_0}
\Big) 
 J^{2} \, ,
\end{align}}

\noindent
and
{\small
\begin{equation}
\Pi_9 = J \left( \left( r^{2} + \mathit{r_0}^{2} \right) \left( r^{2} \omega^{2} + 6 \right) \phi(r) + r \mathit{r_0} \left( r^{2} \omega^{2} - 6 \right) \right) \, .
\end{equation}
}


As in the axial case, there are similarly two algebraic equations for $L$ and $N$ in the polar-led perturbations. 
$L$ is described by\vspace{6pt}
{\small
\begin{align}
L_{2,2}\! \left(r \right) &= 
\Big(\frac{\zeta_1}{4 \pi  \,r^{2} \mathit{r_0}^{5} \left(r^{2}+\mathit{r_0}^{2}\right)^{3}}+\frac{\zeta_2}{2 \left(r^{2}+\mathit{r_0}^{2}\right) \mathit{r_0}^{2}}+\mathit{r_0} \Big) 
\frac{d}{d r}{\phi \mathit{1}}_{2,2}\! \left(r \right)
\nn\\ &
+ 
\frac{\zeta_3}{8 \pi  \omega \left(r^{2}+\mathit{r_0}^{2}\right)^{3} \mathit{r_0}^{6}}+\Big(-1+\frac{\zeta_4}{8 \mathit{r_0}^{6} \pi  \,r^{2} \left(r^{2}+\mathit{r_0}^{2}\right)^{3}}+\frac{\zeta_5}{4 \left(r^{2}+\mathit{r_0}^{2}\right) \mathit{r_0}^{3}}
\nn\\ &
+\frac{\left(r^{2}+\mathit{r_0}^{2}\right) \omega^{2}}{2}\Big) T_{2,2}\! \left(r \right)
- \frac{\zeta_6}{7 \left(r^{2}+\mathit{r_0}^{2}\right)^{4} \omega \,\mathit{r_0}^{6}}-\frac{\zeta_7}{7 \left(r^{2}+\mathit{r_0}^{2}\right)^{4} \mathit{r_0}^{6} \omega^{2}}
\nn\\ &
+\Big(-\frac{\zeta_8}{2 \pi  r \,\mathit{r_0}^{5} \left(r^{2}+\mathit{r_0}^{2}\right)^{4}}-\frac{\zeta_9}{\left(r^{2}+\mathit{r_0}^{2}\right)^{2}}+\frac{2 r \mathit{r_0}}{r^{2}+\mathit{r_0}^{2}}\Big) {\phi \mathit{1}}_{2,2}\! \left(r \right) \, ,   
\end{align}
}
\noindent
where
{\footnotesize
\begin{align}
\zeta_1 &= \Big(
\left(r^{2}+\mathit{r_0}^{2}\right)^{2} \pi  
\left(-12 \mathit{r_0}^{6} + \frac{36}{7} r^{2} \mathit{r_0}^{4} + r^{6} \mathit{r_0}^{2} \omega^{2} + 90 r^{4} \mathit{r_0}^{2} + r^{8} \omega^{2} + \frac{342}{7} r^{6} \right) 
\phi\left( r \right)^2 
\nn\\ &
+ 2 \left(r^{2}+\mathit{r_0}^{2}\right) \Big(
3\left(-\pi^{2}+8\right) \mathit{r_0}^{8} + 12 \pi r \mathit{r_0}^{7} 
+ 3 r^{2} \left(24 + \frac{13 \pi^{2}}{7}\right) \mathit{r_0}^{6} 
+ \frac{558 \pi r^{3} \mathit{r_0}^{5}}{7} 
\nn\\ &
+ 9 r^{4} \left(8 + \frac{37 \pi^{2}}{7}\right) \mathit{r_0}^{4} 
+ r^{5} \pi \left(r^{2} \omega^{2} + \frac{844}{7}\right) \mathit{r_0}^{3} 
+ 3 r^{6} \left(8 + \frac{127 \pi^{2}}{7}\right) \mathit{r_0}^{2} 
\nn\\ &
+ r^{7} \pi \left(r^{2} \omega^{2} + \frac{342}{7}\right) \mathit{r_0} 
+ \frac{108 \pi^{2} r^{8}}{7} \Big) \phi\left( r \right) 
\nn\\ &
+ \Big( 
24 \pi \mathit{r_0}^{9} + 6 r \left(\pi^{2} - 8\right) \mathit{r_0}^{8} 
+ \frac{1200 r^{2} \pi \mathit{r_0}^{7}}{7} 
+ 3 r^{3} \left(-24 + \frac{193 \pi^{2}}{7}\right) \mathit{r_0}^{6} 
+ \frac{1668 r^{4} \pi \mathit{r_0}^{5}}{7} 
\nn\\ &
+ \frac{1248 \pi^{2} r^{5} \mathit{r_0}^{4}}{7} 
+ r^{6} \pi \left(r^{2} \omega^{2} + 158 \right) \mathit{r_0}^{3} 
+ 3 r^{7} \left(8 + \frac{309 \pi^{2}}{7}\right) \mathit{r_0}^{2} 
+ r^{8} \pi \left( r^{2} \omega^{2} + \frac{342}{7} \right) \mathit{r_0} 
\nn\\ &
+ \frac{216 r^{9} \pi^{2}}{7} \Big) \mathit{r_0}
\Big) J^{2} \, ,
\end{align}
}

\vspace{-6pt}
\begin{equation}
\zeta_2 = \omega J r^{2} \left( \mathit{r_0}^{2} \, \phi \left( r \right) + \phi \left( r \right) r^{2} + r \mathit{r_0} \right) \, ,
\end{equation}
\vspace{-6pt}
{\footnotesize\begin{align}
\zeta_3 &= \mathrm{i} \Big( 
\pi \left(r^{2}+\mathit{r_0}^{2}\right)^{2}  \Big(-\frac{180 \mathit{r_0}^{4} \omega^{2}}{7} + \Big( r^{4} \omega^{4} + \frac{96}{7} r^{2} \omega^{2} - 36 \Big) \mathit{r_0}^{2} 
\nn\\ &
+ r^{2} \Big( r^{4} \omega^{4} + \frac{108}{7} r^{2} \omega^{2} - \frac{396}{7} \Big) \Big) r J^{2} \phi(r)^{2} 
\nn\\ &
+ 2 \left(r^{2}+\mathit{r_0}^{2}\right)  \Big( 
- \frac{96}{7} \omega \pi \left( -\frac{7}{96} r^{3} \omega^{2} + J \omega  - \frac{21}{16} r \right) \mathit{r_0}^{7} 
\nn\\ &
- \frac{48 J \omega^{2} r  \left( \pi^{2} - 7 \right) \mathit{r_0}^{6}}{7} 
- \frac{120}{7} \pi \Big(  \Big( r^{2} \omega^{2} + \frac{33}{5} \Big) J - \frac{7 r^{3} \omega \left( r^{2} \omega^{2} + 18 \right)}{60} \Big) \mathit{r_0}^{5} 
\nn\\ &
- 12 \left( \omega^{2} \left( \pi^{2} - 8 \right) r^{2} + 3 \pi^{2} \right)  r J \mathit{r_0}^{4} 
\nn\\ &
+ \pi \left(  \left( r^{4} \omega^{4} + 16 r^{2} \omega^{2} - \frac{1020}{7} \right) J + r^{5} \omega^{3} + 18 r^{3} \omega \right) r^{2} \mathit{r_0}^{3} 
\nn\\ &
- \frac{108}{7}  \left( \omega^{2} \left( \pi^{2} - \frac{28}{9} \right) r^{2} + \frac{16 \pi^{2}}{3} \right)  r^{3} J \mathit{r_0}^{2}
\nn\\ &
+ J r^{4}  \pi \left( r^{4} \omega^{4} + \frac{108}{7} r^{2} \omega^{2} - \frac{396}{7} \right) \mathit{r_0}
- \frac{72}{7} \pi^{2} \left( r^{2} \omega^{2} + \frac{9}{2} \right)  r^{5} J 
\Big) J \phi(r) 
\nn\\ &
+ \Big( 
4 \pi r \mathit{r_0}^{11} \omega^{2} + 24 \omega \pi \left( \frac{r^{3} \omega}{2} + J  \right) \mathit{r_0}^{9} 
- \frac{96 J^{2} \omega^{2}  \pi^{2} \mathit{r_0}^{8}}{7} 
\nn\\ &
+ \frac{348  \omega \pi r \mathit{r_0}^{7}}{7} \left( J^{2} \omega  + \frac{7 r  \left( r^{2} \omega^{2} + 42 \right) J}{174} + \frac{7 r^{4} \omega}{29} \right)
\nn\\ &
- \frac{195   J^{2} \mathit{r_0}^{6}}{7} \left( \omega^{2} \left( \pi^{2} - \frac{168}{65} \right) r^{2} + \frac{138 \pi^{2}}{65} - \frac{336}{65} \right)
\nn\\ &
+ \frac{240 \pi  r \mathit{r_0}^{5}}{7} \left(  \left( r^{2} \omega^{2} - \frac{36}{5} \right) J^{2} + \frac{7 r^{3} \omega  \left( r^{2} \omega^{2} + 24 \right) J}{60} + \frac{7 r^{6} \omega^{2}}{60} \right)
\nn\\ &
- \frac{246   r^{2} J^{2} \mathit{r_0}^{4}}{7} \left( \omega^{2} \left( \pi^{2} - \frac{168}{41} \right) r^{2} + 6 \pi^{2} - \frac{336}{41} \right)
\nn\\ &
+ \pi \Big(  \left( -\frac{1788}{7} + r^{4} \omega^{4} + \frac{128}{7} r^{2} \omega^{2} \right) J 
+ 2 r^{5} \omega^{3} + 36 r^{3} \omega \Big)  r^{3} J \mathit{r_0}^{3} 
\nn\\ &
- \frac{291   r^{4} J^{2} \mathit{r_0}^{2}}{7} \left( \omega^{2} \left( \pi^{2} - \frac{168}{97} \right) r^{2} + \frac{570 \pi^{2}}{97} - \frac{336}{97} \right)
\nn\\ &
+ J^{2} r^{5}  \pi \left( r^{4} \omega^{4} + \frac{108}{7} r^{2} \omega^{2} - \frac{396}{7} \right) \mathit{r_0} 
- \frac{144}{7}  \pi^{2} \Big( r^{2} \omega^{2} + \frac{9}{2} \Big)  r^{6} J^{2}
\Big) \mathit{r_0} 
\Big) \mathit{H1}_{2,2}(r) \, ,
\end{align}}
\begin{align}
\zeta_4 &= \Big( 
\pi \Big( 
r^{10} \omega^{4} 
+ \left( 2 \mathit{r_0}^{2} \omega^{4} + \frac{338}{7} \omega^{2} \right) r^{8} 
+ \left( \omega^{4} \mathit{r_0}^{4} + \frac{860}{7} \mathit{r_0}^{2} \omega^{2} - \frac{432}{7} \right) r^{6} 
\nn\\ &
+ \left( \frac{342}{7} \mathit{r_0}^{4} \omega^{2} - \frac{216}{7} \mathit{r_0}^{2} \right) r^{4} 
+ \left( -\frac{348}{7} \mathit{r_0}^{6} \omega^{2} + 120 \mathit{r_0}^{4} \right) r^{2} 
- 24 \omega^{2} \mathit{r_0}^{8} + 48 \mathit{r_0}^{6} 
\Big) 
\nn\\ & \quad \times
\left( r^{2} + \mathit{r_0}^{2} \right)^{2} \phi(r)^{2} 
\nn\\ &
+ 2 \left( r^{2} + \mathit{r_0}^{2} \right)^{2} \Big( 
r^{9} \omega^{4} \pi \mathit{r_0} + \frac{36 \pi^{2} r^{8} \omega^{2}}{7} 
+ \omega^{2} \pi \mathit{r_0} \left( \mathit{r_0}^{2} \omega^{2} + \frac{338}{7} \right) r^{7} 
\nn\\ &
+ \left( \mathit{r_0}^{2} \left( 48 + \frac{192 \pi^{2}}{7} \right) \omega^{2} - \frac{468 \pi^{2}}{7} \right) r^{6} 
\nn\\ &
+ \frac{850}{7} \pi \left( \mathit{r_0}^{2} \omega^{2} - \frac{216}{425} \right) \mathit{r_0} r^{5} 
+ \left( \mathit{r_0}^{4} \left( 144 + \frac{150 \pi^{2}}{7} \right) \omega^{2} - \frac{864 \pi^{2} \mathit{r_0}^{2}}{7} \right) r^{4} 
\nn\\ &
+ \frac{610}{7} \pi \left( \mathit{r_0}^{2} \omega^{2} - \frac{426}{305} \right) \mathit{r_0}^{3} r^{3} 
- \frac{48}{7} \left( \mathit{r_0}^{2} \left( \pi^{2} - 21 \right) \omega^{2} + \frac{7 \pi^{2}}{2} + 14 \right) \mathit{r_0}^{4} r^{2} 
\nn\\ &
+ 24 \pi \mathit{r_0}^{5} \left( \mathit{r_0}^{2} \omega^{2} - 2 \right) r 
- 6 \mathit{r_0}^{6} \left( \pi^{2} - 8 \right) \left( \mathit{r_0}^{2} \omega^{2} - 2 \right) 
\Big) \phi(r) 
\nn\\ &
+ \mathit{r_0} \Big( 
\pi \mathit{r_0} r^{12} \omega^{4} + \frac{72 r^{11} \omega^{2} \pi^{2}}{7} 
+ 2 \pi \left( \mathit{r_0}^{2} \omega^{2} + \frac{169}{7} \right) \mathit{r_0} \omega^{2} r^{10} 
\nn\\ &
+ \left( \left( 87 \pi^{2} - 24 \right) \mathit{r_0}^{2} \omega^{2} - \frac{936 \pi^{2}}{7} \right) r^{9} 
+ \pi \mathit{r_0} \left( \omega^{4} \mathit{r_0}^{4} - \frac{432}{7} + \frac{1516}{7} \mathit{r_0}^{2} \omega^{2} \right) r^{8} 
\nn\\ &
+ \left( \mathit{r_0}^{4} \left( -168 + \frac{1479 \pi^{2}}{7} \right) \omega^{2} + \mathit{r_0}^{2} \left( -\frac{3372 \pi^{2}}{7} + 96 \right) \right) r^{7} 
\nn\\ &
+ \left( \frac{3070}{7} \pi \mathit{r_0}^{5} \omega^{2} - 336 \pi \mathit{r_0}^{3} \right) r^{6} 
+ \left( -592 \pi \mathit{r_0}^{5} + \frac{3464}{7} \omega^{2} \pi \mathit{r_0}^{7} \right) r^{4} 
\nn\\ &
+ \frac{1503}{7} \left( \mathit{r_0}^{2} \left( \pi^{2} - \frac{280}{167} \right) \omega^{2} - \frac{456 \pi^{2}}{167} + \frac{896}{501} \right) \mathit{r_0}^{4} r^{5} 
\nn\\ &
+ \frac{645}{7} \left( \mathit{r_0}^{2} \left( \pi^{2} - \frac{728}{215} \right) \omega^{2} - \frac{612 \pi^{2}}{215} + \frac{224}{43} \right) \mathit{r_0}^{6} r^{3} 
\nn\\ &
+ \frac{1908}{7} \pi \left( \mathit{r_0}^{2} \omega^{2} - \frac{244}{159} \right) \mathit{r_0}^{7} r^{2} 
+ 12 \mathit{r_0}^{8} \left( \pi^{2} - 8 \right) \left( \mathit{r_0}^{2} \omega^{2} - 2 \right) r 
\nn\\ &
+ 48 \pi \mathit{r_0}^{9} \left( \mathit{r_0}^{2} \omega^{2} - 2 \right) 
\Big) 
\Big)  J^{2} \, ,
\end{align}
\vspace{-16pt}
\begin{align}
\zeta_5 & =  J \Big( \left( r^{4} \omega^{2} 
+ \left( \mathit{r_0}^{2} \omega^{2} + 20 \right) r^{2} 
+ 24 \mathit{r_0}^{2} \right) \left( r^{2} + \mathit{r_0}^{2} \right) \phi\left( r \right) 
\nn\\ &
+ \mathit{r_0} \left( r^{4} \omega^{2} 
+ \left( \mathit{r_0}^{2} \omega^{2} + 20 \right) r^{2} 
+ 28 \mathit{r_0}^{2} \right) r \Big) \omega \, ,
\end{align}
\vspace{-16pt}
\begin{align}
\zeta_6 &= 198 \sqrt{7}\,  J \bigg(
\omega^{2} \left(r^{2}+\mathit{r_0}^{2}\right)^{3}  \left(r^{4} + \frac{58}{33} r^{2} \mathit{r_0}^{2} + \frac{5}{11} \mathit{r_0}^{4}\right) J \phi\left(r\right)^{2} 
\nn\\ &
+ \omega^{2} \left(r^{2}+\mathit{r_0}^{2}\right)^{2} \Big(
r^{6} \pi + \frac{91}{33} r^{4} \pi \mathit{r_0}^{2} + \frac{73}{33} r^{2} \pi \mathit{r_0}^{4} + \frac{5}{11} \pi \mathit{r_0}^{6}
+ 2 r^{5} \mathit{r_0} + \frac{485}{99} r^{3} \mathit{r_0}^{3} 
\nn\\ &
+ \frac{94}{33} r \mathit{r_0}^{5} 
\Big)  J \phi\left(r\right) 
\nn\\ &
+ \bigg(
\frac{15}{11} \omega \left( J \omega  + \frac{2 r}{27} \right) \mathit{r_0}^{9}
+ \frac{47 J \pi r \mathit{r_0}^{8} \omega^{2} }{33}
+ \left( \frac{52}{11} J r^{2} \omega^{2}  + \frac{4}{33} J  + \frac{20}{99} r^{3} \omega \right) \mathit{r_0}^{7} 
\nn\\ &
+ \frac{58 J \pi r^{3} \mathit{r_0}^{6} \omega^{2} }{11}
+ \frac{644}{99} r^{2} \left( J r^{2} \omega^{2}  + \frac{5}{322} r^{3} \omega + \frac{5}{322} J  \right) \mathit{r_0}^{5}
+ \frac{80 J \pi r^{5} \mathit{r_0}^{4} \omega^{2} }{11} 
\nn\\ &
+ \frac{410 J r^{6} \mathit{r_0}^{3} \omega^{2} }{99}
+ \frac{146 J r^{7} \omega^{2}  \pi \mathit{r_0}^{2}}{33}
+ J r^{8} \mathit{r_0} \omega^{2}  
+ J r^{9} \omega^{2}  \pi
\bigg) \mathit{r_0}
\bigg) \mathit{h0}_{3,2}\left(r\right) \, ,
\end{align}
\vspace{6pt}
\begin{align}
\zeta_7 &= 3 \,\mathrm{i} \sqrt{7} \Big(
\omega^{2} \left(r^{2}+\mathit{r_0}^{2}\right)^{3}  r J \left( \left(r^{2} \omega^{2} + 40 \right) \mathit{r_0}^{2} + r^{4} \omega^{2} \right) \phi\left(r\right)^{2} 
\nn\\ &
+ 2 \left(r^{2}+\mathit{r_0}^{2}\right) \Big(
r \omega^{3} \mathit{r_0}^{9} 
+ \left(3 r^{3} \omega^{3} + 30 J \omega^{2}  \right) \mathit{r_0}^{7}
+ 11 J r \omega^{2}  \pi \mathit{r_0}^{6} 
\nn\\ &
+ \left(J r^{4} \omega^{4}  + 3 r^{5} \omega^{3} + 74 J r^{2} \omega^{2}  - 120 J  \right) \mathit{r_0}^{5}
+ 13 J r^{3} \omega^{2}  \pi \mathit{r_0}^{4} 
\nn\\ &
+ \left(2 J r^{6} \omega^{4}  + r^{7} \omega^{3} + \frac{127}{3} J r^{4} \omega^{2}  - 100 J r^{2}  \right) \mathit{r_0}^{3}
- 7 J r^{5} \pi \omega^{2}  \mathit{r_0}^{2} 
\nn\\ &
+ J r^{8} \omega^{4}  \mathit{r_0}
- 9 J r^{7} \omega^{2}  \pi
\Big) \phi\left(r\right) 
\nn\\ &
+ \mathit{r_0} \Big(
\left( 2 r^{2} \omega^{3} + 40 \omega \right) \mathit{r_0}^{9}
+ 24 J \pi \mathit{r_0}^{8} \omega^{2}  
+ \left(-\frac{142}{3} J r \omega^{2}  + \frac{220}{3} r^{2} \omega + 6 r^{4} \omega^{3} \right) \mathit{r_0}^{7} 
\nn\\ &
+ 58 J \pi r^{2} \mathit{r_0}^{6} \omega^{2}  
+ r \left(-240 J  + J r^{4} \omega^{4}  + \frac{98}{3} J r^{2} \omega^{2}  + \frac{100}{3} r^{3} \omega + 6 r^{5} \omega^{3} \right) \mathit{r_0}^{5} 
\nn\\ &
+ 26 J \pi r^{4} \mathit{r_0}^{4} \omega^{2}  
+ 2 r^{3} \left( J r^{4} \omega^{4}  + r^{5} \omega^{3} + \frac{67}{3} J r^{2} \omega^{2}  - 100 J  \right) \mathit{r_0}^{3} 
\nn\\ &
- 26 J \pi r^{6} \mathit{r_0}^{2} \omega^{2}  
+ J r^{9} \mathit{r_0} \omega^{4}  
- 18 J \pi r^{8} \omega^{2}  
\Big)
\Big) J \mathit{h1}_{3,2}\left(r\right) \, ,
\end{align}
\vspace{-16pt}
\begin{align}
\zeta_8 &= \bigg(
\Big(
12 \mathit{r_0}^{6} + \left( r^{4} \omega^{2} + \frac{174}{7} r^{2} \right) \mathit{r_0}^{4} + \left( r^{6} \omega^{2} - \frac{186}{7} r^{4} \right) \mathit{r_0}^{2} 
\nn\\ &
- \frac{108 r^{6}}{7}
\Big) \left(r^{2} + \mathit{r_0}^{2}\right)^{2} \pi \, \phi(r)^2 
+ 2 \left( r^{2} + \mathit{r_0}^{2} \right) \bigg(
\left( 3 \pi^{2} - 24 \right) \mathit{r_0}^{8}
- \frac{120 \pi r \mathit{r_0}^{7}}{7}
\nn\\ &
+ \frac{3 r^{2} \left( \pi^{2} - 168 \right) \mathit{r_0}^{6}}{7}
+ r^{3} \pi \left( r^{2} \omega^{2} - \frac{368}{7} \right) \mathit{r_0}^{5} 
+ \mathit{r_0}^{4} \left( -\frac{195 \pi^{2}}{7} - 72 \right) r^{4}
\nn\\ &
+ \left( r^{2} \omega^{2} - \frac{384}{7} \right) r^{5} \pi \mathit{r_0}^{3}
+ \left( -33 \pi^{2} - 24 \right) \mathit{r_0}^{2} r^{6}
- \frac{108 \pi r^{7} \mathit{r_0}}{7}
\nn\\ &
- \frac{54 \pi^{2} r^{8}}{7}
\bigg) \phi(r) 
+ \bigg(
-24 \pi \mathit{r_0}^{9}
+ \left( -\frac{15 \pi^{2}}{7} + 72 \right) \mathit{r_0}^{8} r
- \frac{1392 \pi r^{2} \mathit{r_0}^{7}}{7}
\nn\\ &
+ \Big( -\frac{390 \pi^{2}}{7} 
+ 144 \Big) \mathit{r_0}^{6} r^{3} 
+ \left( r^{2} \omega^{2} - \frac{1786}{7} \right) r^{4} \pi \mathit{r_0}^{5}
+ \left( -\frac{843 \pi^{2}}{7} + 72 \right) \mathit{r_0}^{4} r^{5}
\nn\\ &
+ r^{6} \pi \left( r^{2} \omega^{2} - 114 \right) \mathit{r_0}^{3} 
- \frac{576 \pi^{2} r^{7} \mathit{r_0}^{2}}{7}
- \frac{108 \pi r^{8} \mathit{r_0}}{7}
- \frac{108 \pi^{2} r^{9}}{7}
\bigg) \mathit{r_0}
\bigg) J^{2} \, ,
\end{align}

\noindent
and

\begin{equation}
\zeta_9 = 
\left( \left( r^{2} + \mathit{r_0}^{2} \right) \phi \left( r \right) + r \mathit{r_0} \right) r \omega J \, .
\end{equation}


The perturbation function $N$ is given by
\begin{align}
N_{2,2}\! \left(r \right) &= 
\left(\frac{\beta_1}{4 \pi  \,r^{2} \mathit{r_0}^{5} \left(r^{2}+\mathit{r_0}^{2}\right)^{3}}+\frac{\beta_2}{2 \left(r^{2}+\mathit{r_0}^{2}\right) \mathit{r_0}^{2}}+\mathit{r_0} \right) \frac{d}{d r}{\phi \mathit{1}}_{2,2}\! \left(r \right)
\nn\\ &
+ \frac{\beta_3}{8 \pi  \omega \left(r^{2}+\mathit{r_0}^{2}\right)^{3} \mathit{r_0}^{6} r}
\nn\\ &
+\left(-1+\frac{\beta_4}{8 \left(r^{2}+\mathit{r_0}^{2}\right)^{3} \pi  \,\mathit{r_0}^{6}}+\frac{\beta_5}{4 \left(r^{2}+\mathit{r_0}^{2}\right) \mathit{r_0}^{3}}+\frac{\left(r^{2}+\mathit{r_0}^{2}\right) \omega^{2}}{2}\right) T_{2,2}\! \left(r \right)
\nn\\ &
- \frac{\beta_6}{7 \left(r^{2}+\mathit{r_0}^{2}\right)^{4} \omega \,\mathit{r_0}^{6}}-\frac{\beta_7}{7 \left(r^{2}+\mathit{r_0}^{2}\right)^{4} \mathit{r_0}^{6} \omega^{2}}
\nn\\ &
+\left(-\frac{\beta_8}{2 \pi  r \,\mathit{r_0}^{5} \left(r^{2}+\mathit{r_0}^{2}\right)^{4}}-\frac{\beta_9}{\left(r^{2}+\mathit{r_0}^{2}\right)^{2}}+\frac{2 r \mathit{r_0}}{r^{2}+\mathit{r_0}^{2}}\right) {\phi \mathit{1}}_{2,2}\! \left(r \right) \, ,  
\end{align}

\noindent
where
\vspace{6pt}
{\footnotesize\begin{align}
\beta_1 &= \Big( 
\left(r^{2}+\mathit{r_0}^{2}\right)^{2} \pi \left(
12 \mathit{r_0}^{6} + \frac{324}{7} r^{2} \mathit{r_0}^{4} + r^{6} \mathit{r_0}^{2} \omega^{2} + \frac{606}{7} r^{4} \mathit{r_0}^{2} 
+ r^{8} \omega^{2} + \frac{198}{7} r^{6}
\right) \phi(r)^2 
\nn\\ &
+ 2 \left(r^{2}+\mathit{r_0}^{2}\right) \Big(
3 \left(\pi^{2}-8\right) \mathit{r_0}^{8} - 12 \pi r \mathit{r_0}^{7} 
+ 3 r^{2} \left(-24 + \frac{47 \pi^{2}}{7}\right) \mathit{r_0}^{6}
+ \frac{134 \pi r^{3} \mathit{r_0}^{5}}{7} 
\nn\\ &
+ 3 r^{4} \left(-24 + \frac{113 \pi^{2}}{7}\right) \mathit{r_0}^{4} 
+ r^{5} \pi \left(r^{2} \omega^{2} + \frac{444}{7}\right) \mathit{r_0}^{3}
+ 3 r^{6} \left(-8 + \frac{85 \pi^{2}}{7}\right) \mathit{r_0}^{2} 
\nn\\ &
+ r^{7} \pi \left(r^{2} \omega^{2} + \frac{198}{7}\right) \mathit{r_0} 
+ \frac{36 \pi^{2} r^{8}}{7}
\Big) \phi(r) 
\nn\\ &
+ \Big( 
-24 \pi \mathit{r_0}^{9} 
+ 6 \left(-\pi^{2} + 8\right) r \mathit{r_0}^{8}
- \frac{120 r^{2} \pi \mathit{r_0}^{7}}{7}
+ 3 r^{3} \left(72 + \frac{61 \pi^{2}}{7}\right) \mathit{r_0}^{6}
+ \frac{52 r^{4} \pi \mathit{r_0}^{5}}{7}
\nn\\ &
+ 12 r^{5} \left(24 + \frac{47 \pi^{2}}{7}\right) \mathit{r_0}^{4}
+ r^{6} \pi \left(\frac{330}{7} + r^{2} \omega^{2}\right) \mathit{r_0}^{3}
+ 3 r^{7} \left(40 + \frac{137 \pi^{2}}{7}\right) \mathit{r_0}^{2}
\nn\\ &
+ \pi r^{8} \left(r^{2} \omega^{2} + \frac{198}{7}\right) \mathit{r_0}
+ \frac{72 r^{9} \pi^{2}}{7}
\Big) \mathit{r_0}
\Big) J^{2} \, ,
\end{align}}
\vspace{-10pt}
\begin{equation}
\beta_2 = 
\omega J r^{2} \left( 
\mathit{r_0}^{2} \phi(r) + r^{2} \phi(r) + r \mathit{r_0} 
\right) \, ,
\end{equation}
\vspace{-16pt}
{\small\begin{align}
\beta_3 &= \mathrm{i} \Big(
\pi \left(r^{2}+\mathit{r_0}^{2}\right)^{2}  J^{2} \Big(
r^{8} \omega^{4} + \left( \mathit{r_0}^{2} \omega^{4} - \frac{36}{7} \omega^{2} \right) r^{6} 
+ \frac{36 \left( 2 \mathit{r_0}^{2} \omega^{2} - 11 \right) r^{4} }{7} 
\nn\\ &
+ 36 \left( \frac{3}{7} \mathit{r_0}^{4} \omega^{2} - \mathit{r_0}^{2} \right) r^{2} 
+ 24 \mathit{r_0}^{6} \omega^{2} \Big) \phi(r)^2
\nn\\ &
+ 2 \left(r^{2}+\mathit{r_0}^{2}\right)  J \Big(
J \pi r^{9} \mathit{r_0} \omega^{4}  
+ \left(-\frac{144}{7} \pi^{2}  \omega^{2} J + \omega^{3} \pi \mathit{r_0}^{3} \right) r^{8}
\nn\\ &
+ J \omega^{2}  \pi \mathit{r_0} \left( \mathit{r_0}^{2} \omega^{2} - \frac{36}{7} \right) r^{7}
\nn\\ &
+ 2 \Big( -\frac{117}{7} J \omega^{2}  \pi^{2} \mathit{r_0}^{2} + \omega^{3} \pi \mathit{r_0}^{5} + 9 \omega \pi \mathit{r_0}^{3} - \frac{162}{7} \pi^{2}  J \Big) r^{6}
\nn\\ &
- \frac{288}{7} \pi  \left( \mathit{r_0}^{2} \omega^{2} + \frac{11}{8} \right) \mathit{r_0} J r^{5} 
\nn\\ &
- \frac{78}{7} \left( -\frac{7 \omega^{3} \pi \mathit{r_0}^{5}}{78} - \frac{42 \omega \pi \mathit{r_0}^{3}}{13} + J \omega^{2}  \left( \pi^{2} + \frac{56}{13} \right) \mathit{r_0}^{2} + \frac{96 \pi^{2}  J}{13} \right) \mathit{r_0}^{2} r^{4}
\nn\\ &
- \frac{544}{7} \pi  \mathit{r_0}^{3} J \left( \mathit{r_0}^{2} \omega^{2} + \frac{171}{136} \right) r^{3}
+ \frac{54}{7} \Big( \frac{7 \omega \pi \mathit{r_0}^{3}}{3} + J \omega^{2}  \left( \pi^{2} - \frac{112}{9} \right) \mathit{r_0}^{2} 
\nn\\ &
- \frac{14 \pi^{2}  J}{3} \Big) \mathit{r_0}^{4} r^{2}
- \frac{264}{7} \pi  \left( \mathit{r_0}^{2} \omega^{2} + \frac{19}{11} \right) \mathit{r_0}^{5} J r 
+ 6 J \omega^{2}  \left( \pi^{2} - 8 \right) \mathit{r_0}^{8}
\Big) \phi(r)
\nn\\ &
+ \mathit{r_0} \Big(
J^{2} r^{10} \omega^{4}  \pi \mathit{r_0}
+ 2 \left( J \omega^{3}  \pi \mathit{r_0}^{3} - \frac{144}{7} J^{2} \omega^{2}  \pi^{2} \right) r^{9}
\nn\\ &
+ \pi \mathit{r_0} \omega^{2} \left( 4 \mathit{r_0}^{4} + J^{2} \omega^{2}  \mathit{r_0}^{2} - \frac{36}{7} J^{2}  \right) r^{8}
\nn\\ &
- \frac{807 }{7} \left( -\frac{28 \omega^{3} \pi \mathit{r_0}^{5}}{807} - \frac{84 \omega \pi \mathit{r_0}^{3}}{269} + J \omega^{2}  \left( \pi^{2} - \frac{392}{269} \right) \mathit{r_0}^{2} + \frac{216 \pi^{2}  J}{269} \right) J r^{7}
\nn\\ &
- \frac{648 \pi \mathit{r_0}  r^{6}}{7} \left( J^{2} \omega^{2}  \mathit{r_0}^{2} - \frac{7}{54} \mathit{r_0}^{6} \omega^{2} + \frac{11}{18} J^{2}  \right)
\nn\\ &
- \frac{930}{7} \Big( -\frac{7 \omega^{3} \pi \mathit{r_0}^{5}}{465} - \frac{112 \omega \pi \mathit{r_0}^{3}}{155} + J \omega^{2}  \left( \pi^{2} - \frac{504}{155} \right) \mathit{r_0}^{2} 
\nn\\ &
+ \frac{57}{31}  J \Big( \pi^{2} - \frac{56}{95} \Big) \Big)  \mathit{r_0}^{2} J r^{5}
- \frac{1376}{7} \pi \left( J^{2} \omega^{2}  \mathit{r_0}^{2} - \frac{21}{344} \mathit{r_0}^{6} \omega^{2} + \frac{279}{344} J^{2}  \right) \mathit{r_0}^{3} r^{4} 
\nn\\ &
- \frac{591 }{7} \left( -\frac{196 \omega \pi \mathit{r_0}^{3}}{197} + J \omega^{2}  \left( \pi^{2} - \frac{840}{197} \right) \mathit{r_0}^{2} 
+ \frac{492}{197} \left( \pi^{2} - \frac{56}{41} \right)  J \right) \mathit{r_0}^{4} J r^{3}
\nn\\ &
- \frac{972 \pi \mathit{r_0}^{5}  r^{2}}{7}  \left( J^{2} \omega^{2}  \mathit{r_0}^{2} - \frac{7}{243} \mathit{r_0}^{6} \omega^{2} + \frac{88}{81} J^{2}  \right)
\nn\\ &
- \frac{180}{7} \left( -\frac{14 \omega \pi \mathit{r_0}^{3}}{15} + J \omega^{2}  \left( \pi^{2} - \frac{56}{15} \right) \mathit{r_0}^{2} + \frac{23}{10} \left( \pi^{2} - \frac{56}{23} \right)  J \right)  \mathit{r_0}^{6} J r
\nn\\ &
- 48 J^{2} \omega^{2} \pi \mathit{r_0}^{9}
\Big)
\Big) \mathit{H1}_{2,2}(r) \, ,
\end{align}}\vspace{6pt}
\begin{align}
\beta_4 & = \Big(
\pi \left(r^{2}+\mathit{r_0}^{2}\right)^{2} \Big(
r^{8} \omega^{4} 
+ \left(2 \mathit{r_0}^{2} \omega^{4} + \frac{194}{7} \omega^{2}\right) r^{6}
\nn\\ &
+ \left( \omega^{4} \mathit{r_0}^{4} + \frac{692}{7} \mathit{r_0}^{2} \omega^{2} - \frac{384}{7} \right) r^{4}
\nn\\ &
+ \left( \frac{606}{7} \mathit{r_0}^{4} \omega^{2} - \frac{648}{7} \mathit{r_0}^{2} \right) r^{2}
+ \frac{108}{7} \mathit{r_0}^{6} \omega^{2} + \frac{24}{7} \mathit{r_0}^{4}
\Big) \phi(r)^2
\nn\\ &
+ 2 \left(r^{2}+\mathit{r_0}^{2}\right)^{2} \Big(
\pi r^{7} \mathit{r_0} \omega^{4} 
- \frac{36}{7} r^{6} \omega^{2} \pi^{2} 
+ \omega^{2} \pi \mathit{r_0} \left( \mathit{r_0}^{2} \omega^{2} + \frac{194}{7} \right) r^{5} 
\nn\\ &
+ \frac{66}{7} \pi^{2} \left( \mathit{r_0}^{2} \omega^{2} - \frac{54}{11} \right) r^{4}
+ \frac{450}{7} \pi \left( \mathit{r_0}^{2} \omega^{2} - \frac{64}{75} \right) \mathit{r_0} r^{3} 
\nn\\ &
+ \left( \frac{156}{7} \mathit{r_0}^{4} \pi^{2} \omega^{2} + \left( -108 \pi^{2} + 96 \right) \mathit{r_0}^{2} \right) r^{2}
+ \frac{186}{7} \pi \mathit{r_0}^{3} \left( \mathit{r_0}^{2} \omega^{2} - \frac{78}{31} \right) r
\nn\\ &
+ \frac{54}{7} \left( \omega^{2} \pi^{2} \mathit{r_0}^{2} - \frac{16}{3} \pi^{2} + \frac{112}{9} \right) \mathit{r_0}^{4}
\Big) \phi(r)
\nn\\ &
+ \mathit{r_0} \Big(
r^{10} \omega^{4} \pi \mathit{r_0}
- \frac{72}{7} r^{9} \omega^{2} \pi^{2}
+ 2 \pi \left( \mathit{r_0}^{2} \omega^{2} + \frac{97}{7} \right) \mathit{r_0} \omega^{2} r^{8}
\nn\\ &
+ \left( \mathit{r_0}^{2} \left( -\frac{51}{7} \pi^{2} + 72 \right) \omega^{2} - \frac{648}{7} \pi^{2} \right) r^{7}
\nn\\ &
+ \pi \mathit{r_0} \left( \omega^{4} \mathit{r_0}^{4} + \frac{596}{7} \mathit{r_0}^{2} \omega^{2} - \frac{384}{7} \right) r^{6}
\nn\\ &
+ \frac{279}{7} \left( \mathit{r_0}^{2} \left( \pi^{2} + \frac{168}{31} \right) \omega^{2} - \frac{260}{31} \pi^{2} - \frac{224}{93} \right) \mathit{r_0}^{2} r^{5}
\nn\\ &
+ \frac{678}{7} \pi \mathit{r_0}^{3} \left( \mathit{r_0}^{2} \omega^{2} - \frac{176}{113} \right) r^{4}
+ \frac{423}{7} \left( \mathit{r_0}^{2} \left( \pi^{2} + \frac{168}{47} \right) \omega^{2} - \frac{304}{47} \pi^{2} - \frac{448}{141} \right) \mathit{r_0}^{4} r^{3}
\nn\\ &
+ \frac{528}{7} \pi \left( \mathit{r_0}^{2} \omega^{2} - \frac{24}{11} \right) \mathit{r_0}^{5} r^{2}
+ \frac{165}{7} \mathit{r_0}^{6} \left( \mathit{r_0}^{2} \left( \pi^{2} + \frac{168}{55} \right) \omega^{2} - \frac{348}{55} \pi^{2} - \frac{224}{55} \right) r
\nn\\ &
+ 36 \pi \mathit{r_0}^{7} \Big( \mathit{r_0}^{2} \omega^{2} - \frac{128}{63} \Big)
\Big) \mathit{r_0}
\Big) J^{2} \, ,
\end{align}
\vspace{-16pt}
\begin{align}
\beta_5 &=  J \omega \Big(
\left( 
r^{4} \omega^{2} + \left( \mathit{r_0}^{2} \omega^{2} + 12 \right) r^{2} + 16 \mathit{r_0}^{2}
\right) \left( r^{2} + \mathit{r_0}^{2} \right) \phi(r)
\nn\\ &
+ \mathit{r_0} r \left( 
r^{4} \omega^{2} + \left( \mathit{r_0}^{2} \omega^{2} + 12 \right) r^{2} + 20 \mathit{r_0}^{2}
\right)
\Big) \, ,
\end{align}
\vspace{-16pt}
\begin{align}
\beta_6 & = 18 \sqrt{7}  \Big(
\omega^{2} \left(r^{2} + \mathit{r_0}^{2}\right)^{3} \left( r^{4} + 6 r^{2} \mathit{r_0}^{2} + \frac{5}{3} \mathit{r_0}^{4} \right)  J \, \phi(r)^2 
\nn\\ &
+ \omega \left(r^{2} + \mathit{r_0}^{2}\right)^{2} \Big(
- \frac{4 \mathit{r_0}^{7}}{3} 
+ \frac{5 J \omega  \pi \mathit{r_0}^{6}}{3} 
+ \frac{2 \left(17 J r \omega  - 4 r^{2}\right) \mathit{r_0}^{5}}{3} 
+ \frac{23 J r^{2} \omega  \pi \mathit{r_0}^{4}}{3} 
\nn\\ &
+ \frac{\mathit{r_0}^{3}}{3} \left( \frac{125}{3} J r^{3} \omega  - 4 r^{4} \right) 
+ 7 J r^{4} \omega  \pi \mathit{r_0}^{2} 
+ 2 J r^{5} \omega  \mathit{r_0} 
+ J r^{6} \omega  \pi 
\Big) \phi(r)
\nn\\ &
+ \Big(
\Big( 5 J \omega^{2}  - \frac{2}{9} r \omega \Big) \mathit{r_0}^{9} 
+ \frac{17 J \pi r \mathit{r_0}^{8} \omega^{2} }{3} 
+ \frac{2}{3} \Big( 23 J r^{2} \omega^{2}  + \frac{4}{3} J  - \frac{8}{3} r^{3} \omega \Big) \mathit{r_0}^{7} 
\nn\\ &
+ 18 J \pi r^{3} \mathit{r_0}^{6} \omega^{2}  
+ \frac{2  \mathit{r_0}^{5}}{3} \left( J r^{2}  + \frac{82}{3} J r^{4} \omega^{2}  - \frac{13}{3} r^{5} \omega \right)
+ 20 J \pi r^{5} \mathit{r_0}^{4} \omega^{2}  
\nn\\ &
+ \frac{80 \omega r^{6}  \mathit{r_0}^{3}}{9} \left( J \omega  - \frac{3 r}{20} \right)
+ \frac{26 J r^{7} \omega^{2}  \pi \mathit{r_0}^{2}}{3} 
+ J r^{8} \mathit{r_0} \omega^{2}  
+ J r^{9} \omega^{2}  \pi 
\Big) \mathit{r_0}
\Big) J \mathit{h0}_{3,2}(r) \, ,
\end{align}
\vspace{6pt}
\begin{align}
\beta_7 &= 3 \, \mathrm{i} \sqrt{7} \,  J \Big(
\omega^{2} \left(r^{2}+\mathit{r_0}^{2}\right)^{3}  \left( \left(r^{2} \omega^{2} + 16\right) \mathit{r_0}^{2} + r^{4} \omega^{2} - 24 r^{2} \right) r J \, \phi(r)^{2}
\nn\\ &
+ 2 \left(r^{2}+\mathit{r_0}^{2}\right) \Big(
r \omega^{3} \mathit{r_0}^{9}
+ \left( 3 r^{3} \omega^{3} + 14 J \omega^{2}  \right) \mathit{r_0}^{7}
- J r \omega^{2}  \pi \mathit{r_0}^{6}
\nn\\ &
+ \left( J r^{4} \omega^{4}  + 3 r^{5} \omega^{3} + 18 J r^{2} \omega^{2}  - 80 J  \right) \mathit{r_0}^{5}
- 23 J r^{3} \omega^{2}  \pi \mathit{r_0}^{4}
\nn\\ &
+ \left( 2 J r^{6} \omega^{4}  + r^{7} \omega^{3} - \frac{65}{3} J r^{4} \omega^{2}  - 60 J r^{2}  \right) \mathit{r_0}^{3}
- 43 J r^{5} \omega^{2}  \pi \mathit{r_0}^{2}
\nn\\ &
+ J r^{6} \omega^{2}  \left(r^{2} \omega^{2} - 24\right) \mathit{r_0}
- 21 J r^{7} \omega^{2}  \pi
\Big) \phi(r)
\nn\\ &
+ \Big(
\Big( 2 r^{2} \omega^{3} + \frac{80}{3} \omega \Big) \mathit{r_0}^{9}
+ 8 J \pi \mathit{r_0}^{8} \omega^{2} 
+ \Big( -\frac{46}{3} J r \omega^{2}  + \frac{140}{3} r^{2} \omega + 6 r^{4} \omega^{3} \Big) \mathit{r_0}^{7}
\nn\\ &
- 14 J \pi r^{2} \mathit{r_0}^{6} \omega^{2} 
+ r \left( -160 J  + J r^{4} \omega^{4}  + \frac{26}{3} J r^{2} \omega^{2}  + 20 r^{3} \omega + 6 r^{5} \omega^{3} \right) \mathit{r_0}^{5}
\nn\\ &
- 94 J \pi r^{4} \mathit{r_0}^{4} \omega^{2} 
+ 2 r^{3} \left( J r^{4} \omega^{4}  + r^{5} \omega^{3} - \frac{53}{3} J r^{2} \omega^{2}  - 60 J  \right) \mathit{r_0}^{3}
\nn\\ &
- 114 J \pi r^{6} \mathit{r_0}^{2} \omega^{2} 
+ J r^{7} \omega^{2}  \left(r^{2} \omega^{2} - 24\right) \mathit{r_0}
- 42 J \pi r^{8} \omega^{2} 
\Big) \mathit{r_0}
\Big) \mathit{h1}_{3,2}(r) \, ,
\end{align}
\vspace{-16pt}
\begin{align}
\beta_8 & = \Big(
\left(r^{2}+\mathit{r_0}^{2}\right)^{2} \pi \Big(
-12 \mathit{r_0}^{6} 
+ \left( r^{4} \omega^{2} - \frac{114}{7} r^{2} \right) \mathit{r_0}^{4}
+ \left( r^{6} \omega^{2} - \frac{162}{7} r^{4} \right) \mathit{r_0}^{2}
\nn\\ &
+ \frac{36 r^{6}}{7}
\Big) \phi(r)^{2}
+ 2 \left(r^{2}+\mathit{r_0}^{2}\right) \Big(
\left(-3 \pi^{2} + 24\right) \mathit{r_0}^{8}
+ \frac{104 \pi r \mathit{r_0}^{7}}{7}
\nn\\ &
+ \left(-\frac{99 \pi^{2}}{7} + 72\right) \mathit{r_0}^{6} r^{2}
+ r^{3} \pi \left(r^{2} \omega^{2} + 24\right) \mathit{r_0}^{5}
+ \left(-\frac{201 \pi^{2}}{7} + 72\right) \mathit{r_0}^{4} r^{4}
\nn\\ &
+ \pi r^{5} \Big( r^{2} \omega^{2} + \frac{72}{7} \Big) \mathit{r_0}^{3}
+ \left( -15 \pi^{2} + 24 \right) \mathit{r_0}^{2} r^{6}
+ \frac{36 \pi r^{7} \mathit{r_0}}{7}
+ \frac{18 \pi^{2} r^{8}}{7}
\Big) \phi(r)
\nn\\ &
+ \Big(
24 \pi \mathit{r_0}^{9}
+ \Big( \frac{69 \pi^{2}}{7} - 24 \Big) \mathit{r_0}^{8} r
+ \frac{184 \pi r^{2} \mathit{r_0}^{7}}{7}
+ \frac{6 r^{3} \left( \pi^{2} - 168 \right) \mathit{r_0}^{6}}{7}
\nn\\ &
+ \left( r^{2} \omega^{2} + \frac{342}{7} \right) \pi r^{4} \mathit{r_0}^{5}
+ \left( -\frac{159 \pi^{2}}{7} - 216 \right) \mathit{r_0}^{4} r^{5}
+ \left( r^{2} \omega^{2} + \frac{234}{7} \right) \pi r^{6} \mathit{r_0}^{3}
\nn\\ &
+ \Big( -\frac{60 \pi^{2}}{7} - 96 \Big) \mathit{r_0}^{2} r^{7}
+ \frac{36 \pi r^{8} \mathit{r_0}}{7}
+ \frac{36 \pi^{2} r^{9}}{7}
\Big) \mathit{r_0}
\Big)  J^{2} \, ,
\end{align}

\noindent
and

\begin{equation}
\beta_9 = \left( 
\left( r^{2} + \mathit{r_0}^{2} \right) \phi(r) 
+ r \mathit{r_0} 
\right) r \omega J \, .
\end{equation}


Lastly, the differential equations for the axial metric functions $\mathit{h0}$ and $\mathit{h1}$ are rather simple.
For $\mathit{h0}$, we obtain
\begin{align}
\frac{d}{d r}\mathit{h0}_{3,2}\! \left(r \right) & = 
\frac{J  r \sqrt{7} }{7 \mathit{r_0}^{2} \left(r^{2}+\mathit{r_0}^{2}\right)} 
\left(\left(r^{2}+\mathit{r_0}^{2}\right) \phi \! \left(r \right)+r \mathit{r_0} \right)
\frac{d}{d r}{\phi \mathit{1}}_{2,2}\! \left(r \right)
\nn\\ &
+ \frac{\mathrm{i} \left(r^{2} \omega^{2}+6\right) \sqrt{7}\, J  }{14 \left(r^{2}+\mathit{r_0}^{2}\right) \omega \,\mathit{r_0}^{3}}
\left(\left(r^{2}+\mathit{r_0}^{2}\right) \phi \! \left(r \right)+r \mathit{r_0} \right) \mathit{H1}_{2,2}\! \left(r \right)
\nn\\ &
+  
\frac{\sqrt{7}\, J  }{14 \left(r^{2}+\mathit{r_0}^{2}\right) \mathit{r_0}^{3}}
(r \,\omega^{2} (r^{2}+\mathit{r_0}^{2})^{2} \phi \! \left(r \right)+(r^{2} \omega^{2}+4) \mathit{r_0}^{3}+\mathit{r_0} \,r^{4} \omega^{2}) T_{2,2}\! \left(r \right)
\nn\\ &
+ 
\left(\frac{2 J }{\omega \left(r^{2}+\mathit{r_0}^{2}\right)^{2}}+\frac{2 r}{r^{2}+\mathit{r_0}^{2}}\right) \mathit{h0}_{3,2}\! \left(r \right) 
\nn\\ &
+\frac{\lambda_1}{\left(r^{2}+\mathit{r_0}^{2}\right)^{2} \omega^{2} \mathit{r_0}^{3}}-\frac{2 \left(\left(r^{2}+\mathit{r_0}^{2}\right) \phi \! \left(r \right)+r \mathit{r_0} \right) \sqrt{7}\, J  }{7 \left(r^{2}+\mathit{r_0}^{2}\right)^{2}}  {\phi \mathit{1}}_{2,2}\! \left(r \right) \, , 
\end{align}

\noindent
where
\begin{align}
\lambda_1 &= 6 \,\mathrm{i} \Big( 
- \left( r^{2} \omega^{2} + \mathit{r_0}^{2} \omega^{2} + 10 \right)  \left( r^{2} + \mathit{r_0}^{2} \right) J \, \phi(r) 
- \mathit{r_0} \Big( 
\frac{\omega^{3} \mathit{r_0}^{6}}{6} 
+ \frac{(r^{2} \omega^{3} - 5 \omega) \mathit{r_0}^{4}}{3} 
\nn\\ &
+ r \omega \Big( \frac{r^{3} \omega^{2}}{6} + J \omega  - \frac{5r}{3} \Big) \mathit{r_0}^{2} 
+ J r  \left( r^{2} \omega^{2} + 10 \right) 
\Big) 
\Big) \mathit{h1}_{3,2}(r) \, ,
\end{align}
and for $\mathit{h1}$, we obtain
\begin{align}
\frac{d}{d r}\mathit{h1}_{3,2}\! \left(r \right) & = 
-\frac{\mathrm{i}  \omega \sqrt{7}\, J  }{7 \mathit{r_0}^{3}}
\left(\left(r^{2}+\mathit{r_0}^{2}\right) \phi \! \left(r \right)+r \mathit{r_0} \right)
T_{2,2}\! \left(r \right)
\nn\\ &
+ \frac{ \mathrm{i}}{\left(r^{2}+\mathit{r_0}^{2}\right) \mathit{r_0}^{3}}
\left(-6  \left(r^{2}+\mathit{r_0}^{2}\right) J \phi \! \left(r \right)-r^{2} \mathit{r_0}^{3} \omega -\mathit{r_0}^{5} \omega -6 J r \mathit{r_0}  \right)
\mathit{h0}_{3,2}\! \left(r \right) 
\nn\\ &
+\frac{10 J  }{\left(r^{2}+\mathit{r_0}^{2}\right)^{2} \omega} \,  \mathit{h1}_{3,2}\! \left(r \right) \, .
\end{align}

Here, compared to the axial perturbation sector, 
the polar perturbation sector shows an opposite dependence on the order of rotation.
In all the ODEs of the polar metric functions
$\mathit{H1}(r),T(r),L(r),N(r)$, and of the scalar function $\phi\mathit{1}(r),$
up to second-order terms from rotation enter the equations.
On the other hand, equations for the axial metric functions
$\mathit{h0}(r)$ and $\mathit{h1}(r)$ only have up to the first-order terms in rotation.

\begin{adjustwidth}{-\extralength}{0cm}

\reftitle{References}





\PublishersNote{}
\end{adjustwidth}
\end{document}